\documentclass{elsarticle}

\usepackage[reviewcopy]{adndt}
\usepackage{longtable}

\usepackage{amsmath}

\usepackage{amssymb}

\biboptions{square,sort&compress}
\bibpunct[]{[}{]}{,}{n}{}{;}
\begin{document}

\begin{frontmatter}

\journal{Atomic Data and Nuclear Data Tables}


\title{Discovery of palladium, antimony, tellurium, iodine, and xenon isotopes}

\author{J. Kathawa}
\author{C. Fry}
\author{M. Thoennessen\corref{cor1}}\ead{thoennessen@nscl.msu.edu}

 \cortext[cor1]{Corresponding author.}

 \address{National Superconducting Cyclotron Laboratory and \\ Department of Physics and Astronomy, Michigan State University, \\ East Lansing, MI 48824, USA}

\begin{abstract}
Currently, thirty-eight palladium, thirty-eight antimony, thirty-nine tellurium, thirty-eight iodine, and forty xenon isotopes have been observed and the discovery of these isotopes is discussed here. For each isotope a brief synopsis of the first refereed publication, including the production and identification method, is presented.
\end{abstract}

\end{frontmatter}





\newpage
\tableofcontents
\listofDtables

\vskip5pc

\section{Introduction}\label{s:intro}

The discovery of palladium, antimony, tellurium, iodine, and xenon isotopes is discussed as part of the series summarizing the discovery of isotopes, beginning with the cerium isotopes in 2009 \cite{2009Gin01}. Guidelines for assigning credit for discovery are (1) clear identification, either through decay-curves and relationships to other known isotopes, particle or $\gamma$-ray spectra, or unique mass and Z-identification, and (2) publication of the discovery in a refereed journal. The authors and year of the first publication, the laboratory where the isotopes were produced as well as the production and identification methods are discussed. When appropriate, references to conference proceedings, internal reports, and theses are included. When a discovery includes a half-life measurement the measured value is compared to the currently adopted value taken from the NUBASE evaluation \cite{2003Aud01} which is based on the ENSDF database \cite{2008ENS01}. In cases where the reported half-life differed significantly from the adopted half-life (up to approximately a factor of two), we searched the subsequent literature for indications that the measurement was erroneous. If that was not the case we credited the authors with the discovery in spite of the inaccurate half-life. All reported half-lives inconsistent with the presently adopted half-life for the ground state were compared to isomers half-lives and accepted as discoveries if appropriate following the criterium described above.

The first criterium excludes measurements of half-lives of a given element without mass identification. This affects mostly isotopes first observed in fission where decay curves of chemically separated elements were measured without the capability to determine their mass. Also the four-parameter measurements (see, for example, Ref. \cite{1970Joh01}) were, in general, not considered because the mass identification was only $\pm$1 mass unit.

The second criterium affects especially the isotopes studied within the Manhattan Project. Although an overview of the results was published in 1946 \cite{1946TPP01}, most of the papers were only published in the Plutonium Project Records of the Manhattan Project Technical Series, Vol. 9A, ��Radiochemistry and the Fission Products,'' in three books by Wiley in 1951 \cite{1951Cor01}. We considered this first unclassified publication to be equivalent to a refereed paper.

The initial literature search was performed using the databases ENSDF \cite{2008ENS01} and NSR \cite{2008NSR01} of the National Nuclear Data Center at Brookhaven National Laboratory. These databases are complete and reliable back to the early 1960's. For earlier references, several editions of the Table of Isotopes were used \cite{1940Liv01,1944Sea01,1948Sea01,1953Hol02,1958Str01,1967Led01}. A good reference for the discovery of the stable isotopes was the second edition of Aston's book ``Mass Spectra and Isotopes'' \cite{1942Ast01}.

\section{Discovery of $^{91-128}$Pd}

Thirty-eight palladium isotopes from A = 91--128 have been discovered so far; these include 6 stable, 12 neutron-deficient and 20 neutron-rich isotopes. According to the HFB-14 model \cite{2007Gor01}, on the neutron-rich side the bound isotopes should reach at least up to $^{148}$Pd while on the neutron deficient side six more isotopes should be particle stable ($^{85-90}$Pd). Thus, about 26 isotopes have yet to be discovered corresponding to 41\% of all possible palladium isotopes.

\begin{figure}
	\centering
	\includegraphics[scale=.7]{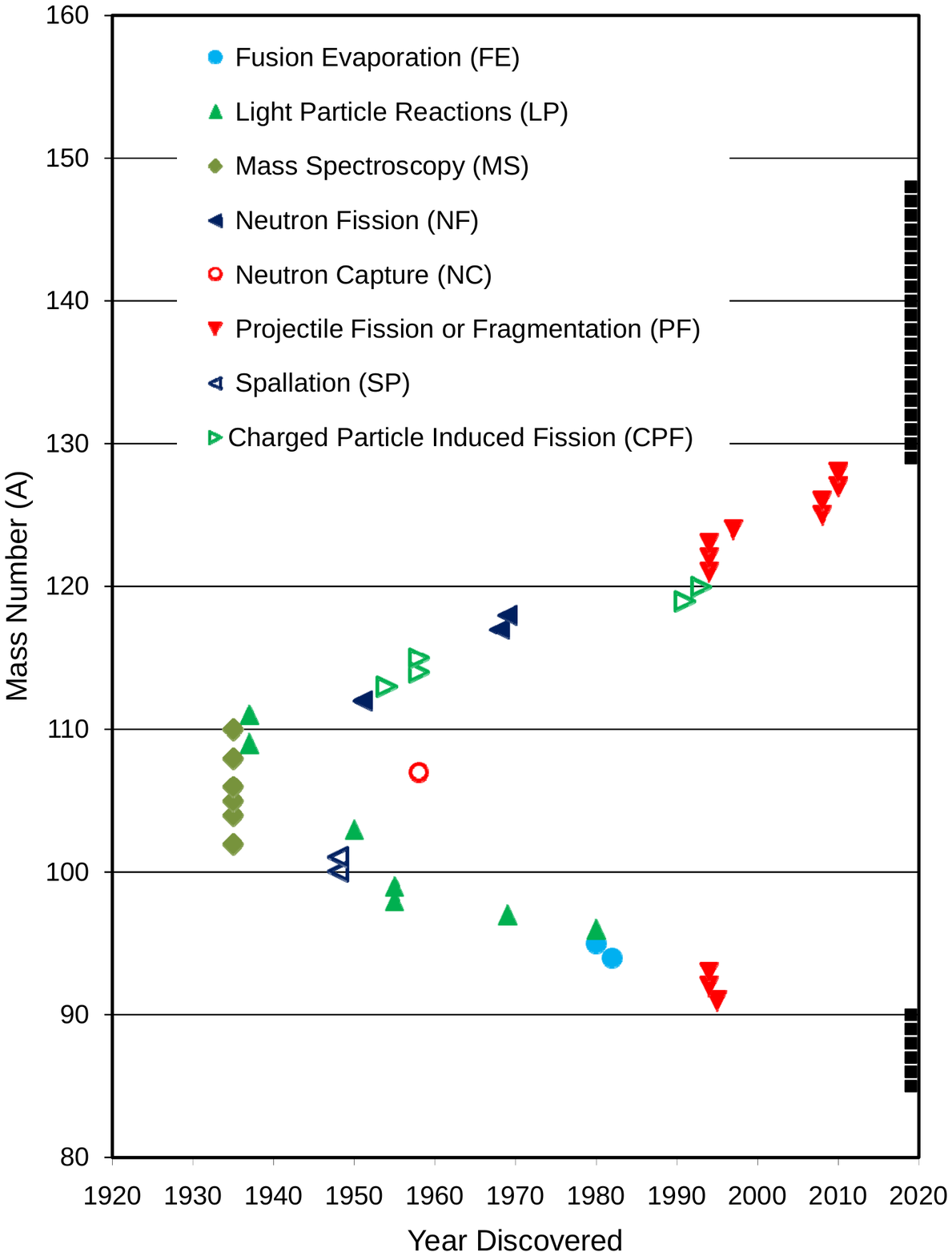}
	\caption{Palladium isotopes as a function of time when they were discovered. The different production methods are indicated. The solid black squares on the right hand side of the plot are isotopes predicted to be bound by the HFB-14 model.}
\label{f:year-pd}
\end{figure}

Figure \ref{f:year-pd} summarizes the year of first discovery for all palladium isotopes identified by the method of discovery. The range of isotopes predicted to exist is indicated on the right side of the figure. The radioactive palladium isotopes were produced using fusion evaporation reactions (FE), light-particle reactions (LP), neutron induced fission (NF), spallation (SP), neutron capture reactions (NC), charged particle induced fission (CPF), and projectile fragmentation or fission (PF). The stable isotopes were identified using mass spectroscopy (MS). 

\subsection*{$^{91}$Pd}\vspace{0.0cm}
Rykaczewski et al.\ discovered $^{91}$Pd in their 1995 paper ``Identification of new nuclei at and beyond the proton drip line near the doubly magic nucleus $^{100}$Sn'' \cite{1995Ryk01}. A 63 MeV/nucleon $^{112}$Sn beam from the GANIL cyclotron complex bombarded a natural nickel target. $^{87}$Ru was identified with the Alpha and LISE3 spectrometers. ``The obtained data have allowed also for the identification of six other new nuclei, namely $^{103}$Sb, $^{104}$Sb, $^{98}$In, $^{91}$Pd, $^{89}$Rh, and $^{87}$Ru, which are clearly isolated from the neighboring heavier isotopes in the mass spectra of [the figure].''

\subsection*{$^{92,93}$Pd}\vspace{0.0cm}
In ``Identification of new nuclei near the proton drip line,'' Hencheck et al.\ reported the discovery of $^{92}$Pd and $^{93}$Pd \cite{1994Hen01}. A $^{106}$Cd beam accelerated to 60 MeV/u at the National Superconducting Cyclotron Laboratory (NSCL) at Michigan State University bombarded a natural nickel target. The isotopes were analyzed with the A1900 projectile fragment separator and identified event-by-event with measurements of the magnetic rigidity, time of flight, energy-loss, and total energy. ``A number of new nuclides were identified including $^{88}$Ru, $^{90,91,92,93}$Rh, $^{92,93}$Pd, and $^{94,95}$Ag.''

\subsection*{$^{94}$Pd}\vspace{0.0cm}
In 1982 the article, ``Investigations of very neutron-deficient isotopes below $^{100}$Sn in $^{40}$Ca-induced reactions,'' Kurcewicz et al.\ reported the first observation of $^{94}$Pd \cite{1982Kur01}. A 4.0 MeV/u $^{40}$Ca from the GSI heavy-ion accelerator UNILAC bombarded a $^{58}$Ni target to produce $^{94}$Pd in the fusion evaporation reaction $^{58}$Ni($^{40}$Ca,4n). X- and $\gamma$-rays were measured with two Ge(Li) detectors following on-line mass separation. ``The half-life of (9$\pm$0.5)~s for $^{94}$Pd decay resulted from the analysis of the decay-curves of Rh, K X-rays and of the $\gamma$-transitions with energies of 54.6 and 558.3 keV.'' This half-life is the currently accepted value.

\subsection*{$^{95}$Pd}\vspace{0.0cm}
In ``$\beta$ delayed proton emission from a long-lived high-spin isomer in $^{95}$Pd'', Nolte and Hick described the discovery of $^{95}$Pd in 1980 \cite{1980Nol01}. Enriched $^{58}$Ni targets were bombarded with a 135 MeV $^{40}$Ca beam from the Munich MP tandem and $^{95}$Pd was produced in the (n2p) fusion-evaporation reaction. Gamma-rays and delayed protons were measured with a coaxial Ge(Li) and Si surface barrier detector, respectively. ``A long-lived high-spin isomeric state in $^{95}$Pd with a half-life of 14 $\pm$ 1~s and with J$^{\pi}$ = 21/2$^+$ has been identified. It has been found that this state is a $\beta$ delayed proton precursor.'' This half-life is consistent with the presently accepted value of 13.3(3)~s.

\subsection*{$^{96}$Pd}\vspace{0.0cm}
In 1980 Aras et al.\ reported the first observation of $^{96}$Pd in ``Decay of the new closed-shell nuclide 2.0 min $^{96}$Pd'' \cite{1980Ara01}. The Maryland Cylcotron was used to bombard an enriched $^{96}$Ru target with 60 MeV $\alpha$ particles and $^{96}$Pd was produced in the ($\alpha$,4n) reaction. Gamma-ray spectra and decay curves were measured with a hyperpure Ge detector. ``The closed-shell nuclide $^{96}$Pd has been found to decay with a 2.0 min half-life to a 1$^+$ state at 177 keV in $^{96}$Rh.'' This half-life is consistent with the presently accepted value of 122(2)~s.

\subsection*{$^{97}$Pd}\vspace{0.0cm}
$^{97}$Pd was first identified in 1969 by Aten and Kapteyn in ``Palladium-97'' \cite{1969Ate01}. An enriched $^{96}$Ru target was irradiated with 15$-$22 MeV $^{3}$He beams and $^{97}$Pd was formed in the ($^3$He,2n) reaction. Decay curves and $\gamma$-ray spectra were measured following chemical separation. ``There is, however, also an activity present in these samples with a half-life of 3.3$\pm$0.3 min. Presumably we are dealing with $^{97}$Pd, as, at least with 15 MeV He$^{3+}$-ions, the energy would not be sufficient for a ($^{3}$He,3n) process.'' This half-life agrees with the accepted value of 3.1(9)~min.

\subsection*{$^{98,99}$Pd}\vspace{0.0cm}
``Formation and properties of neutron-deficient isotopes of rhodium and palladium'' by Aten Jr and De Vries-Hamerling describes the identification of $^{98}$Pd and $^{99}$Pd in 1955 \cite{1955Ate01}. Ruthenium targets were bombarded with 24 and 52 MeV $\alpha$ particles from the Amsterdam cyclotron. Beta- and $\gamma$-ray spectra were measured following chemical separation. ``We have since found by means of milking experiments that the palladium-mother of the 9-minutes rhodium can also be obtained from ruthenium irradiated with 24 MeV helium ions, which rules out the mass number 96, as the latter can only be formed by an ($\alpha$,4n) process. Therefore it seems likely that the 9-minutes period is due to $^{98}$Rh and its 17-minutes mother to $^{98}$Pd... A number of experiments have been carried out in which the 4.5 hours rhodium, supposed to be $^{99}$Rh, is milked from the palladium fraction, obtained by irradiating ruthenium with 52 MeV helium ions... The half-life of the palladium mother ($^{99}$Pd if the mass of the 4.5 hours rhodium is correct) was found to be 24 minutes, with an error which may be between 5 and 10 minutes.''  The 17~min and 24~min half-lives for $^{98}$Pd and $^{99}$Pd agree with the accepted half-lives of 17.7(3)~min and 21.4(2)~min, respectively. Earlier,  Aten Jr and De Vries-Hamerling had measured a 15(3)~min half-life but could only assign it to either $^{96}$Pd or $^{98}$Pd \cite{1953Ate01}.

\subsection*{$^{100,101}$Pd}\vspace{0.0cm}
Lindner and Perlman discovered $^{100}$Pd and $^{101}$Pd in 1948 in ``Neutron deficient isotopes of tellurium and antimony'' \cite{1948Lin02}. A 50 MeV deuteron beam from the Berkeley 184-inch cyclotron bombarded a thin rhodium metal foil. Beta-decay curves as well as X- and $\gamma$-ray spectra were measured following chemical separation. ``4.0-day Pd$^{100}$: After the decay of 9-hr. Pd$^{101}$, the half-life for Pd$^{100}$ was determined by removing the rhodium isotopes which had grown and, following the decay after the 19.4-hr. Rh$^{100}$, again came to equilibrium... 9-hr. Pd$^{101}$: ...The positron decay was measured directly using the spectrometer; the decay was determined indirectly by the rate of decrease in amount of 4.3-day Rh$^{101}$ which grew into the palladium fraction, and the x-ray decay curve showed this component when corrected for the growth of 19.4-hr. Rh$^{100}$.'' These half-lives of 9~h and 4.0~days for $^{101}$Pd and $^{100}$Pd agree with the accepted half-lives of 8.47(6)~h and 3.63(9)~days, respectively.

\subsection*{$^{102}$Pd}\vspace{0.0cm}
Dempster reported the observation of stable $^{102}$Pd in the 1935 paper ``Isotopic constitution of palladium and gold'' \cite{1935Dem02}. Palladium ions were produced with a high-frequency spark and analyzed with the Chicago mass spectrograph. ``With the mass-spectrograph it was found that palladium consists of six isotopes with atomic masses 102, 104, 105, 106, 108, 110.''

\subsection*{$^{103}$Pd}\vspace{0.0cm}
In 1950 Mei et al.\ described the observation of $^{103}$Pd in ``The disintegration of ruthenium 103 and palladium 103'' \cite{1950Mei02}. A rhodium metal target was bombarded with 23-MeV $\alpha$-particles from the Indiana University cyclotron. Gamma- and beta-ray spectra were measured following chemical separation. ``Pd$^{103}$ disintegrates by orbital electron capture. The spectrum consists of an electron line at 369 kev, shown to be due to L electrons from a highly converted gamma-ray of 40.4-kev energy, together with Auger electrons.'' The measured half-life of 17~d is consistent with the presently accepted value of 16.991(19)~d.

\subsection*{$^{104-106}$Pd}\vspace{0.0cm}
Dempster reported the observation of stable $^{104}$Pd, $^{105}$Pd and $^{106}$Pd in the 1935 paper ``Isotopic constitution of palladium and gold'' \cite{1935Dem02}. Palladium ions were produced with a high-frequency spark and analyzed with the Chicago mass spectrograph. ``With the mass-spectrograph it was found that palladium consists of six isotopes with atomic masses 102, 104, 105, 106, 108, 110.''

\subsection*{$^{107}$Pd}\vspace{0.0cm}
The identification of $^{107}$Pd was described by Schindewolf in the 1958 paper ``Mass assignments of 23-sec Pd$^{107m}$ and 4.8-min Pd$^{109m}$; search for $<$3-sec Pd$^{105m}$'' \cite{1958Sch01}. Enriched $^{106}$Pd targets were irradiated with slow and fast neutrons produced by bombarding beryllium with 15 MeV deuterons from the M.I.T. cyclotron and $^{107}$Pd was formed in neutron-capture reactions. $^{107}$Pd was also observed in (n,2n) reactions off enriched $^{108}$Pd. Gamma- and beta-rays were measured with a NaI(Tl) scintillation counter and a $\beta$-proportional counter, respectively. ``The formation of the 23-sec and 4.8-min Pd isomers by neutron activation of Pd$^{106}$ and Pd$^{108}$ samples leads to the mass assignments 23-sec Pd$^{107m}$ and 4.8-min Pd$^{109m}$.'' This half-life for the isomeric state of $^{106}$Pd agrees with the presently accepted value of 21.3(5)~s. Previously, the 23(2)~s half-life had been incorrectly assigned to $^{105m}$Pd \cite{1952Fla02}. The observation of the long-lived ground-state had been reported in 1949 but only in a classified report \cite{1949Par02}.

\subsection*{$^{108}$Pd}\vspace{0.0cm}
Dempster reported the observation of stable $^{108}$Pd in the 1935 paper ``Isotopic constitution of palladium and gold'' \cite{1935Dem02}. Palladium ions were produced with a high-frequency spark and analyzed with the Chicago mass spectrograph. ``With the mass-spectrograph it was found that palladium consists of six isotopes with atomic masses 102, 104, 105, 106, 108, 110.''

\subsection*{$^{109}$Pd}\vspace{0.0cm}
In ``Radioactive isotopes of palladium and silver from palladium'', Kraus and Cork described the observation of $^{109}$Pd in 1937 \cite{1937Kra01}. $^{109}$Pd was produced by bombarding palladium with 6.3 MeV deuterons from the University of Michigan cyclotron and irradiating palladium with fast and slow neutrons. Decay curves were measured with Lauritsen quartz fiber electroscopes and a Wulf string electrometer equipped with an ionization chamber following chemical separation. ``The 13-hr.\ palladium activity could be due to either isotopes 107 or 109. The assignment to isotope 109 is preferred when one considers the ratio of activities for fast and slow neutrons and the relative abundance of the parent isotopes.'' This half-life is consistent with the accepted value of 13.7012(24)~h. Previously, a 12 h half-life of palladium was reported without a mass assignment \cite{1935Ama01}.

\subsection*{$^{110}$Pd}\vspace{0.0cm}
Dempster reported the observation of stable $^{110}$Pd in the 1935 paper ``Isotopic constitution of palladium and gold'' \cite{1935Dem02}. Palladium ions were produced with a high-frequency spark and analyzed with the Chicago mass spectrograph. ``With the mass-spectrograph it was found that palladium consists of six isotopes with atomic masses 102, 104, 105, 106, 108, 110.''

\subsection*{$^{111}$Pd}\vspace{0.0cm}
In ``Radioactive isotopes of palladium and silver from palladium'', Kraus and Cork reported the discovery of $^{111}$Pd in 1937 \cite{1937Kra01}. $^{111}$Pd was produced by bombarding palladium with 6.3 MeV deuterons from the University of Michigan cyclotron and irradiating palladium with fast and slow neutrons. Decay curves were measured with Lauritsen quartz fiber electroscopes and a Wulf string electrometer equipped with an ionization chamber following chemical separation. ``If one of the observed periods is due to the isotope of mass 111 then by beta-decay it should produce a radioactive silver since there is no stable silver of mass 111. Subsequent precipitation of silver from the palladium precipitates proved this to be the case. Moreover, by making these subsequent separations at different times after the original bombardment, it appears to be quite certain that this silver activity (180-hr.\ half-life) must be built up from the 17-min. palladium and not the 13-hr. palladium.'' The half-life of 17~min is within a factor of two of the accepted value of 23.4(2)~min and Kraus and Cork identified the decay of the $^{111}$Ag correctly. Previously, a 15 min half-life of palladium was reported without a mass assignment \cite{1935Ama01}.

\subsection*{$^{112}$Pd}\vspace{0.0cm}
$^{112}$Pd was identified by Seiler in ``Palladium-silver chains in fission'' as part of the Plutonium Project published in 1951 \cite{1951Sei01}. Uranium metal was irradiated with neutrons from the Argonne Heavy-water pile. Absorption- and decay-curves as well as $\gamma$- and $\beta$-rays were measured following chemical separation. ``The work reported here shows that the so-called ``17-h isotope of palladium; is actually two isotopes with half-lives of 21 hr and 13 hr and with the following relationship:... 21h Pd$^{112}\stackrel{\beta}{\longrightarrow}$3.2h Ag$^{112}\stackrel{\beta}{\longrightarrow}$stable Cd$^{112}$.'' This half-life agrees with the presently adopted value of 21.03(5)~h. The previous assignment of the 17~h half-life to $^{112}$Pd \cite{1940Nis02,1941Seg01} mentioned in the quote was credited with the discovery of $^{112}$Pd because it was due to a mixture of two isotopes.

\subsection*{$^{113}$Pd}\vspace{0.0cm}
In the 1954 article, ``A new isotope of palladium, 1.5-minute Pd$^{113}$'', Hicks and Gilbert described the observation of $^{113}$Pd \cite{1954Hic01}. A uranium foil was irradiated with 190 MeV deuterons from the Berkeley 184-in.\ cyclotron. The resulting activity was measured with a chlorine-quenched Geiger tube following chemical separation. ``A new isotope, 1.5-minute Pd$^{113}$, has been isolated from the fission products of natural uranium bombarded with 190-Mev deuterons. The mass assignment and half-life were determined by successive milkings of the 5.3-hour Ag$^{113}$ daughter.'' This half-life is consistent with the presently accepted value of 93(5)~s.

\subsection*{$^{114,115}$Pd}\vspace{0.0cm}
Alexander et al.\ discovered in 1958 $^{114}$Pd and $^{115}$Pd as reported in ``Short-lived isotopes of Pd and Ag of masses 113-117'' \cite{1958Ale01}. Uranium was bombarded with 15 MeV deuterons from the M.I.T. cyclotron. Gamma- and beta-ray spectra were measured with NaI(Tl) scintillation spectrometers following chemical separation. ``The 0.56-Mev $\gamma$ ray associated with the decay of 2.4-min Pd$^{114}$ is ascribed to a $\gamma$-transition following $\beta$ decay of 5-sec Ag$^{114}$... The half-life of $^{115}$Pd was found by periodic extraction and counting of 21.1-min Ag$^{115}$ to be 50$\pm$12 seconds. A more accurate value of 44$\pm$3 seconds was found by identical period extraction and subsequent purification and counting of 2.2-day Cd$^{115}$ after decay of 21.1-min Ag$^{115}$. The half-period chosen is 45$\pm$3 sec.'' These half-lives of 2.4(1)~min and 45(3)~s agree with the presently accepted values of 2.42(6)~min and 50(3)~s for the ground state of $^{114}$Pd and an isomeric state of $^{115}$Pd, respectively.

\subsection*{$^{116}$Pd}\vspace{0.0cm}
Evidence for $^{116}$Pd was observed by Cheifetz et al.\ in the 1970 paper ``Experimental information concerning deformation of neutron rich nuclei in the A $\sim$ 100 region'' \cite{1970Che01}. Fission fragments from the spontaneous fission of $^{252}$Cf were observed in coincidence with X- and $\gamma$-rays measured in Ge(Li) detectors. The observed $\gamma$-rays were not discussed in detail for each isotopes and the results were presented in a table. For $^{116}$Pd levels at 340.6 (2$^+$) and 878.6 (4$^+$) keV were reported which agree with the accepted values of 340.26(8) (2$^+$) and 877.58(12) (4$^+$) keV, respectively. The first measurement of the ground state half-life was reported less than a month later by Aron et al.\ \cite{1970Aro01}.

\subsection*{$^{117}$Pd}\vspace{0.0cm}
$^{117}$Pd was identified by Weiss et al.\ in 1968, as reported in ``Identification and yield of 5.0-sec $^{117}$Pd in the thermal-neutron fission of $^{235}$U'' \cite{1968Wei02}. A uranium solution enriched in $^{235}$U was irradiated with neutrons from the Vallecitos Nuclear Test Reactor. Decay curves of the resulting activities were measured with a gas-flow proportional counter following chemical separation of the fission fragments. ``The counting rate of $^{117}$Cd for twelve palladium appears in [the figure]. A linear relationship between the logarithm of the counting rate and the time of separation is evident. From a least-squares analysis of these data, a half-life of $5.0_{-0.7}^{+0.5}$ sec for $^{117}$Pd was computed.'' This half-life agrees with the presently accepted value of 4.3(3)~s.

\subsection*{$^{118}$Pd}\vspace{0.0cm}
In 1969 Weiss et al.\ described the observation of $^{118}$Pd in ''Nuclear charge distribution in symmetric fission of $^{235}$U with thermal neutrons: Yields of $^{117}$Ag, $^{118}$Ag, and $^{118}$Pd'' \cite{1969Wei01}. A uranium solution enriched in $^{235}$U was irradiated with neutrons from the Vallecitos Nuclear Test Reactor.  Decay curves of the resulting activities were measured following chemical separation of the fission fragments. ``The corrected counting rate of $^{118}$Cd for nine Pd separations, made 1-10 sec after fission, appears in [the figure]. Separation time is defined as starting at the end of irradiation and ending at the midpoint of filtration of the fission solution through the Cu bed. The relationship between the logarithm of the counting rate and the separation time is linear. Analysis of the data by the method of least squares gives a half-life of 3.1$\pm$0.3 sec for $^{118}$Pd.'' This half-life is close to the accepted value of 1.9(1)~s.

\subsection*{$^{119}$Pd}\vspace{0.0cm}
In ``First observation of the beta decay of $^{117}$Pd and the discovery of a new isotope $^{119}$Pd'', Penttil\"a et al.\ identified $^{119}$Pd in 1991 \cite{1991Pen02}. A $^{238}$U target was bombarded with 20 MeV protons and fission fragments were separated with the on-line mass separator IGISOL. Gamma-, X- and beta-rays were measured with a HPGe, a planar Ge and a plastic scintillator detector, respectively. ``For $^{119}$Pd, the existence of two beta-decaying states with the same half-life but of opposite parity cannot be excluded. The measured half-life for $^{119}$Pd is 0.92$\pm$0.13~s.'' This half-life is consistent with the currently accepted value of 0.92(1)~s.

\subsection*{$^{120}$Pd}\vspace{0.0cm}
In ``Gamow-Teller decay of $^{118}$Pd and of the new isotope $^{120}$Pd'', Janas et al.\ reported the discovery of $^{120}$Pd in 1993 \cite{1993Jan01}. A $^{238}$U target was bombarded with 20 MeV protons from the Jyv\"askyl\"a MC-20 cyclotron and fission fragments were separated with the on-line mass separator IGISOL. Gamma-, X- and beta-rays were measured with an intrinsic germanium, a planar germanium and a plastic scintillator detector, respectively. ``The decay of $^{118}$Pd was reinvestigated in detail, and evidence for the new isotope $^{120}$Pd, with a half-life $T_{1/2}$=0.5$\pm$0.1~s, was found.'' This half-life is the currently accepted value.

\subsection*{$^{121-123}$Pd}\vspace{0.0cm}
In 1994, Bernas et al.\ published the discovery of $^{121}$Pd, $^{122}$Pd, and $^{123}$Pd in ``Projectile fission at relativistic velocities: A novel and powerful source of neutron-rich isotopes well suited for in-flight isotopic separation'' \cite{1994Ber01}. The isotopes were produced using projectile fission of $^{238}$U at 750 MeV/nucleon on a lead target at GSI, Germany. ``Forward emitted fragments from $^{80}$Zn up to $^{155}$Ce were analyzed with the Fragment Separator (FRS) and unambiguously identified by their energy-loss and time-of-flight.'' This experiment yielded 211, 79, and 12 counts of $^{121}$Pd, $^{122}$Pd, and $^{123}$Pd, respectively.

\subsection*{$^{124}$Pd}\vspace{0.0cm}
$^{124}$Pd was discovered by Bernas et al.\ in 1997, as reported in ``Discovery and cross-section measurement of 58 new fission products in projectile-fission of 750$\cdot$A MeV $^{238}$U'' \cite{1997Ber01}. The experiment was performed using projectile fission of $^{238}$U at 750~MeV/nucleon on a beryllium target at GSI in Germany. ``Fission fragments were separated using the fragment separator FRS tuned in an achromatic mode and identified by event-by-event measurements of $\Delta$E-B$\rho$-ToF and trajectory.'' During the experiment, 19 individual counts were recorded for $^{124}$Pd.

\subsection*{$^{125,126}$Pd}\vspace{0.0cm}
The discovery of $^{125}$Pd and $^{126}$Pd was reported in the 2008 article ``Identification of new isotopes $^{125}$Pd and $^{126}$Pd produced by in-flight fission of 345 MeV/nucleon $^{238}$U: First results from the RIKEN RI beam factory '' by Ohnishi et al.\ \cite{2008Ohn01}. The experiment was performed at the RI Beam Factory at RIKEN, where the new isotopes were created by in-flight fission of a 345 MeV/nucleon $^{238}$U beam on a beryllium target. The isotopes were separated and identified with the BigRIPS superconducting in-flight separator. ``We observed the production of two new isotopes $^{125}$Pd and $^{126}$Pd, even though the uranium beam intensity was far from the goal for the RIBF and the total observation time was only about a day.''

\subsection*{$^{127,128}$Pd}\vspace{0.0cm}
The discovery of $^{127}$Pd and $^{128}$Pd were reported in the 2010 article ``Identification of 45 new neutron-rich isotopes produced by in-flight fission of a $^{238}$U beam at 345 MeV/nucleon,'' by Ohnishi et al.\ \cite{2010Ohn01}. The experiment was performed at the RI Beam Factory at RIKEN, where the new isotopes were created by in-flight fission of a 345 MeV/nucleon $^{238}$U beam on a beryllium target. The isotopes were separated and identified with the BigRIPS superconducting in-flight separator. The results for the new isotopes discovered in this study were summarized in a table. Individual counts of 70 and 13 for $^{127}$Pd and $^{128}$Pd, respectively, were recorded.

\section{Discovery of $^{103-140}$Sb}

Thirty-eight antimony isotopes from A = 103--140 have been discovered so far; these include 2 stable, 18 neutron-deficient and 18 neutron-rich isotopes. According to the HFB-14 model \cite{2007Gor01}, $^{172}$Sb should be the last odd-odd particle stable neutron-rich nucleus while the even-odd particle stable neutron-rich nuclei should continue through $^{177}$Sb. The proton dripline has been crossed with the observation of the proton emitter $^{105}$Sb, however, $^{102}$Sb, $^{101}$Sb, and $^{100}$Sb could still have half-lives longer than 10$^{-9}$~s \cite{2004Tho01}. Thus, about 37 isotopes have yet to be discovered corresponding to 49\% of all possible antimony isotopes.

\begin{figure}
	\centering
	\includegraphics[scale=.7]{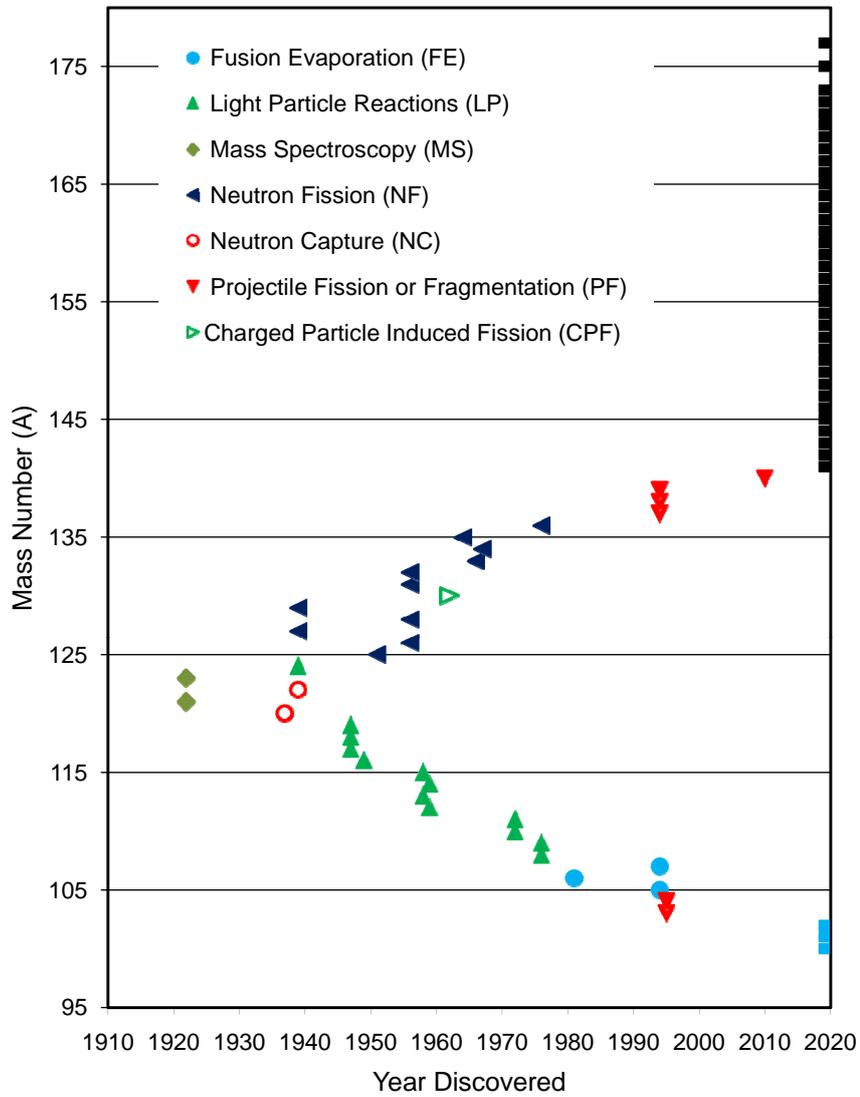}
	\caption{Antimony isotopes as a function of time when they were discovered. The different production methods are indicated. The solid black squares on the right hand side of the plot are isotopes predicted to be bound by the HFB-14 model. On the proton-rich side the light blue squares correspond to unbound isotopes predicted to have half-life larger than $\sim 10^{-9}$~s.}
\label{f:year-sb}
\end{figure}

Figure \ref{f:year-sb} summarizes the year of first discovery for all antimony isotopes identified by the method of discovery. The range of isotopes predicted to exist is indicated on the right side of the figure. The radioactive antimony isotopes were produced using fusion evaporation reactions (FE), light-particle reactions (LP), neutron induced fission (NF), charged-particle induced fission (CPF), neutron capture (NC), and projectile fragmentation or fission (PF). The stable isotopes were identified using mass spectroscopy (MS). In the following, the discovery of each antimony isotope is discussed in detail.

\subsection*{$^{103,104}$Sb}\vspace{0.0cm}
Rykaczewski et al.\ discovered $^{103}$Sb and $^{104}$Sb in their 1995 paper ``Identification of new nuclei at and beyond the proton drip line near the doubly magic nucleus $^{100}$Sn'' \cite{1995Ryk01}. A 63 MeV/nucleon $^{112}$Sn beam from the GANIL cyclotron complex bombarded a natural nickel target. $^{103}$Sb and $^{104}$Sb were identified with the Alpha and LISE3 spectrometers. ``The obtained data have allowed also for the identification of six other new nuclei, namely $^{103}$Sb, $^{104}$Sb, $^{98}$In, $^{91}$Pd, $^{89}$Rh, and $^{87}$Ru, which are clearly isolated from the neighboring heavier isotopes in the mass spectra of [the figure].''

\subsection*{$^{105}$Sb}\vspace{0.0cm}
In the 1994 paper ``Evidence for the ground-state proton decay of $^{105}$Sb,'' Tighe et al.\ reported the discovery of $^{105}$Sb \cite{1994Tig01}. Enriched $^{50}$Cr$_{2}$O$_{3}$ targets were bombarded with 220 and 260~MeV $^{58}$Ni beams from the Berkeley 88-in.\ cyclotron. Evaporation residues were deposited by a helium jet on a slow moving tape in the center of a low-energy proton detector ball consisting of six gas-$\Delta$E, gas-$\Delta$E, Si-E triple telescopes. ``Thus, this 478~keV group must originate from a direct proton emitter produced in the $^{58}$Ni$+^{50}$Cr compound nuclear reaction forming $^{108}$Te... we assign the group at 478~keV to the ground-state proton decay of $^{105}$Sb.'' A previous search for $^{105}$Sb was unsuccessful \cite{1987Gil01}.

\subsection*{$^{106}$Sb}\vspace{0.0cm}
$^{106}$Sb was discovered by Plochocki et al.\ in 1981 as reported in ``Measurements of proton separation-energies close to the proton drip line in the antimony-cesium region'' \cite{1981Plo01}. A $^{58}$Ni target was bombarded with a 290~MeV $^{58}$Ni beam from the GSI UNILAC accelerator forming $^{114}$Cs in the (1p1n) fusion-evaporation reaction. $^{114}$Cs was separated with the GSI on-line mass separator and $^{106}$Sb was then populated by $\alpha$-decays. Subsequent proton emission was measured with a telescope consisting of surface barrier silicon detectors. ``In a continuation of these studies, we have now determined S$_p$ values for $^{114}$Cs, $^{110}$I and $^{106}$Sb.'' The separation energy for $^{106}$Sb was determined to be 930$\pm$210~keV.

\subsection*{$^{107}$Sb}\vspace{0.0cm}
The discovery of $^{107}$Sb was reported in the 1994 article ``In-beam spectroscopy at the proton-drip line. First observation of excited states in $^{106}$Sb and $^{107}$Sb'' by Seweryniak et al.\ \cite{1994Sew01}. A 270~MeV $^{58}$Ni beam from the Roskilde tandem accelerator bombarded an enriched $^{54}$Fe target. $^{107}$Sb was populated in the (1p1$\alpha$) fusion-evaporation reaction and identified with the NORDBALL detector array. ``Evaporation residues were identified by detecting protons and $\alpha$ particles in a 4$\pi$ charged particle multi-detector set-up and neutrons in a 1$\pi$ neutron detector wall in coincidence with $\gamma$ rays. Excited states of the proton drip line nucleus $^{106}$Sb and of $^{107}$Sb were identified for the first time.'' At the time of this observation only the $\alpha$-decay energy of the parent nucleus $^{111}$I was known \cite{1979Sch02}. The half-life of the ground state was measured three years later \cite{1997Shi01}.

\subsection*{$^{108,109}$Sb}\vspace{0.0cm}
$^{108}$Sb and $^{109}$Sb were discovered by Oxorn et al.\ in 1976 as reported in ``Decay of $^{108,109,110}$Sb'' \cite{1976Oxo01}. Enriched $^{112}$Sn targets were bombarded with 25$-$65~MeV protons from the McGill cyclotron forming $^{108}$Sb and $^{109}$Sb in (p,5n) and (p,4n) reactions, respectively. Gamma-ray singles and coincidences were recorded with two Ge(Li) detectors. ``New isotopes $^{109,108}$Sb are identified and their decay properties are discussed. The decay half-life of $^{109}$Sb is measured to be 18.3$\pm$0.5~s and that of $^{108}$Sb to be 7.0$\pm$0.5~s.'' These half-lives are consistent with the accepted values of 7.4(3)~s and 17.0(7)~s for $^{108}$Sb and $^{109}$Sb, respectively.

\subsection*{$^{110}$Sb}\vspace{0.0cm}
In 1972, Miyano et al.\ published the first identification of $^{110}$Sb in ``The decay of $^{110}$Sb'' \cite{1972Miy01}. The Tokyo frequency modulated cyclotron was used to bombard an enriched $^{112}$SnO$_2$ targets with a 52-MeV proton beam. Beta- and gamma-rays were measured with a plastic scintillator and a Ge(Li) detector, respectively. ``This was decomposed into three components of half-lives 56~sec, 21.3$\pm$1.5~s and 5.4~sec. The 56-sec component is due to the decay of $^{112}$Sb. From this decay curve, the half-life of $^{110}$Sb was determined to be 21.3$\pm$1.5~sec.'' This half-life is consistent with the accepted half-life of 23.0(4)~s. Miyano et al.\ did not consider their result a new discovery referring to a conference proceeding by Sunier et al.\ \cite{1970Sun01}. These results were published in a referred paper two months later \cite{1972Sin01}.

\subsection*{$^{111}$Sb}\vspace{0.0cm}
In the 1972 paper ``Decay of the light antimony isotopes $^{110,111,112,113,114}$Sb following the $^{112,114}$Sn(p,xn) reactions,'' Singh et al.\ reported the discovery of $^{111}$Sb \cite{1972Sin01}. Enriched $^{112}$Sn targets were bombarded with 30~MeV protons from the UCLA cyclotron forming $^{111}$Sb in the (p,2n) reaction. Beta-ray spectra were measured with a single wedge-gap magnetic spectrometer and a stack of four Si(Li) detectors and $\gamma$-rays were recorded with two Ge(Li) and one NaI(Tl) detector. ``The $\gamma$-spectrum of $^{111}$Sb, recorded for 120~sec following the irradiation of $^{112}$Sn with 30~MeV protons is shown in [the figure]. The transitions belonging to $^{111}$Sb are clearly separated from $^{112,114}$Sb and other weaker contaminants.'' The reported half-life of 74.1(14)~s is in agreement with the currently accepted value of 75(1)~s.

\subsection*{$^{112}$Sb}\vspace{0.0cm}
$^{112}$Sb was identified in 1959 by Selinov et al.\ in ``The new isotopes Sb$^{112}$ and Sb$^{114}$ and the identification of Sb$^{113}$ and Sb$^{115}$'' \cite{1959Sel01}. The Moscow 120-centimeter phasotron bombarded enriched $^{112}$Sn targets with 7$-$30~MeV protons to produce $^{112}$Sb in (p,n) charge exchange reactions. Antimony was chemically separated from the target and $\gamma$ spectra were measured with a scintillation spectrometer. ``In addition, we discovered two new isotopes: Sb$^{112}$ with T$_{1/2}$ = 0.9 $\pm$ 0.1 min, and Sb$^{114}$, with T$_{1/2}$ = 3.3 $\pm$ 0.3 min.'' The reported half-life for $^{112}$Sb is consistent with the presently adopted value of 51.4(10)~s.

\subsection*{$^{113}$Sb}\vspace{0.0cm}
The observation of $^{113}$Sb was reported in the 1958 article ``New samarium isotopes,'' by Selinov et al.\ \cite{1958Sel01}. Enriched $^{112}$Sn targets were bombarded with 10~MeV deuterons producing $^{113}$Sb in the (d,n) reaction. Decay curves as well as $\beta$- and $\gamma$-ray spectra were measured following chemical separation. In the translation antimony was apparently translated as samarium: ``The new samarium isotopes may be formed as a result of (d,n) reactions or (d,2n) reactions; however, it is more probable that these isotopes result from the reaction Sn$^{112}$(d,n)Sb$^{113}$ or Sn$^{114}$(d,n)Sb$^{115}$.'' The reported half-life of 7.0(5)~min for $^{113}$Sb is consistent with the currently adopted value of 6.67(7)~min.

\subsection*{$^{114}$Sb}\vspace{0.0cm}
$^{114}$Sb was identified in 1959 by Selinov et al.\ in ``The new isotopes Sb$^{112}$ and Sb$^{114}$ and the identification of Sb$^{113}$ and Sb$^{115}$'' \cite{1959Sel01}. The Moscow 120-centimeter phasotron bombarded enriched $^{114}$Sn targets with 7$-$30~MeV protons to produce $^{114}$Sb in (p,n) charge exchange reactions. Antimony was chemically separated from the target and $\gamma$ spectra were measured with a scintillation spectrometer. ``In addition, we discovered two new isotopes: Sb$^{112}$ with T$_{1/2}$ = 0.9 $\pm$ 0.1 min, and Sb$^{114}$, with T$_{1/2}$ = 3.3 $\pm$ 0.3 min.'' The reported half-life for $^{114}$Sb is consistent with the presently accepted value of 3.49(3)~min.

\subsection*{$^{115}$Sb}\vspace{0.0cm}
The observation of $^{115}$Sb was reported in the 1958 article ``New samarium isotopes,'' by Selinov et al.\ \cite{1958Sel01}. Enriched $^{114}$Sn targets were bombarded with 10~MeV deuterons producing $^{115}$Sb in the (d,n) reaction. Decay curves as well as $\beta$- and $\gamma$-ray spectra were measured following chemical separation. In the translation antimony was apparently translated as samarium: ``The new samarium isotopes may be formed as a result of (d,n) reactions or (d,2n) reactions; however, it is more probable that these isotopes result from the reaction Sn$^{112}$(d,n)Sb$^{113}$ or Sn$^{114}$(d,n)Sb$^{115}$.'' The reported half-life for $^{115}$Sb of 31(1)~min is consistent with the currently accepted value of 32.1(3)~min. A 60~min half-life listed in the 1958 Table of Isotopes \cite{1958Str01} was based on a thesis \cite{1957Rho01} and evidently incorrect.

\subsection*{$^{116}$Sb}\vspace{0.0cm}
In 1949, Temmer published the first identification of $^{116}$Sb in ``Excitation functions for ($\alpha$,n), ($\alpha$,2n), and ($\alpha$,3n) reactions on indium.'' \cite{1949Tem01}. The Berkeley 60-inch cyclotron was used to bombard a $^{115}$In target with $\alpha$ particles up to 37~MeV and $^{116}$Sb was formed in ($\alpha$,3n) reactions. Activities were chemically identified as antimony and masses were determined by Calutron analysis. ``The periods found for Sb$^{118}$ and Sb$^{117}$ (5.1 hours and 2.8 hours) agree with those reported by Coleman and Pool. Sb$^{116}$ decays with a 60-minute half-life.'' This half-life agrees with the currently accepted value of 60.3(6)~min.

\subsection*{$^{117-119}$Sb}\vspace{0.0cm}
$^{117}$Sb, $^{118}$Sb and $^{119}$Sb were discovered by Coleman and Pool in 1947 as reported in ``X-Ray emitting isotopes of radioactive Sb and Sn'' \cite{1947Col01}. Tin and indium were bombarded with 10~MeV deuterons and 20~MeV $\alpha$-particles forming $^{117,119}$Sb and $^{118}$Sb, respectively. Characteristic x-rays were photographed with a pair of Cauchois cameras following chemical separation. ``Three new x-ray emitting activities in Sb with half-lives of 2.8~hours, 5.1~hours and 39~hours have been found by the use of the curved crystal camera in conjunction with the x-ray decay curves. The assignments are $^{117}$Sb, $^{118}$Sb, and $^{119}$Sb respectively.'' The half-life for $^{117}$Sb agrees with the currently accepted value of 2.80(1)~h. The half-life for $^{118}$Sb is the currently accepted value and the half-life for $^{119}$Sb is included in the calculation of the weighted average for the accepted value of 38.19(22)~h.

\subsection*{$^{120}$Sb}\vspace{0.0cm}
The first observation of $^{120}$Sb was reported in the 1937 paper ``Herstellung neuer Isotope durch Kernphotoeffekt'' by Bothe and Gentner \cite{1937Bot01}. $^{120}$Sb was produced in photo-nuclear reactions with lithium $\gamma$-rays. ``Antimon (Sb$^{121}$ + Sb$^{123}$) zeigte eine Halbwertzeit von 13 min. Es wurde noch nicht nach der bekannten Halbwertzeit von 2,5 Tagen gesucht, welche man durch Anlagerung erh\"alt; daher ist die Zuordnung des 13 min-Abfalls zu Sb$^{120}$ noch nicht ganz sicher.'' [Antimony (Sb$^{121}$ + Sb$^{123}$) showed a half-life of 13~min. We have not yet searched for the known half-life of 2.5 days, which can be produced by neutron capture; thus the assignment of the 13 min decay to Sb$^{120}$ is not yet completely certain.] This half-life agrees with the presently adopted value of 15.89(4)~min.

\subsection*{$^{121}$Sb}\vspace{0.0cm}
Aston discovered stable $^{121}$Sb in 1922 as reported in ``The isotopes of antimony'' \cite{1922Ast04}. A pure antimony trimethyl sample was used in the Cavendish mass spectrograph. ``The element was characterized by two intense lines at 121 and 123. The first is the more intense by perhaps 10 to 20 per cent. If sufficient exposure is given two faint companions are visible at 122, 124, but the general evidence suggests that these are due to hydrogen addition products.  The isotopic nature of the lines 121, 123 is amply confirmed by the appearance of similar pairs 15 and 30 units higher, due to molecules of their monomethides and dimethides.''

\subsection*{$^{122}$Sb}\vspace{0.0cm}
Livingood and Seaborg reported the discovery of $^{122}$Sb in the 1939 article ``Radioactive antimony from I+n and Sn+D'' \cite{1939Liv04}. Tin samples were irradiated with 5~MeV deuterons. The subsequent antimony activities were followed for two years after chemical separations.. ``Of these five activities only two can be immediately identified: the 17-minute period is due to Sn$^{119}$(d,n)Sn$^{120}$, as previously reported, while the 2.5-day period must be due to Sn$^{122}$(d,2n)Sb$^{122}$ or to Sn$^{120}$(d,$\gamma$)Sb$^{122}$.'' The reported half-life for $^{122}$Sb of 2.5~d is consistent with the accepted half-life of 2.7238(2)~d. This half-life had previously been measured without a mass assignment \cite{1935Ama01,1937Liv03}.

\subsection*{$^{123}$Sb}\vspace{0.0cm}
Aston discovered stable $^{123}$Sb in 1922 as reported in ``The isotopes of antimony'' \cite{1922Ast04}. A pure antimony trimethyl sample was used in the Cavendish mass spectrograph. ``The element was characterized by two intense lines at 121 and 123. The first is the more intense by perhaps 10 to 20 per cent. If sufficient exposure is given two faint companions are visible at 122, 124, but the general evidence suggests that these are due to hydrogen addition products.  The isotopic nature of the lines 121, 123 is amply confirmed by the appearance of similar pairs 15 and 30 units higher, due to molecules of their monomethides and dimethides.''

\subsection*{$^{124}$Sb}\vspace{0.0cm}
Livingood and Seaborg reported the discovery of $^{124}$Sb in the 1939 article ``Radioactive antimony from I+n and Sn+D'' \cite{1939Liv04}. Sodium iodide was irradiated with fast neutrons produced by bombarding lithium with 8~MeV deuterons. The subsequent antimony activities were followed for 140~days after chemical separations. ``Since there is but one stable form of iodine the reaction must be I$^{127}$(n,$\alpha$)Sb$^{124}$... The present experiment shows conclusively that the 60-day period is associated with Sb$^{124}$ and makes it practically certain that the 2.5-day period belongs to Sb$^{122}$.'' The reported half-life of 60~d is consistent with the presently adopted half-life of 60.20(3)~d. This half-life had previously been measured without a mass assignment \cite{1937Liv03}.

\subsection*{$^{125}$Sb}\vspace{0.0cm}
The observation of $^{125}$Sb was published in 1951 by Stanley and Glendenin as part of the Plutonium Project: ``Study of long-lived antimony in fission (III)'' \cite{1951Sta01}. Uranium was irradiated with neutrons in the Clinton Pile. Absorption and decay curves, as well as $\beta$- and $\gamma$-ray spectra were measured following chemical separation. A 2.7-year half-life was measured which was also observed in neutron irradiation of tin: ``A long-lived antimony activity was observed which was identical in radiation characteristics with the antimony activity from fission. Since neutron-irradiated tin can give rise to only one $\beta$-emitting isotope of antimony, namely, the isotope with mass 125, the long-lived antimony isotope is assigned a mass of 125.'' This half-life agrees with the currently adopted value of 2.75856(25)~y. Previously, a 250$-$300~day halflife was observed without mass assignment \cite{1951Cam01} and a 2.7~y half-life was assigned to either 121, 123 or 125 \cite{1951Lea01}. It should be mentioned that in 1949 Kerr et al.\ \cite{1949Ker01} measured spectra of $^{125}$Sb referring to the known half-life of $^{125}$Sb from the Plutonium Project work.

\subsection*{$^{126}$Sb}\vspace{0.0cm}
The 1956 article ``Ein neues Antimonisotop, $^{126}$Sb'' by Fr\"{a}nz et al.\ described the discovery of $^{126}$Sb \cite{1956Fra02}. A tellurium target was bombarded with 28~MeV deuterons from the Buenos Aires synchrocyclotron and $^{126}$Sb was formed in the reaction $^{128}$Te(d,$\alpha$). Following chemical separation $\beta$- and $\gamma$-ray spectra were measured with a Geiger-M\"uller counter and a single-channel scintillation spectrometer, respectively. ``Wegen der genauen \"Ubereinstimmung der st\"arksten $\gamma$-Linie des 18.8 min-Antimons mit dem ersten angeregten Niveau von $^{126}$Te, bzw. mit der den K-Einfang begleitenden Photonenstrahlung von 650~keV des $^{126}$J, kann man diesem Isotop die Massenzahl 126 zuschreiben.'' [Because of the the exact agreement of the strongest $\gamma$-line of the 18.8~min antimony with the first excited state of $^{126}$Te, or with the K-capture of the accompanying 650~keV radiation of $^{126}$I, it is possible to assign this isotope to mass 126.] This half-life agrees with the presently adopted value of 19.15(8)~m for the isomeric state. A previously measured half-life of 9~h \cite{1951Bar02} was evidently incorrect.

\subsection*{$^{127}$Sb}\vspace{0.0cm}
The discovery of $^{127}$Sb was reported in the 1939 article ``An investigation of the products of the disintegration of uranium by neutrons,'' by Abelson \cite{1939Abe01}. Uranium samples were irradiated with neutrons produced by bombarding beryllium with 8~MeV deuterons from the Berkeley cyclotron. Absorption and decay curves were measured following chemical separation. ``Beta-ray absorption curves of this body and of the lower Te$^{127}$ isomer are identical. These curves are shown in [the figure]. The 10-hour substance is obtained as the daughter of an 80-hour antimony isotope.'' This half-life is close to the accepted half-life of 3.85(5)~d.

\subsection*{$^{128}$Sb}\vspace{0.0cm}
In 1956, Fr\"anz et al.\ identified $^{128}$Sn in ``Die beiden Antimonisomere mit der Massenzahl 128'' \cite{1956Fra01}. Neutrons produced by bombarding beryllium with 28 MeV deuterons from the Buenos Aires synchrocylotron induced fission of uranium. Gamma-rays were detected following chemical separation. ``Das Antimonisotop von 10,3 min Halbwertszeit l\"a\ss t sich von seiner Muttersubstanz, dem Spaltzinn von 57 min Halbwertszeit, sehr rein abtrennen... Man mu\ss\  daher der Isobarenreihe 57 min-Zinn $\rightarrow$ 10,3 min-Antimon die Massenzahl 128 zuordnen.'' [The 10.3~m antimon isotope can be easily separated from the mother substance tin with a half-life of 57~min... Therefore the isobar chain: 57~m tin $\rightarrow$ 10.3~m antimony has to be assigned to mass 128.] In addition, Fr\"anz et al.\ observed a 9.6~h half-life by bombarding tellurium with deuterons or neutrons which they also assigned to $^{128}$Sb. These half-lives agree with the currently accepted values of 9.01(4)~h and 10.4(2)~min for the ground state and the isomeric state of $^{128}$Sb, respectively. In an earlier paper Fr\"anz and Carminatti had assigned the 10.3~min half-life incorrectly to $^{130}$Sn \cite{1955Fra01}. Even earlier a 10~min half-life had been reported without a mass assignment \cite{1951Bar02}.

\subsection*{$^{129}$Sb}\vspace{0.0cm}
The discovery of $^{129}$Sb was reported in the 1939 article ``An investigation of the products of the disintegration of uranium by neutrons,'' by Abelson \cite{1939Abe01}. Uranium samples were irradiated with neutrons produced by bombarding beryllium with 8~MeV deuterons from the Berkeley cyclotron. Absorption and decay curves were measured following chemical separation. ``Among the active fragments of uranium fission is an antimony isotope whose half-life is 4.2~hours... Hence, the most accurate determination of the half-life of the 4.2-hour body is obtained through making periodic extractions of tellurium from antimony and then following decay curves of the separated tellurium fractions to determine the amount of seventy-minute body present at the time of extraction.'' The 70~min half-life was shown to be $^{129}$Te. The reported half-life of 4.2~h for $^{129}$Sb is in agreement with the presently adopted value of 4.40(1)~h. Earlier, Abelson had reported the 4.6~h half-life without a mass assignment \cite{1939Abe02}.

\subsection*{$^{130}$Sb}\vspace{0.0cm}
Hagebo et al.\ reported the discovery of $^{130}$Sb in the 1962 article ``Radiochemical studies of isotopes of antimony and tin in the mass region 127-130'' \cite{1962Hag01}. Enriched $^{130}$Te targets were irradiated with neutrons produced by bombarding beryllium with 21~MeV deuterons from the Amsterdam cyclotron and $^{130}$Sb was formed in (n,p) charge exchange reactions. Decay curves were measured with a liquid $\beta$-counter or an end-window Geiger-M\"{u}ller tube. ``A first analysis of the gross decay curve of antimony samples obtained from the fast neutron irradiation of the tellurium-130 target disclosed half-lives of 10-15 min, about 5 hr and in addition traces of a long-lived tail. The last two components were evidently due to the decay of the 4.6 hr $^{129}$Sb formed by the (n,pn) reaction in tellurium-130. Corrections for the 129-chain and also for the activity formed by reactions in the 3.8 per cent tellurium-128 in the target were performed in a way similar to that described for the antimony-128 decay curve. The final analysis of the antimony decay curves gave two components of 7.1$\pm$0.4 min and 33$\pm$2 min.'' These half-lives are close to the presently accepted values of 6.3(2)~min and 39.5(8)~min for an isomeric and the ground state of $^{130}$Sb, respectively. A 40~min half-life had been reported by Abelson without a mass assignment \cite{1939Abe02}. Earlier assignments of a 10~min half-life to $^{130}$Sb \cite{1955Fra01,1956Pap01} were later reassigned to $^{128}$Sb \cite{1956Fra01}.


\subsection*{$^{131,132}$Sb}\vspace{0.0cm}
In 1956, Pappas and Wiles identified $^{131}$Sb and $^{132}$Sb in ``New short-lived isotopes of tin found in fission'' \cite{1956Pap01}. $^{131}$Sb and $^{132}$Sb were produced via fission of uranium with thermal neutrons. Beta-decay curves were measured with an inverted externally quenched end-window Geiger-M\"uller tube following chemical separation. ``The method of curve analysis used was to fit to the experimental points a constructed curve for the decay of the 23.1-min Sb$^{131}-$25-min Te$^{131}$ pair, subtracting whatever long-lived background (as 4.6-h Sb$^{129}-$70-min Te$^{129}$) was necessary to obtain a reasonable fit. Construction of the curve was based on the genetic relationships shown in [the table], and on equal counting efficiency for antimony-131 and tellurium-131, as demonstrated earlier by Pappas. This constructed curve was then subtracted from the experimental points to give successively the 10-min (Sb$^{130}$) and 2-min (Sb$^{132}$) components.'' The measured half-lives of 23.1~min  ($^{131}$Sb) and 1.9(2)~min ($^{132}$Sb) are close to the presently accepted values of 23.03(4)~min and 2.79(7)~min, respectively. Pappas and Wiles did not consider their measurements new discoveries referring to internal reports \cite{1951Coo01,1953Pap01} and a conference proceeding \cite{1954Pap01}. A 5~min half-life had been reported by Abelson first without a mass assignment \cite{1939Abe02} and then later he assigned it to either $^{132}$Sn, $^{134}$Sb, or $^{136}$Sb \cite{1939Abe03}. This half-life is close to the 4.15(5)~min isomeric state of $^{132}$Sb.

\subsection*{$^{133}$Sb}\vspace{0.0cm}
Strom et al.\ reported the discovery of $^{131}$Sb in the 1966 paper ``Nuclear-charge distribution of fission-product chains of mass numbers 131-133'' \cite{1966Str01}. $^{133}$Sb was produced in thermal-neutron induced fission of $^{235}$U at the Stanford 10-kW reactor. Decay curves were measured following chemical separation ``For $^{133}$Sb, the decay curve given by [the figure] indicates that the yield of $\sim$55-sec tin precursor must be low. No contribution from precursors is evident. The half-life obtained was found to be 2.67$\pm$0.33~min.'' This half-life agrees with the currently accepted value of 2.34(5)~min. Previously, half-lives of 4.1~min \cite{1953Pap01} and 4.2~min \cite{1951Coo01} were only reported in internal reports.

\subsection*{$^{134}$Sb}\vspace{0.0cm}
The discovery of $^{134}$Sb was reported in the 1967 article ``A new antimony delayed neutron precursor,'' by Tomlinson and Hurdus \cite{1967Tom01}. $^{134}$Sb was produced by thermal-neutron induced fission of $^{235}$U at the Harwell LIDO reactor. Decay curves were measured with BF$_3$ counters surrounded by paraffin following rapid chemical separation. ``The half-life obtained by Strom et al. for the half-life of $^{134}$Sb is 11$\pm$1~s. Within experimental error, this latter value is identical to the half-life found in the present work and the mass of the delayed neutron precursor is accordingly assigned to 134.'' The measured half-life of 11.3(3)~s is in agreement with the currently accepted half-life of 10.07(5)~s. The reference to Strom et al.\ mentioned in the quote listed as ``to be published'' was submitted two months later \cite{1968Del01}.

\subsection*{$^{135}$Sb}\vspace{0.0cm}
Bemis et al.\ reported the observation of $^{135}$Sb in the 1964 article ``Half life of $^{135}$Sb and evidence on the half life of $^{134}$Sb'' \cite{1964Bem01}. Uranium samples were irradiated with thermal neutrons in the MIT Reactor. Decay curves were measured with a BF$_3$-filled proportional counter surrounded with a paraffin moderator following chemical separation. ``The $^{135}$Sb decay curve is shown in [the figure]. Uncertainties in activities arise almost entirely from statistical uncertainties in the $^{135}$I activities, as the net counting rates were
often lower than the background rate. Time uncertainties include possible errors in correlating the stop-watch with the master clock controlling the pneumatic-tube operation and uncertainties in the time of SbH$_3$ generation, the duration of which was usually limited to $\sim$2~sec by closing a valve. From the data of [the figure], we derive a half life of 1.9$^{+0.9}_{-0.5}$~sec for $^{135}$Sb'' This half-life is consistent with the accepted value of 1.679(15)~s.

\subsection*{$^{136}$Sb}\vspace{0.0cm}
$^{136}$Sb was discovered by Lund and Rudstam in 1976 as reported in ``Delayed-neutron activities produced in fission: Mass range 122-146'' \cite{1976Lun01}. $^{136}$Sb was produced via neutron fission in a uranium target at the Studsvik R2-0 reactor and separated with the OSIRIS on-line mass-separator facility. Thirty $^3$He neutron counters were used to measure the delayed neutron activities. ``The short-lived activity is found to have a half-life of 0.82 $\pm$ 0.02~sec. Antimony and tellurium are both expected to be delayed-neutron precursors. Isomeric states of even-mass Te isotopes are not expected, however, and $^{136}$Sb is therefore the most probable assignment of this short-lived activity.'' This half-life agrees with the currently accepted value of 0.923(14)~s.

\subsection*{$^{137-139}$Sb}\vspace{0.0cm}
In 1994, Bernas et al.\ published the discovery of $^{137}$Sb, $^{138}$Sb, and $^{139}$Sb in ``Projectile fission at relativistic velocities: A novel and powerful source of neutron-rich isotopes well suited for in-flight isotopic separation'' \cite{1994Ber01}. The isotopes were produced using projectile fission of $^{238}$U at 750 MeV/nucleon on a lead target at GSI, Germany. ``Forward emitted fragments from $^{80}$Zn up to $^{155}$Ce were analyzed with the Fragment Separator (FRS) and unambiguously identified by their energy-loss and time-of-flight.'' This experiment yielded 548, 50, and 8 counts of $^{137}$Sb, $^{138}$Sb, and $^{139}$Sb, respectively.

\subsection*{$^{140}$Sb}\vspace{0.0cm}
The discovery of $^{140}$Sb was reported in the 2010 article ``Identification of 45 new neutron-rich isotopes produced by in-flight fission of a $^{238}$U beam at 345 MeV/nucleon,'' by Ohnishi et al.\ \cite{2010Ohn01}. The experiment was performed at the RI Beam Factory at RIKEN, where the new isotopes were created by in-flight fission of a 345 MeV/nucleon $^{238}$U beam on a lead target. The isotopes were separated and identified with the BigRIPS superconducting in-flight separator. The results for the new isotopes discovered in this study were summarized in a table. Twelve individual counts for $^{140}$Sb were recorded.

\section{Discovery of $^{105-143}$Te}

Thirty-nine tellurium isotopes from A = 105--143 have been discovered so far; these include 8 stable, 16 neutron-deficient and 15 neutron-rich isotopes.
According to the HFB-14 model \cite{2007Gor01}, $^{175}$Te should be the last odd-even particle stable neutron-rich nucleus while the even-even particle stable neutron-rich nuclei should continue through $^{178}$Te while on the neutron deficient side one more isotopes should be particle stable ($^{104}$Te). In addition, four more isotopes ($^{100-103}$Te) could possibly still have half-lives longer than 10$^{-9}$~s \cite{2004Tho01}. Thus, about 39 isotopes have yet to be discovered corresponding to 50\% of all possible tellurium isotopes.

\begin{figure}
	\centering
	\includegraphics[scale=.7]{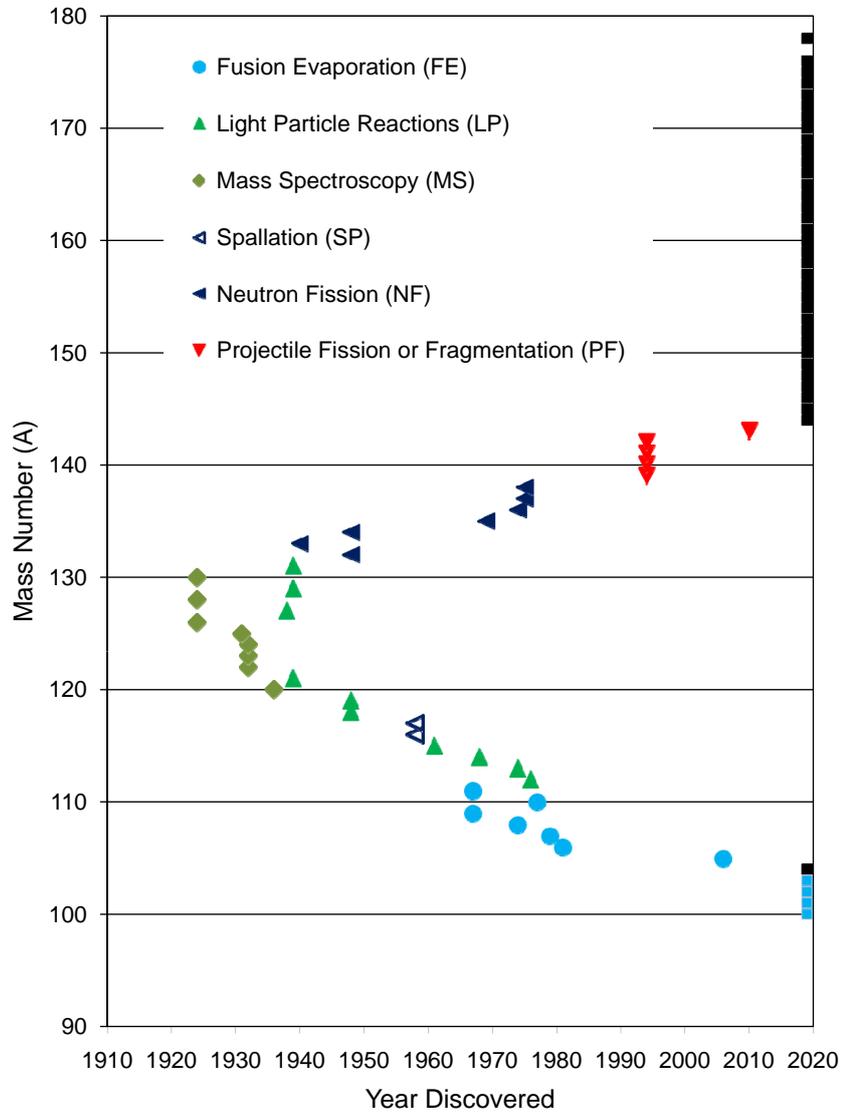}
	\caption{Tellurium isotopes as a function of time when they were discovered. The different production methods are indicated. The solid black squares on the right hand side of the plot are isotopes predicted to be bound by the HFB-14 model. On the proton-rich side the light blue squares correspond to unbound isotopes predicted to have half-lives larger than $\sim 10^{-9}$~s.}
\label{f:year-te}
\end{figure}

Figure \ref{f:year-te} summarizes the year of first discovery for all tellurium isotopes identified by the method of discovery. The range of isotopes predicted to exist is indicated on the right side of the figure. The radioactive tellurium isotopes were produced using fusion evaporation reactions (FE), light-particle reactions (LP), neutron induced fission (NF), spallation reactions (SP), and projectile fragmentation or fission (PF). The stable isotope was identified using mass spectroscopy (MS). In the following, the discovery of each tellurium isotope is discussed in detail.

\subsection*{$^{105}$Te}\vspace{0.0cm}
In 2006, Seweryniak et al.\ published the first identification of $^{105}$Te in ``$\alpha$ decay of $^{105}$Te'' \cite{2006Sew01}. A $^{50}$Cr target was bombarded with 204, 214, and 224~MeV $^{58}$Ni beams accelerated by the Argonne Tandem Linac Accelerator System and $^{105}$Te was formed in the (3n) fusion-evaporation reaction. Recoil products were separated with the Argonne Fragment Analyzer FMA and implanted in a double-sided strip detector which also recorded subsequent $\alpha$-decays. ``The $^{105}$Te half-life was obtained from the decay times of the $^{105}$Te events using the maximum likelihood method. As a result, an energy of E$_\alpha$ = 4720(50)~keV, corresponding to Q$_\alpha$ = 4900(50)~keV, and a half-life of T$_{1/2}$ = 0.7(-0.17 +0.25)~$\mu$s were extracted.'' This half-life is consistent with the currently accepted value of 0.62(7)~$\mu$s. Two months later Liddick et al.\ independently reported a half-life of 620(70)~ns for $^{105}$Te \cite{2006Lid01}.

\subsection*{$^{106}$Te}\vspace{0.0cm}
Schardt et al.\ discovered $^{106}$Te as reported in ``Alpha decay of neutron-deficient isotopes with 52$\leq$Z$\leq$55, including the new isotopes $^{106}$Te($T_{1/2}$=60 $\mu$s) and $^{110}$Xe'' \cite{1981Sch01}. A $^{58}$Ni target was bombarded with a $^{58}$Ni beam from the GSI UNILAC accelerator forming $^{110}$Xe in the (2p4n) fusion-evaporation reaction. $^{110}$Xe was separated with the GSI on-line mass separator and $^{106}$Te was then populated by $\alpha$-decay which were measured with surface-barrier detector telescopes. ``The high-energy $\alpha$-lines in the mass-110 spectrum at 3737$\pm$30~keV and 4160$\pm$30~keV are ascribed to the new isotopes $^{110}$Xe and $^{106}$Te, respectively, based on the mass assignment from the magnetic separation and on $\alpha$-energy systematics.'' The reported half-life of 60$^{+30}_{-10}~\mu$s is consistent with the accepted value of 70(17)~$\mu$s.

\subsection*{$^{107}$Te}\vspace{0.0cm}
In the 1979 paper ``Alpha decay studies of tellurium, iodine, xenon and cesium isotopes'' Schardt et al.\ described the observation of $^{107}$Te \cite{1979Sch02}. A 290~MeV $^{58}$Ni beam from the GSI UNILAC accelerator bombarded a $^{58}$Ni target to produce $^{111}$Xe in the (2p3n) fusion-evaporation reaction. $^{111}$Xe was separated with the GSI on-line mass separator and $^{107}$Te was then populated by $\alpha$-decay which were measured with a detector telescope. ``Therefore we assign the first two lines to the decay of $^{111}$Xe and the 3.833~MeV line, which is fed by the two $^{111}$Xe decays, to $^{107}$Te. Using the signals of both detectors the half-life of $^{107}$Te was determined by the method of maximum likelihood to be 3.6$^{+0.6}_{-0.4}$~ms.'' This value agrees with the currently adopted half-life of 3.1(1)~ms. An earlier reported 2.2(2)~s half-life \cite{1965Mac02} was later reassigned to $^{108}$Te \cite{1974Bog02,1977Kir01}.

\subsection*{$^{108}$Te}\vspace{0.0cm}
$^{108}$Te was observed by Bogdanov et al.\ in the 1974 paper ``Delayed protons from Te$^{109}$ and the $\beta$-decay strength function'' \cite{1974Bog02}. $^{108}$Te was produced in the (4n) fusion-evaporation reaction by bombarding an enriched $^{96}$Ru target with a 137~MeV $^{16}$O beam from the Dubna cyclotron. Reaction products were transported by a gas-jet in front of detectors measuring $\beta$-rays, $\alpha$ and proton decays. ``Different assumptions concerning the number of evaporated neutrons, namely 4 (Te$^{108}$) and 5 (Te$^{107}$), yield for $\epsilon$ the values 7.0$\pm$0.8 and 3.0$\pm$0.6 MeV. The second value is low, and the identification of this isotope as Te$^{108}$ is preferable.'' The reported half-life of 2.0(2)~s is consistent with the accepted half-life of 2.1(1)~s. Previously reported half-lives of 2.2(2)~s and 5.3(4)~s assigned to $^{107}$Te and $^{108}$Te \cite{1965Mac02,1965Sii02} were later reassigned to $^{108}$Te and $^{109}$Te, respectively \cite{1973Bog01,1977Kir01}.

\subsection*{$^{109}$Te}\vspace{0.0cm}
In the 1967 paper ``Proton emitters among Te isotopes'', Karnaukhov et al.\ identified $^{109}$Te \cite{1967Kar01}. $^{92}$Mo and $^{94}$Mo targets were bombarded with a $^{20}$Ne beam from the Dubna 300~cm heavy-ion cyclotron forming $^{109}$Te in the (3n) and (5n) fusion-evaporation reactions, respectively. $^{109}$Te was identified by measuring excitation functions of proton emitters which were detected with a charged-particle spectrometer consisting of a flat proportional counter and a silicon surface barrier detector. ``Proton activities were observed with the half-lives 4.2$\pm$0.2, 19$\pm$0.7, 13$\pm$2, and 60-80 sec. The first two emitters are probably Te isotopes with the respective mass numbers 109 and 111.'' The reported half-life of 4.2(2)~s for $^{109}$Te is part of the weighted average of the accepted value of 4.6(3)~s. Essential the identical paper was submitted to Nuclear Physics A about five months later \cite{1967Kar02}. A previous assignment of a 5.3(4)~s half-life to $^{108}$Te \cite{1965Mac02,1965Sii02} was later changed to $^{109}$Te \cite{1973Bog01,1977Kir01}.

\subsection*{$^{110}$Te}\vspace{0.0cm}
Kirchner et al.\ reported the first observation of $^{110}$Te in the paper ``New neutron-deficient tellurium, iodine, and xenon isotopes produced by reactions of 290 MeV $^{58}$Ni ions on $^{58}$Ni and $^{63}$Cu targets'' in 1977 \cite{1977Kir01}. An enriched $^{58}$Ni target was bombarded with a 290~MeV $^{58}$Ni beam forming $^{110}$Te in (4p2n) or ($\alpha$2p) fusion-evaporation reactions. Beta-, gamma-, and X-rays, as well as protons and $\alpha$ particles were measured following mass separation with the GSI on-line mass separator facility. ``We wish to report in this letter the identification of the new neutron-deficient isotopes $^{108-110}$Te, $^{110-114}$I, and $^{114}$Xe.'' The reported half-life of 18.4(8)~s is in good agreement with the accepted value of 18.6(8)~s.

\subsection*{$^{111}$Te}\vspace{0.0cm}
In the 1967 paper ``Proton emitters among Te isotopes'', Karnaukhov et al.\ identified $^{111}$Te \cite{1967Kar01}. $^{92}$Mo and $^{94}$Mo targets were bombarded with $^{22}$Ne and $^{20}$Ne beams from the Dubna 300~cm heavy-ion cyclotron, respectively, forming $^{111}$Te in the (3n) fusion-evaporation reaction. $^{111}$Te was identified by measuring excitation functions of proton emitters which were detected with a charged-particle spectrometer consisting of a flat proportional counter and a silicon surface barrier detector. ``Proton activities were observed with the half-lives 4.2$\pm$0.2, 19$\pm$0.7, 13$\pm$2, and 60-80 sec. The first two emitters are probably Te isotopes with the respective mass numbers 109 and 111.'' The reported half-life of 19.0(7)~s for $^{111}$Te is part of the weighted average of the accepted value of 19.3(4)~s. Essential the identical paper was submitted to Nuclear Physics A about five months later \cite{1967Kar02}.

\subsection*{$^{112}$Te}\vspace{0.0cm}
Wigmans et al.\ described the first observation of $^{112}$Te in the 1976 article ``Decay of $^{112,113,114}$Te and $^{115}$Te$^{g+m}$'' \cite{1976Wig01}. Natural tin foils were bombarded with 25$-$40~MeV $^3$He beams from the Amsterdam cyclotron and $^{112}$Te was formed in the reaction $^{112}$Sn($^3$He,3n). Gamma-ray singles and coincidence spectra were measured with Ge(Li) detectors after mass separation. ``The production of pure Te sources with the isotope separator offered the possibility to assign two newly found activities uniquely to the isotopes $^{112}$Te and $^{113}$Te.'' The reported half-life of 2.0(2)~min corresponds to the presently accepted value.

\subsection*{$^{113}$Te}\vspace{0.0cm}
In the 1974 paper ``Sur la d\'esint\'egration d'un isotope nouveau: le tellure 133 (T$_{1/2}$=2,0~min)'' Charvet et al.\ reported the observation of $^{113}$Te \cite{1974Cha01}. A $^{112}$Sn target was irradiated with 48~MeV $\alpha$ particles from the Lyon synchrocyclotron forming $^{113}$Te in the ($\alpha$,3n) reaction. Gamma- and X-rays were measured with a Ge(Li) and a Si(Li) detector, respectively. ``A new tellurium activity with T$_{1/2}$ = 2.0$\pm$0.2~min has been produced by irradiating $^{112}$Sn with 45~MeV $\alpha$-particles. The yield curve and the decay rate of $^{113}$Sb causes this activity to be assigned to $^{113}$Te.'' This half-life agrees with the presently adopted value of 1.7(2)~min. Later in the year Burminskii et al.\ independently reported a half-life of 68(15)~s for $^{113}$Te \cite{1974Bur02}. Also, two years later Wigmans et al.\ claimed the discovery of $^{113}$Te although they were aware of the work by Charvet et al.

\subsection*{$^{114}$Te}\vspace{0.0cm}
Rahmouni reported the identification of $^{114}$Te in the 1968 paper ``Nouvelles transitions des \'etains pairs 120,118,116,114 et de l'antimone impair-impair 116'' \cite{1968Rah01}. Natural antimony targets were irradiated with 155~MeV protons from the Orsay synchrocyclotron forming $^{114}$Te in (p,xn) reactions. Gamma-ray coincidences were measured with Ge(Li) and NaI(Tl) detectors following chemical separation. ``Par irradiation de l'antimoine en protons, nous avons form\'e $^{114}$Te. La p\'eriode est de T = 17$\pm$0.5~mn.'' [Irradiating antimony with protons, we formed $^{114}$Te. The half-life is T =17$\pm$0.5~min]. This half-life is close to the currently accepted value of 15.2(7)~min.

\subsection*{$^{115}$Te}\vspace{0.0cm}
$^{115}$Te was discovered by Selinov et al.\ in 1961 as reported in ``New isotope Te$^{115}$'' \cite{1961Sel01}. An enriched $^{112}$Sn target was irradiated with 21~MeV $\alpha$ particles from the Moscow cyclotron. Resulting activities were measured with end-window counters and a one channel scintillation $\gamma$ spectrometer following chemical separation. ``In order to make a positive identification of this isotope, we carried out a fractional separation (in intervals of 5 min) of the isotope Sb$^{115}$ (T = 32~min) by precipitating the antimony with hydrogen sulfide from the solution containing the tellurium. After the activity of the antimony decreased, it was found that the half-life of Te$^{115}$, which is the parent of Sb$^{115}$, is 6.0$\pm$0.5~min.'' This half-life agrees with the accepted half-life of 5.8(2)~min.

\subsection*{$^{116,117}$Te}\vspace{0.0cm}
Kuznetsova et al.\ reported the observation of $^{116}$Te and $^{117}$Te in 1958 in ``The light isotope of tellurium'' \cite{1958Kuz01}. Metallic antimony was bombarded with protons from the Dubna synchrocyclotron. The resulting activity was measured with an end-window MST-40 counter following chemical separation. ``From the decrease with time the radioactivity of the daughter products it was found that the half-value periods of Te$^{116}$ and Te$^{117}$ are 2.5~hours and 1.7~hours, respectively.'' The half-life for $^{116}$Te agrees well with the presently accepted value of 2.49(4)~h, while the half-life of $^{117}$Te is within a factor of two of the current value of 62(2)~min. It should be mentioned that a few years later Vartanov et al.\ quote the Kuznetsova et al.\ results with a half-life of 1.17~h for $^{117}$Te \cite{1961Var01}.


\subsection*{$^{118,119}$Te}\vspace{0.0cm}
Lindner and Perlman discovered $^{118}$Te and $^{119}$Te in 1948 in ``Neutron deficient isotopes of tellurium and antimony'' \cite{1948Lin01}. 40 and 200~MeV deuteron beams accelerated by the Berkeley 184-inch cyclotron bombarded antimony targets. X-, $\beta$-, and $\gamma$-rays were measured following chemical separation. ``The most reasonable assignment of the 6.0-day tellurium-3.5-minute antimony isobars is mass number 118... The assignment of the 4.5-day tellurium-39-hour antimony isobars to mass number 119 is most reasonable in view of their production with 40-Mev deuterons on antimony, their decay characteristics, and the possibilities open.'' The reported values are in agreement with the accepted values of 6.00(2)~d and 4.70(4)~d for $^{118}$Te and $^{119}$Te, respectively.

\subsection*{$^{120}$Te}\vspace{0.0cm}
The discovery of stable $^{120}$Te was reported by Dempster in the 1936 paper ``The isotopic constitution of strontium and tellurium'' \cite{1936Dem02}. Mass spectra of charged atoms from a spark between a palladium electrode and a tellurium electrode were recorded. ``The new isotope at 120 cannot be ascribed to tin as the other strong tin isotopes are not present. In a chart of the elements it extends the limit of stability of the tellurium nucleus to make it more in line with the lightest isotopes of tin, xenon, and barium.''

\subsection*{$^{121}$Te}\vspace{0.0cm}
$^{121}$Te was observed by Seaborg et al.\ as reported in ``Radioactive tellurium: Further production and separation of isomers'' in 1939 \cite{1939Sea02}. Antimony was bombarded with 8~MeV deuterons and 4~MeV protons. Electrons were measured following chemical separation. ``The activation of antimony with eight-Mev deuterons or with four-Mev protons gives in both cases a new chemically identified tellurium with the
half-life about 120 days. The only reasonable reactions which would lead to unstable tellurium isotopes are Sb$^{121}$(d,2n)Te$^{121}$ and Sb$^{121}$(p,n)Te$^{121}$.'' This half-life is close to the accepted value of 164.2(8)~d. A 31-d half-life reported in a conference abstract \cite{1938Tap01} was evidently incorrect.

\subsection*{$^{122-124}$Te}\vspace{0.0cm}
Stable $^{122}$Te, $^{123}$Te, and $^{124}$Te were discovered by Bainbridge in 1932 in ``The constitution of tellurium'' \cite{1932Bai01}. Kahlbaum tellurium was placed in ``wells'' in an aluminum cathode in front of the Franklin Institute mass spectrograph. ``The isotopic constitution of tellurium has been determined by the method previously described. Aston's analysis was confirmed and extended. Additional isotopes of mass numbers 124, 123 and 122 were found with indications of an extremely faint isotope at 127.''

\subsection*{$^{125}$Te}\vspace{0.0cm}
In 1931, Aston described the observation of stable $^{125}$Te in ``The isotopic constitution and atomic weights of selenium, bromine, boron, tungsten, antimony, osmium, ruthenium, tellurium, germanium, rhenium and chlorine'' \cite{1931Ast05}. A pure sample of tellurium chloride was vaporized and analyzed with the Cavendish mass spectrograph. ``The lines of tellurium have now been obtained, the previous analysis confirmed and a new faint isotope at 125 discovered.''

\subsection*{$^{126}$Te}\vspace{0.0cm}
Aston discovered stable $^{126}$Te in 1924 as reported in ``The mass-spectra of cadmium, tellurium, and bismuth'' \cite{1924Ast03}. Metallic tellurium was ground into the anode mixture in front of the Cavendish mass spectrograph. ``Tellurium gives three lines of mass numbers-126, 128, 130. The intensities of the two latter appear about equal and double that of the first. I have repeated this result with an anode containing tellurium and lithium fluoride, and have no reason to doubt that these are all genuine isotopes.''

\subsection*{$^{127}$Te}\vspace{0.0cm}
In 1938, Livingood and Seaborg published the first identification of $^{127}$Te in ``Radioactive iodine isotopes'' \cite{1938Liv06}. Iodine was irradiated with fast neutrons which were produced by bombarding lithium with deuterons. Electrons and $\gamma$-rays were measured following chemical separation. ``The antimony fraction of this same bombardment was inactive, while the tellurium precipitate exhibited a 10-hour half-life which can be ascribed definitely to Te$^{127}$.'' This half-life agrees with the presently accepted half-life of 9.35(7)~h. The 10-h half-life had previously been assigned to either $^{127}$Te or $^{129}$Te in a conference abstract \cite{1938Tap01}.

\subsection*{$^{128}$Te}\vspace{0.0cm}
Aston discovered stable $^{128}$Te in 1924 as reported in ``The mass-spectra of cadmium, tellurium, and bismuth'' \cite{1924Ast03}. Metallic tellurium was ground into the anode mixture in front of the Cavendish mass spectrograph. ``Tellurium gives three lines of mass numbers-126, 128, 130. The intensities of the two latter appear about equal and double that of the first. I have repeated this result with an anode containing tellurium and lithium fluoride, and have no reason to doubt that these are all genuine isotopes.''

\subsection*{$^{129}$Te}\vspace{0.0cm}
In 1939, Seaborg et al.\ reported the first identification of $^{129}$Te in ``Radioactive tellurium: further purification and separation of isomers'' \cite{1939Sea02}. $^{129}$Te was produced as a result of deuteron bombardment of tellurium. ``Isomer separations show that a 70-minute tellurium activity grows from this 30-day period. The 70-minute activity is also undoubtedly produced directly by deuteron bombardment, but the presence of so many other activities makes it difficult to observe; however, an electron emitting activity with half-life about one hour seems to be present in the deuteron activated samples. If our electron-emitting 70-minute activity is to be identified with the 60-minute period that is produced by gamma-rays, the only remaining possibility for the isotopic assignment is Te$^{129}$.'' The reported half-lives of 70~min and  30~d agree with the currently accepted values of 69.6(3)~min and 33.6(1)~d for the ground state and the isomeric state, respectively. Previously, half-lives of 45~min \cite{1935Ama01} and 70~min \cite{1939Abe02} had been measured without a mass assignment and a 60~min half-life was assigned to either $^{129}$Te or $^{131}$Te \cite{1937Bot03}. Also, a 66~min half-life had been assigned to either $^{127}$Te or $^{129}$Te, and the 31-d half-life had been assigned to $^{121}$Te in a conference abstract \cite{1938Tap01}.

\subsection*{$^{130}$Te}\vspace{0.0cm}
Aston discovered stable $^{130}$Te in 1924 as reported in ``The mass-spectra of cadmium, tellurium, and bismuth'' \cite{1924Ast03}. Metallic tellurium was ground into the anode mixture in front of the Cavendish mass spectrograph. ``Tellurium gives three lines of mass numbers-126, 128, 130. The intensities of the two latter appear about equal and double that of the first. I have repeated this result with an anode containing tellurium and lithium fluoride, and have no reason to doubt that these are all genuine isotopes.''

\subsection*{$^{131}$Te}\vspace{0.0cm}
In 1939, Seaborg et al.\ reported the first identification of $^{131}$Te in ``Radioactive tellurium: Further purification and separation of isomers'' \cite{1939Sea02}. $^{131}$Te was produced as a result of deuteron bombardment of tellurium. ``The 1.2-day and 25-minute activities, which are both directly produced by deuteron bombardment of tellurium, are isomeric and isomer separation experiments show that the 25-minute period grows from the 1.2-day activity; it is in fact possible to observe, by successive extractions of iodine, the growth of the eight-day iodine from the 25-minute tellurium activity obtained by extraction from its parent isomer. Bothe and Gentner did not find a 25-minute activity when exposing tellurium to gamma-rays; this activation could produce Te$^{129}$ but not Te$^{131}$, so it appears that the 1.2-day and 25-minute isomers must belong to Te$^{131}$.'' The reported 25~min and 1.2~d half-lives agree with the presently accepted values of 25.0(1)~min and 30(2)~h for the ground state and the isomeric state, respectively. An 8-d half-life reported in a conference abstract \cite{1938Tap01} was evidently incorrect.

\subsection*{$^{132}$Te}\vspace{0.0cm}
In 1948, Katcoff et al.\ published the first identification of $^{132}$Te in ``Ranges in air and mass identification of plutonium fission fragments'' \cite{1948Kat01}. Plutonium foils were irradiated with neutrons in the Los Alamos homogeneous pile. Fission recoils were stopped in a series of fourteen zapon lacquer films. Differential range curves were extracted, the films chemically separated and the activities measured with a Geiger-M\"uller tube. ``In three separate experiments the range of fragments that result in 77-hr.\ Te was found to be greater than the range of those whose mass is known to be 133. This places an upper limit of 132 on the mass of 77-hr.\ Te, its 5-min.\ Sb parent, and its 2.4-hr.\ I daughter. Masses lower than 132 are ruled out because they are assigned to other well-known isotopes. These facts combined with some secondary considerations (e.g., fission yield) establish the mass assignment at 132.'' The half-life of $^{132}$Te agrees with the accepted value of 3.204(13)~d. A 78~h \cite{1940Pol02} half-life had been measured earlier without a mass assignment. Abelson first reported a 72~hr without a mass assignment \cite{1939Abe02} and later assigned a 77~hr to either $^{132}$Te, $^{134}$Te, or $^{136}$Te \cite{1939Abe03}. In 1944 Joliot had assigned a 77~h half-life to an isotope of tellurium with A$>$131 \cite{1944Jol02}. The track length of the 77-h tellurium fragment was measured by Suzor noting a mass assignment of 132 in brackets \cite{1947Suz01,1947Suz02}.

\subsection*{$^{133}$Te}\vspace{0.0cm}
Wu described the identification of $^{133}$Te in the 1940 article ``Identification of two radioactive xenons from uranium fission'' \cite{1940Wu01}. Barium and cesium targets were irradiated with neutrons produced by bombarding beryllium with 16~MeV deuterons. Resulting activities were measured with an ionization chamber following chemical separation. ``Based on these results, the two chains found in uranium and thorium fission may be identified as $_{52}$Te$^{133}$ 60~min.\ $\rightarrow_{53}$I$^{133}$ 22~hr.\ $_{54}$Xe$^{133}$ 5~days $\rightarrow_{55}$Cs$^{133}$...'' The 60-min tellurium activity itself was observed by bombarding uranium with neutrons produced by the Berkeley cyclotron and reported earlier in 1940 \cite{1940Seg01}. The quoted half-life of 60~min for $^{133}$Te agrees with the 55.4(4)~min isomeric state. A 1~hr half-life had been reported earlier without a mass assignment \cite{1939Abe02,1939Abe01,1939Abe03}.

\subsection*{$^{134}$Te}\vspace{0.0cm}
In 1948, Katcoff et al.\ published the first identification of $^{134}$Te in ``Ranges in air and mass identification of plutonium fission fragments'' \cite{1948Kat01}. Plutonium foils were irradiated with neutrons in the Los Alamos homogeneous pile. Fission recoils were stopped in a series of fourteen zapon lacquer films. Differential range curves were extracted, the films chemically separated and the activities measured with a Geiger-M\"uller tube. ``The same three experiments showed that fragments resulting in 43-min.\ Te have, within experimental error, very nearly the same range as those of mass 133. For the same reasons that were presented above, the mass of 43-min.\ Te and its 54-min.\ I daughter must be greater than 133. Since mass 135 is definitely assigned to other isotopes, 134 is very probably the correct mass number. Assignment to a greater mass would require series distortion of the range-mass curve.'' The half-life agrees with the accepted value of 41.8(8)~min. Previously, Abelson had reported a 40~min \cite{1939Abe02,1939Abe01} and later a 43~min \cite{1939Abe03} half-life without a mass assignment.

\subsection*{$^{135}$Te}\vspace{0.0cm}
$^{135}$Te was observed by Denschlag in the 1969 paper ``Independent yield of $^{135}$I in the thermal neutron fission of $^{235}$U, The half-life of $^{135}$Te'' \cite{1969Den01}. Uranyl nitrate (90\% $^{235}$U) was irradiated with thermal neutrons in the Mainz TRIGA-Reactor. Gamma-ray activities of the fission fragments were measured with a NaI(Tl) crystal following chemical separation. ``The independent yield of $^{135}$I in the thermal neutron fission of $^{235}$U (47$\pm$2\% of chain yield) and the half-life of $^{135}$Te (18$\pm$2~sec) have been determined using a rapid chemical isolation of fission iodine via ion exchange on a preformed AgCl precipitate.'' The reported half-life of 18(2)~s $^{135}$Te is included in the calculation of the weighted average of the currently adopted value of 19.0(2)~s. A 29.5~s half-life was mentioned by Herrmann \cite{1964Her01} quoting a 1962 conference proceeding by Greendale et al.\ \cite{1962Gre01}.

\subsection*{$^{136}$Te}\vspace{0.0cm}
$^{136}$Te was observed by Grapengiesser et al.\ in the 1974 paper ``Survey of short-lived fission products obtained using the isotope-separator-on-line facility at Studsvik'' \cite{1974Gra01}. $^{136}$Te was produced by neutron induced fission and identified at the OSIRIS isotope-separator online facility. Separated fission products were then carried by a tape system to a counting station with a plastic scintillator. In the first long table, the half-life of $^{136}$Te is quoted as 24(2)~s, which is close to the currently accepted value of 17.63(8)~s. The observation was not considered a new discovery, quoting a conference proceeding \cite{1969Sch01}. Earlier an upper limit of 20~s was deduced for the half-life of $^{136}$Te \cite{1949Sta01}. Also, an approximate value of $\sim$33~s differs by almost a factor of two and was not determined directly depending on the chemical behavior of secondary iodine products \cite{1967Wun01}.

\subsection*{$^{137,138}$Te}\vspace{0.0cm}
The first observations of $^{137}$Te and $^{138}$Te were described in ``The P$_n$ values of the $^{235}$U(n$_{th}$,f) produced precursors in the mass chains 90, 91, 93-95, 99, 134 and 137-139,'' in 1975 by Asghar et al.\ \cite{1975Asg01}. $^{235}$U targets were irradiated with neutrons from the Grenoble high flux reactor. Beta-ray decay curves were measured following mass separation with the Lohengrin mass separator. ``The present work led to three new periods corresponding to the new isotopes of selenium..., strontium... and tellurium ($^{138}$Te, 1.3$\pm$0.3~sec)'' $^{137}$Te was not considered a new discovery referring to a conference proceeding \cite{1969Sch01}. The reported half-lives of 2.1(5)~s for $^{137}$Te and 1.3(3)~s for $^{138}$Te agree with the presently adopted values of 2.49(5)~s and 1.4(4)~s, respectively.

\subsection*{$^{139-142}$Te}\vspace{0.0cm}
In 1994, Bernas et al.\ published the discovery of $^{139}$Te, $^{140}$Te, $^{141}$Te, and $^{142}$Te in ``Projectile fission at relativistic velocities: a novel and powerful source of neutron-rich isotopes well suited for in-flight isotopic separation'' \cite{1994Ber01}. The isotopes were produced using projectile fission of $^{238}$U at 750 MeV/nucleon on a lead target at GSI, Germany. ``Forward emitted fragments from $^{80}$Zn up to $^{155}$Ce were analyzed with the Fragment Separator (FRS) and unambiguously identified by their energy-loss and time-of-flight.'' This experiment yielded 1549, 264, 39, and 5 counts of $^{139}$Te, $^{140}$Te, $^{141}$Te, and $^{142}$Te, respectively.

\subsection*{$^{143}$Te}\vspace{0.0cm}
The discovery of $^{143}$Te was reported in the 2010 article ``Identification of 45 new neutron-rich isotopes produced by in-flight fission of a $^{238}$U beam at 345 MeV/nucleon,'' by Ohnishi et al.\ \cite{2010Ohn01}. The experiment was performed at the RI Beam Factory at RIKEN, where the new isotopes were created by in-flight fission of a 345 MeV/nucleon $^{238}$U beam on a lead target. The isotopes were separated and identified with the BigRIPS superconducting in-flight separator. The results for the new isotopes discovered in this study were summarized in a table. Eight individual counts for $^{143}$Te were recorded.

\section{Discovery of $^{108-145}$I}

Thirty-eight iodine isotopes from A = 108--145 have been discovered so far; these include 1 stable 19 neutron-deficient and 18 neutron-rich isotopes. According to the HFB-14 model \cite{2007Gor01}, $^{174}$I should be the last odd-odd particle stable neutron-rich nucleus while the odd-even particle stable neutron-rich nuclei should continue through $^{179}$I. The proton dripline has been crossed with the observation of proton emission from $^{109}$. However, four more isotopes $^{104-107}$I could possibly still have half-lives longer than 10$^{-9}$~s \cite{2004Tho01}. Thus, about 36 isotopes have yet to be discovered corresponding to 49\% of all possible iodine isotopes.

\begin{figure}
	\centering
     \includegraphics[scale=.7]{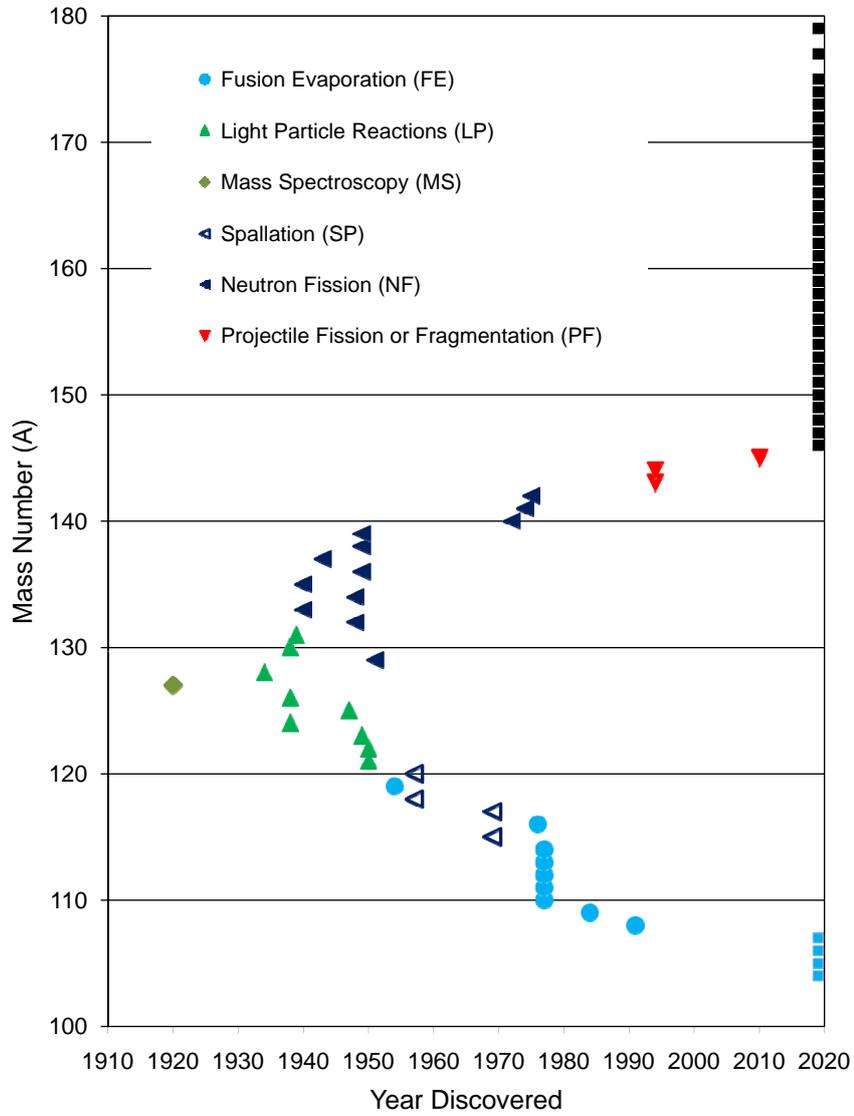}
	\caption{Iodine isotopes as a function of time when they were discovered. The different production methods are indicated. The solid black squares on the right hand side of the plot are isotopes predicted to be bound by the HFB-14 model. On the proton-rich side the light blue squares correspond to unbound isotopes predicted to have half-lives larger than $\sim 10^{-9}$~s.}
\label{f:year-i}
\end{figure}

Figure \ref{f:year-i} summarizes the year of first discovery for all iodine isotopes identified by the method of discovery. The range of isotopes predicted to exist is indicated on the right side of the figure. The radioactive iodine isotopes were produced using fusion evaporation reactions (FE), light-particle reactions (LP), neutron induced fission (NF), spallation (SP), and projectile fragmentation or fission (PF). The stable isotope was identified using mass spectroscopy (MS). In the following, the discovery of each iodine isotope is discussed in detail.

\subsection*{$^{108}$I}\vspace{0.0cm}
Page et al.\ reported the first observation of $^{108}$I in the paper ``Evidence for the alpha decay of $^{108}$I'' in 1991 \cite{1991Pag01}. A $^{54}$Fe target was bombarded with a 260~MeV $^{58}$Ni beam forming $^{108}$I in the (p3n) fusion-evaporation reaction. Recoil products were separated with the Daresbury recoil mass spectrometer (RMS) and implanted into the Residue Implantation Detection System (RIDS) which also recorded subsequent $\alpha$ decays. ``These l-values would suggest a half life of $\sim$50 ms for $^{108}$I, assuming a reduced width of 1.5 as was measured for $^{110}$I. This estimated half life would be consistent with the lower limit of 10~ms measured for these decay lines.'' This reported half-life is close to the accepted value of 36(6)~ms. A previous search for the proton decay of $^{108}$I was not successful \cite{1987Gil01}.

\subsection*{$^{109}$I}\vspace{0.0cm}
Faestermann et al.\ published ``Evidence for proton radioactivity of $^{113}$Cs and $^{109}$I'' in 1984 describing the first observation of $^{109}$I \cite{1984Fae01}. Enriched $^{58}$Ni targets were bombarded with a 250~MeV $^{58}$Ni beam from the Munich MP Tandem-linear accelerator. Evaporation residues were collected on a catcher foil and charged particles were measured with a parallel plate avalanche counter and a Bragg curve spectroscopy ionization chamber. ``...the proton line of 0.83 MeV observed with the $^{58}$Ni + $^{54}$Fe reaction can only be due to the decay of $^{109}$I or $^{105}$Sb. We consider the former as more probable because the production cross section is similar to that for $^{113}$Cs, thus favoring also the (p2n) evaporation channel. In conclusion we assign the observed 0.98 MeV proton line definitely to the decay of $^{113}$Cs and the 0.83 MeV line tentatively to $^{109}$I.'' The reported half-life of 125~$\mu$s is consistent with the presently accepted value of 103(5)~$\mu$s.

\subsection*{$^{110-114}$I}\vspace{0.0cm}
The discoveries of $^{110}$I, $^{111}$I, $^{112}$I, $^{113}$I and $^{114}$I were reported by Kirchner et al.\ in the 1977 publication ``New neutron-deficient tellurium, iodine, and xenon isotopes produced by reactions of 290~MeV $^{58}$Ni ions on $^{58}$Ni and $^{63}$Cu targets'' \cite{1977Kir01}. $^{58}$Ni and $^{63}$Cu targets were bombarded with 290~MeV $^{58}$Ni beams forming $^{110-113}$I and $^{114}$I in (3p(0-3)n) and (4p3n) fusion-evaporation reactions, respectively. Beta-, gamma-, and X-rays, as well as protons and $\alpha$ particles were measured following mass separation with the GSI on-line mass separator facility. ``We wish to report in this letter the identification of the new neutron-deficient isotopes $^{108-110}$Te, $^{110-114}$I, and $^{114}$Xe.'' The reported half-lives of 0.70(6)~s, 3.3(2)~s and 5.8(5)~s are consistent with the accepted half-lives of 0.65(2)~s, 3.42(11)~s and 6.6(2)~s for $^{110}$I, $^{112}$I and $^{113}$I, respectively. The measured half-lives of 2.5(2)~s and 2.1(2)~s are the currently accepted values for $^{111}$I and $^{114}$I, respectively.

\subsection*{$^{115}$I}\vspace{0.0cm}
Hansen et al.\ reported the observation of $^{115}$I in the paper ``Decay characteristics of short-lived radio-nuclides studied by on-line isotope separator techniques'' in 1969 \cite{1969Han01}. Protons of 600~MeV from the CERN synchrocyclotron bombarded a molten tin target and reaction products were separated using the on-line isotope separator ISOLDE. The decay of $^{115}$I was not specifically discussed but the results were presented in a table. The measured half-life of 1.3(2)~m corresponds to the presently adopted value.

\subsection*{$^{116}$I}\vspace{0.0cm}
The observation of $^{116}$I was reported in ``On-line mass separator investigation of the new isotope 2.9-sec $^{116}$I'' in 1976 by Gowdy et al.\ \cite{1976Gow01}. A 83~MeV $^{16}$O beam from the Oak Ridge isochronous cyclotron bombarded a $^{103}$Rh target forming $^{116}$I in the (3n) fusion-evaporation reaction. Reaction products were separated and analyzed with the UNISOR on-line isotope separator. ``Using on-line mass separation $^{116}$I has been identified and its half-life determined to be 2.91$\pm$0.15~sec'' This half-life is the presently adopted value. A 2.9(2)~s half-life had been reported by the UNISOR collaboration in a first overview of their program \cite{1974Ham01}. Previously only an upper limit of $\ll$0.5~min had been reported \cite{1969Han01} and a measurement of a $\sim$2~min half-life \cite{1969Spe01} was evidently incorrect.

\subsection*{$^{117}$I}\vspace{0.0cm}
Hansen et al.\ reported the observation of $^{117}$I in the paper ``Decay characteristics of short-lived radio-nuclides studied by on-line isotope separator techniques'' in 1969 \cite{1969Han01}. Protons of 600~MeV from the CERN synchrocyclotron bombarded a molten tin target and reaction products were separated using the on-line isotope separator ISOLDE. The decay of $^{117}$I was not specifically discussed but the results were presented in a table. The measured half-life of 2.4(1)~m agrees with the currently accepted value of 2.22(4)~min. Previous reports of $\sim$10~min \cite{1960Zai01}, 14.5~m \cite{1965But01} and 6.5(1)~m \cite{1965And01} for the half-life of $^{117}$I were evidently incorrect. One month later, Sergolle et al.\ independently measured a 2.4(3)~min half-life \cite{1969Ser01} and six month later Ladenbauer-Bellis et al.\ reported a 2.7(2)~min half-life \cite{1969Lad01} quoting neither Hansen et al.\ nor Sergolle et al.

\subsection*{$^{118}$I}\vspace{0.0cm}
In 1957 Aagaard et al.\ described the observation of $^{118}$I in the paper ``Measurements on electromagnetically separated radioactive isotopes of iodine'' \cite{1957Aag01}. Cesium chloride was irradiated with 170~MeV protons in the Uppsala synchrocyclotron and $^{118}$I was produced in spallation reactions. An electromagnetic isotope separator was used following chemical separation and the resulting activities were counted with a Geiger-M\"uller tube. ``In order to check some previously uncertain mass assignments the decay curves of mass numbers 118$-$122 were measured. The main results are summarized in [the table].'' The measured $\sim$10~min half-life for $^{118}$I agrees with currently adopted values of 13.7(5)~min and 8.5(5)~min for the ground state and the isomeric state, respectively.

\subsection*{$^{119}$I}\vspace{0.0cm}
The identification of $^{119}$I was reported by Rossi et al.\ in 1954 in ``The acceleration of nitrogen-14 (+6) ions in a 60-inch cyclotron'' \cite{1954Ros01}. A 120~MeV $^{14}$N beam from the Berkeley 60-in.\ cyclotron bombarded palladium foils forming $^{119}$I in the fusion evaporation reactions $^{110}$Pd($^{14}$N,5n) or $^{108}$Pd($^{14}$N,3n). Decay curves were measured following chemical separation. ``The 18-minute activity may be the same as the $\sim$30-minute radio-iodine found by Marquez and Perlman \cite{1950Mar01} from high-energy helium-ion bombardments of antimony, and assigned by them to either I$^{119}$ or I$^{120}$. The 5-day tail in the decay curve may correspond to 4.5-day Te$^{119}$, in which case the 18-minute radio-iodine probably is I$^{119}$.'' The reported half-life of 18(1)~min is consistent with the accepted half-life of 19.1(4)~min.

\subsection*{$^{120}$I}\vspace{0.0cm}
In 1957 Aagaard et al.\ described the observation of $^{120}$I in the paper ``Measurements on electromagnetically separated radioactive isotopes of iodine'' \cite{1957Aag01}. Cesium chloride was irradiated with 170~MeV protons in the Uppsala synchrocyclotron and $^{120}$I was produced in spallation reactions. An electromagnetic isotope separator was used following chemical separation and the resulting activities were counted with a Geiger-M\"uller tube. ``Neutron deficient iodine isotopes down to mass number 118 were separated from spallation products of caesium. The following mass assignments are reported: $^{119}$I (19~min), $^{120}$I (1.4~hr), and $^{121}$I (2.0~hr).'' The measured half-life for $^{120}$I is close to the currently adopted values of 81.6(2)~min. Previously, a $\sim$30~min half-life was assigned to either $^{119}$I or $^{120}$I \cite{1950Mar01}.

\subsection*{$^{121,122}$I}\vspace{0.0cm}
The discovery of $^{121}$I and $^{122}$I was reported by Marquez and Perlman in the 1950 article ``Neutron deficient isotopes of iodine'' \cite{1950Mar01}. Antimony targets were irradiated with 45$-$360~MeV $\alpha$ particles from the Berkeley 174-in.\ cyclotron. Electrons, X-rays, and $\gamma$-rays were measured following chemical separation. ``1.8-HR. I$^{121}$: An activity with 1.8-hr. half-life with a 1.2-Mev positron and conversion electrons of 185~kev appeared in irradiation of antimony with 60-, 100-, and 360-Mev helium ions. Its decay is followed by the appearance of 17-day Te$^{121}$ in approximately the proper yield for a parent-daughter relationship... 4-MIN. I$^{122}$: ...More rapid chemistry showed a 4-min.\ iodine activity, and a single yield determination at 45~MeV showed it to be in the expected range as indicated in [the figure]... The assignment of this activity to I$^{122}$ is also based on the fact that it was the only new activity beyond I$^{123}$, I$^{124}$, I$^{125}$, and I$^{126}$ to appear at 45~Mev.'' The reported half-lives of 4~min ($^{121}$I) and 1.8~hr ($^{122}$I) are consistent with the presently accepted values of 3.63(6)~min and 2.12(1)~h, respectively.

\subsection*{$^{123}$I}\vspace{0.0cm}
Mitchell et al.\ published ``Disintegration of I$^{124}$ and I$^{126}$'' in 1949 describing the first observation of $^{123}$I \cite{1949Mit01}. Metallic antimony was bombarded by 23~MeV $\alpha$-particles from the Indiana University cyclotron. The iodine was chemically separated and counted with an end-window Geiger-M\"{u}ller counter. ``An active state of I, of 13.0$\pm$0.5~hr.\ half-life, which decays by K-capture and the emission of an internally converted gamma-ray of energy 0.159~Mev, has been found. This activity is assigned to I$^{123}$.'' This half-life agrees with the currently adopted value of 13.2235(19)~h.

\subsection*{$^{124}$I}\vspace{0.0cm}
In 1938, Livingood and Seaborg identified $^{124}$I in ``Radioactive iodine isotopes'' \cite{1938Liv05}. Antimony targets were bombarded with 16~MeV $\alpha$ particles from the Berkeley cyclotron and $^{124}$I was formed in the $^{121}$Sb($\alpha$,n) reaction. Reaction products were chemically separated and the resulting activities were observed with a Lauritsen-type quartz-fiber electroscope. ``There are but two stable antimony isotopes, Sb$^{121}$ and Sb$^{123}$, which by the ($\alpha$,n) reaction would lead to radioactive I$^{124}$ and I$^{126}$. Since prolonged exposure of iodine to fast neutrons yields only the 13-day period [through I$^{127}$(n,2n)I$^{126}$] but not the 4-day activity, it is certain that this latter must be due to I$^{124}$.'' This half-life is in agreement with the accepted value of 4.1760(3)~d.

\subsection*{$^{125}$I}\vspace{0.0cm}
In the 1947 article ``A note on the long-lived radio-iodine'' Glendenin and Edwards identified $^{125}$I \cite{1947Gle01}. Tellurium targets were bombarded with 14~MeV deuterons from the M.I.T. cyclotron to produce $^{125}$I in the reactions $^{125}$Te(d,2n) and $^{124}$Te(d,n). Decay curves and absorption spectra were measured with a Kr-filled Geiger counter following chemical separation. ``It can be shown by various considerations that the mass number of the long-lived iodine is more probably 125 than 129 as proposed by Reid and Keston. The formation of I$^{125}$ directly in the Te targets at the cyclotron is to be expected from the d,2n reaction on Te$^{125}$ (6 percent abundance) and the (d,n) reaction on Te$^{124}$ (4.5 percent abundance).'' The 56~d reported half-life is close to the accepted half-life of 59.407(10)~d. As mentioned in the quote Reid and Keston had assigned this half-life incorrectly to $^{129}$Te a year earlier \cite{1946Rei01}.

\subsection*{$^{126}$I}\vspace{0.0cm}
In 1938, Livingood and Seaborg published the first identification of $^{126}$I in ``Radioactive iodine isotopes'' \cite{1938Liv06}. Iodine was irradiated with fast neutrons produced by bombarding lithium with deuterons. The resulting activities were measured following chemical separation. ``We have irradiated iodine with the fast neutrons from a lithium plus deuterons source and confirm the 13-day period reported for the same bombardment by Tape and Cork, who surmised it to be due to I$^{126}$. We have chemically identified this activity as an iodine isotope, so that the assignment to I$^{126}$ appears to be definite.'' This half-life agrees with the accepted half-life of 12.93(5)~d. The work by Tape and Cork mentioned in the quote was only published in a conference abstract \cite{1938Tap01}.

\subsection*{$^{127}$I}\vspace{0.0cm}
Aston discovered stable $^{127}$I in 1920 as reported in ``The constitution of the elements'' \cite{1920Ast05}. CH$_3$I vapor was used in the Cavendish mass spectrograph to identify $^{127}$I. ``Fortunately, iodine (atomic weight 126.92) gave definite and unmistakable effects. It proves to be a simple element of mass 127-a result satisfactorily confirmed by a single line at 142 corresponding to CH$_3$I, the vapour used in the experiments.''

\subsection*{$^{128}$I}\vspace{0.0cm}
In 1934 Fermi et al.\ reported the first observation of $^{128}$I in the paper ``Artificial radioactivity produced by neutron bombardment'' \cite{1934Fer01}. A neutron source of beryllium powder and radon was used to irradiate iodine and ammonium iodide. Activities were then followed with Geiger-M\"uller counters. ``53$-$Iodine$-$: We irradiated both the element and ammonium iodide. Both showed a strong activity, decaying with a period of 30~minutes.'' In a table the half-life was assigned to $^{128}$I. This half-life is close the currently adopted value of 24.99(2)~min.

\subsection*{$^{129}$I}\vspace{0.0cm}
The identification of $^{129}$I was reported by Katcoff in the 1951 publication ``Half-life of I$^{129}$ and the age of the elements'' \cite{1951Kat06}. $^{129}$I was produced by irradiating a uranium slug in the Oak Ridge pile for four years. Activities were measured with proportional counters following chemical separation and the half-life was deduced from  measuring the $^{129}$I to $^{127}$I ratio in a 60$^\circ$ sector type mass spectrometer. Nine radioactive methyl iodide samples were measured. ``The average of the nine values for the half-life of I$^{129}$ is 1.72$\times$10$^7$~years with an over-all estimated probable error of 5 percent.'' This half-life agrees with the currently adopted half-life of 1.57(4)$\times$10$^7$~y. The existence of $^{129}$I had been known for awhile but no properties had been measured \cite{1951Lea02}. An estimated half-life of 10$^8$~y was reported four years earlier \cite{1947Kat02} and in 1949 a 3(1)$\times$10$^7$~y half-life was measured and described in an internal report \cite{1949Par03}.

\subsection*{$^{130}$I}\vspace{0.0cm}
In 1938, Livingood and Seaborg identified $^{130}$I in ``Radioactive iodine isotopes'' \cite{1938Liv05}. Tellurium targets were bombarded with 8~MeV deuterons from the Berkeley cyclotron forming $^{130}$I in the reactions $^{130}$Te(d,2n) and $^{128}$Te(d,$\gamma$). Reaction products were chemically separated and the resulting activities were then observed with a Lauritsen-type quartz-fiber electroscope. ``We have not been able to find any trace of this 13-hour period growing in deuteron-activated tellurium, nor do we find it after neutron bombardments of tellurium. We therefore feel confident that it must be assigned to I$^{130}$.'' The experimental half-life of 12.6(10)~h is in agreement with the accepted half-life of 12.36(1)~h.

\subsection*{$^{131}$I}\vspace{0.0cm}
In 1939, Seaborg et al.\ identified $^{131}$I in ``Radioactive tellurium: further purification and separation of isomers'' \cite{1939Sea02}. Tellurium was bombarded with 8~MeV deuterons and an 8-day iodine activity was measured following chemical separation as reported in an earlier paper \cite{1938Liv06}. However, it was not possible to decide if the activity was due to $^{129}$I or $^{131}$I. This question was resolved with the assignment of the parent activities to $^{131}$Te: ``The 1.2-day and 25-minute activities, which are both directly produced by deuteron bombardment of tellurium, are isomeric and isomer separation experiments show that the 25-minute period grows from the 1.2-day activity; it is in fact possible to observe, by successive extractions of iodine, the growth of the eight-day iodine from the 25-minute tellurium activity obtained by extraction from its parent isomer.'' The reported half-life of 8~d for $^{131}$I agrees with the accepted half-life of 8.0252(6)~d. The 8-day half-life had been reported earlier without a mass assignment \cite{1939Abe02}.

\subsection*{$^{132}$I}\vspace{0.0cm}
In 1948, Katcoff et al.\ published the first identification of $^{132}$I in ``Ranges in air and mass identification of plutonium fission fragments'' \cite{1948Kat01}. Plutonium foils were irradiated with neutrons in the Los Alamos homogeneous pile. Fission recoils were stopped in a series of fourteen zapon lacquer films. Differential range curves were extracted, the films chemically separated and the activities measured with a Geiger-M\"uller tube. ``In three separate experiments the range of fragments that result in 77-hr.\ Te was found to be greater than the range of those whose mass is known to be 133. This places an upper limit of 132 on the mass of 77-hr.\ Te, its 5-min.\ Sb parent, and its 2.4-hr.\ I daughter. Masses lower than 132 are ruled out because they are assigned to other well-known isotopes. These facts combined with some secondary considerations (e.g., fission yield) establish the mass assignment at 132.'' The half-life of $^{132}$I agrees with the accepted value of 2.295(13)~h. In 1939/1940 a 2.3~hr \cite{1939Hah04}, a 2.5~h \cite{1939Abe02}, and a 2.4~h \cite{1940Pol02} half-life were reported without a mass assignment. Also, Abelson assigned the 2.4~h half-life to either $^{132}$I, $^{134}$I, or $^{136}$I \cite{1939Abe03}.

\subsection*{$^{133}$I}\vspace{0.0cm}
Wu described the identification of $^{133}$I in the 1940 article ``Identification of two radioactive xenons from uranium fission'' \cite{1940Wu01}. Barium and cesium targets were irradiated with neutrons produced by bombarding beryllium with 16~MeV deuterons. Resulting activities were measured with an ionization chamber following chemical separation. ``Based on these results, the two chains found in uranium and thorium fission may be identified as $_{52}$Te$^{133}$ 60~min.\ $\rightarrow_{53}$I$^{133}$ 22~hr.\ $\rightarrow_{54}$Xe$^{133}$ 5~days $\rightarrow_{55}$Cs$^{133}$...'' The 22-h iodine activity itself was observed by bombarding uranium with neutrons produced by the Berkeley cyclotron and reported earlier in 1940 \cite{1940Seg01}. The quoted half-life 22~h for $^{133}$I agrees with the presently adopted value of 20.8(1)~h. The 22~hr half-life had been reported earlier without a mass assignment \cite{1939Abe02,1939Abe01,1939Abe03}.

\subsection*{$^{134}$I}\vspace{0.0cm}
In 1948, Katcoff et al.\ published the first identification of $^{134}$I in ``Ranges in air and mass identification of plutonium fission fragments'' \cite{1948Kat01}. Plutonium foils were irradiated with neutrons in the Los Alamos homogeneous pile. Fission recoils were stopped in a series of fourteen zapon lacquer films. Differential range curves were extracted, the films chemically separated and the activities measured with a Geiger-M\"uller tube. ``From the range-mass curve drawn for well-known masses, definite assignments of 92, 93, and 132 were given to 3.5-hr.\ Y, 10-hr.\ Y, and 77-hr.\ Te, respectively. Highly probable assignments of 94 and 134 were given to 20-min.\ Y and 54-min.\ I, respectively.'' This half-life for $^{134}$I agrees with the currently accepted value of 52.5(2)~min for $^{134}$I. The 54~min half-life had been reported earlier without a mass assignment \cite{1939Abe02,1939Abe01,1939Abe03}.

\subsection*{$^{135}$I}\vspace{0.0cm}
Wu described the identification of $^{135}$I in the 1940 article ``Identification of two radioactive xenons from uranium fission'' \cite{1940Wu01}. Barium and cesium targets were irradiated with neutrons produced by bombarding beryllium with 16~MeV deuterons. Resulting activities were measured with an ionization chamber following chemical separation. ``Based on these results, the two chains found in uranium and thorium fission may be identified as... and $_{52}$Te$^{135}\sim$10~min.\ $\rightarrow_{53}$I$^{135}$ 6.6~hr.\ $\rightarrow_{54}$Xe$^{135}$ 9.4~hr.\ $\rightarrow_{55}$Cs$^{135}$.'' The 6.6-h iodine activity itself was observed by bombarding uranium with neutrons produced by the Berkeley cyclotron and reported earlier in 1940 \cite{1940Seg01}. The quoted half-life 6.6~h for $^{133}$I agrees with the presently adopted value of 6.57(2)~h.

\subsection*{$^{136}$I}\vspace{0.0cm}
In 1948, Stanley and Katcoff identified $^{136}$I in the paper ``The properties of 86-second I$^{(136)}$'' \cite{1949Sta01}. Uranium and plutonium were irradiated in the Los Alamos homogeneous pile. The resulting activities were measured with a bell-type Geiger counter with a mica window following chemical separation. ``From similar considerations of fission yield, 86-sec.\ I is probably not isomeric with these either. Mass numbers 133 and 134 are also improbable for another reason: 60-min Te$^{133}$ and 43-min.\ Te$^{184}$ cannot decay to 86-sec.\ I with a fission yield greater than 0.05 percent as shown in [the table]. The possibility that 86-sec.\ I decays to a long-lived isomeric state of 3.4-min.\ Xe$^{137}$ or of 17-min.\ Xe$^{138}$ has not been ruled out. Nevertheless the most probable assignment of 86-sec.\ I is to mass 136.'' This half-life of 86(1)~s is consistent with the accepted half-life of 83.4(10)~s. Strassmann and Hahn had reported a 1.8(4)~min half-life without a mass assignment \cite{1940Str01}.

\subsection*{$^{137}$I}\vspace{0.0cm}
The identification of $^{137}$I was determined by Riezler in the 1947 article ``Aktivierung von Xenon durch Neutronen'' \cite{1943Rie01}. In 1940 Strassmann and Hahn observed a new activity of iodine in neutron-induced fission of uranium with a half-life of 30(6)~s \cite{1940Str01}. Three years later Seelmann-Eggebert and Born established that this 30~s activity decays to a 3.8~min activity of Xenon \cite{1943See01}. Riezler used neutron irradiation of xenon to identify the origin of this 3.8~min half-life \cite{1943See01}. ``Die Strahlung des 3.4-Minuten-K\"orpers ist sehr hart, 2~mm Aluminium lassen noch 25\% durch. Es ist anzunehmen, da\ss\ diese Aktivit\"at mit dem von H.\ J.\ Born und W.\ Seelman-Eggebert bei der Uranspaltung gefundenen 3.8-Minuten-K\"orper identisch ist. Als Massenzahl kommt dann nur 137 in Frage.'' [The radiation of the 3.4~min emitter is very hard; 25\% are transmitted through 2~mm aluminum. It can be assumed that this activity is identical to the 3.8~min emitter that H.\ J.\ Born and W.\ Seelman-Eggebert found in the fission of uranium. Only mass number 137 is then reasonable.] Although Riezler did not measure or mention $^{137}$I directly his mass identification of $^{137}$Xe directly implied the identification of $^{137}$I as acknowledged later by Snell et al.\ \cite{1947Sne01} and Sugarman \cite{1949Sug01}. The reported half-life of 30(6)~s agrees with the currently accepted half-life of 24.5(2)~s.

\subsection*{$^{138,139}$I}\vspace{0.0cm}
The discoveries of $^{138}$I and $^{139}$I were reported by Sugarman in the 1949 publication ``Short-lived halogen fission products'' \cite{1949Sug01}. Uranyl nitrate was irradiated to produce $^{138}$I and $^{139}$I by neutron induced fission. Decay curves were measured following chemical separation. ``Isolation of Cs$^{138}$ both from the gas separated from AgI precipitates and from the precipitates directly led to a half-life determination for I$^{138}$ of 5.9$\pm$0.4~sec. Extractions of Ba$^{149}$ yielded a value of 2.7$\pm$0.1~sec. for the half-life of I$^{139}$.'' These half-lives are close to the presently accepted values of 6.23(3)~s and 2.28(11)~s for $^{138}$I and $^{139}$I, respectively.

\subsection*{$^{140}$I}\vspace{0.0cm}
``Hauptkomponenten unter den Vorl\"aufern verz\"ogerter Neutronen bei der Spaltung von Uran-235 durch thermische Neutronen'' reported the first observation of $^{140}$I by Sch\"ussler and Herrmann in 1972 \cite{1972Sch01}. $^{140}$I was produced by irradiating $^{235}$U with thermal neutrons in the Mainz reactor. Neutron decay curves were measured with five $^3$He filled proportional counters following chemical separation. ``A new delayed-neutron precursor, 0.8~sec $^{140}$I, has been identified and another one, 1.1~sec $^{98,99}$Y, has been confirmed.'' This half-life for $^{140}$I agrees with the presently adopted value of 860(40)~ms.

\subsection*{$^{141}$I}\vspace{0.0cm}
In 1974, Kratz and Herrmann discovered $^{141}$I as published in ``Delayed-neutron emission from short-lived Br and I isotopes'' \cite{1974Kra01}. Enriched $^{235}$U was irradiated with thermal neutrons in the Mainz Triga reactor to produce $^{141}$I. Neutron decay curves were measured with five $^3$He filled proportional counters following chemical separation. ``In addition to the known precursors $^{137}$I, $^{138}$I, $^{139}$I and $^{140}$I, the iodine fraction, [the figure] shows a component of 0.41~sec half-life attributed to $^{141}$I.'' This half-life of 0.41(8)~s agrees with the presently adopted value of 0.43(2)~s. Two months later Rudstam et al.\ \cite{1974Rud01} independently reported a half-life of 0.44~s quoting a compilation \cite{1973Tom01} and a conference proceeding \cite{1973Rud01}. The compilation by Tomlinson quoted the half-life as 0.44(6)~s based on an internal report \cite{1970Her01} and stated that $^{141}$I had also been identified in a conference proceedings \cite{1969Sch01} and in a paper ``to be published'' \cite{1973Sch03}.

\subsection*{$^{142}$I}\vspace{0.0cm}
The observation of $^{142}$I was reported by Kratz et al.\ in the article ``Gamma-ray emission from short-lived bromine and iodine isotopes'' in 1975 \cite{1975Kra01}. $^{142}$I was produced by irradiating $^{235}$U with thermal neutrons in the Mainz Triga reactor. Gamma-ray spectra were measured with two Ge(Li) detectors following chemical separation. ``In the xenon region there is still a slight evidence for a $\simeq$0.2 sec growth of $^{142}$Xe from $^{142}$I.'' This half-life is the currently accepted value.

\subsection*{$^{143,144}$I}\vspace{0.0cm}
In 1994, Bernas et al.\ published the discovery of $^{143}$I and $^{144}$I in ``Projectile fission at relativistic velocities: a novel and powerful source of neutron-rich isotopes well suited for in-flight isotopic separation'' \cite{1994Ber01}. The isotopes were produced using projectile fission of $^{238}$U at 750 MeV/nucleon on a lead target at GSI, Germany. ``Forward emitted fragments from $^{80}$Zn up to $^{155}$Ce were analyzed with the Fragment Separator (FRS) and unambiguously identified by their energy-loss and time-of-flight.'' This experiment yielded 49 and 7 counts of $^{143}$I and $^{144}$I, respectively.

\subsection*{$^{145}$I}\vspace{0.0cm}
The discovery of $^{145}$I was reported in the 2010 article ``Identification of 45 new neutron-rich isotopes produced by in-flight fission of a $^{238}$U beam at 345 MeV/nucleon,'' by Ohnishi et al.\ \cite{2010Ohn01}. The experiment was performed at the RI Beam Factory at RIKEN, where the new isotopes were created by in-flight fission of a 345 MeV/nucleon $^{238}$U beam on a lead target. The isotopes were separated and identified with the BigRIPS superconducting in-flight separator. The results for the new isotopes discovered in this study were summarized in a table. 57 individual counts for $^{145}$I were recorded.

\section{Discovery of $^{109-148}$Xe}

Forty xenon isotopes from A = 109--148 have been discovered so far; these include 9 stable, 17 neutron-deficient and 14 neutron-rich isotopes.
According to the HFB-14 model \cite{2007Gor01}, $^{179}$Xe should be the last odd-even particle stable neutron-rich nucleus while the even-even particle stable neutron-rich nuclei should continue through $^{184}$Xe. At the proton dripline one more isotope ($^{108}$Xe) is predicted to be particle stable. In addition, $^{103-107}$Xe could possibly still have half-lives longer than 10$^{-9}$~s \cite{2004Tho01}. Thus, about 40 isotopes have yet to be discovered corresponding to 50\% of all possible xenon isotopes.

\begin{figure}
	\centering
	\includegraphics[scale=.7]{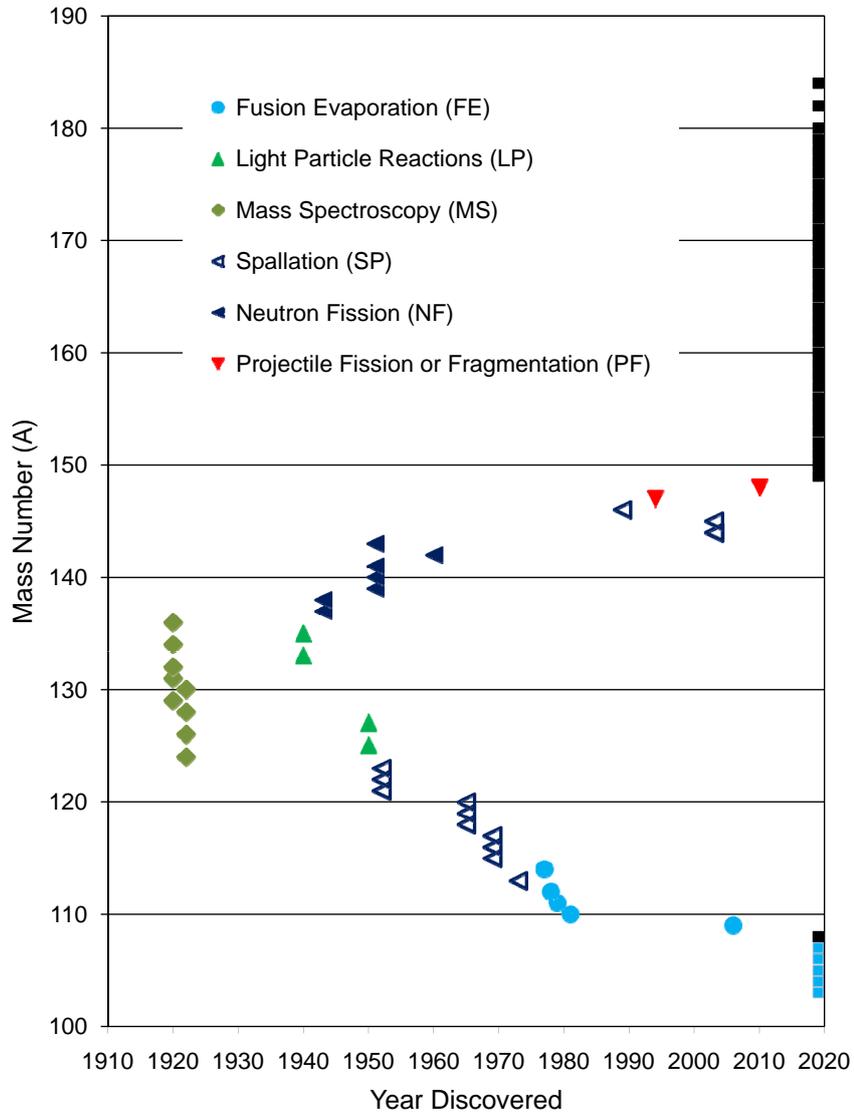}
	\caption{Xenon isotopes as a function of time when they were discovered. The different production methods are indicated. The solid black squares on the right hand side of the plot are isotopes predicted to be bound by the HFB-14 model. On the proton-rich side the light blue square corresponds to an unbound isotope predicted to have a lifetime larger than $\sim 10^{-9}$~s.}
\label{f:year-xe}
\end{figure}

Figure \ref{f:year-xe} summarizes the year of first discovery for all xenon isotopes identified by the method of discovery. The range of isotopes predicted to exist is indicated on the right side of the figure. The radioactive xenon isotopes were produced using fusion evaporation reactions (FE), light-particle reactions (LP), neutron induced fission (NF), spallation reactions (SP), and projectile fragmentation or fission (PF). The stable isotopes were identified using mass spectroscopy (MS). In the following, the discovery of each xenon isotope is discussed in detail.

\subsection*{$^{109}$Xe}\vspace{0.0cm}
In 2006, Liddick et al.\ reported the observation of $^{109}$Xe in the paper ``Discovery of $^{109}$Xe and $^{105}$Te: Superallowed $\alpha$ decay near the doubly magic $^{100}$Sn'' \cite{2006Lid01}. A $^{54}$Fe target was bombarded with 220$-$225~MeV $^{58}$Ni beams and $^{109}$Xe was formed in the (3n) fusion-evaporation reaction. Reaction products were separated with the Holifield Radioactive Ion Beam Facility Recoil Mass Spectrometer and implanted into a double-sided silicon strip detector which also recorded subsequent $\alpha$ decays. ``A total of 100 $\alpha-\alpha$ decay events were attributed to the $^{109}$Xe$\rightarrow^{105}$Te$\rightarrow^{101}$Sn decay chain... Using the method of [the reference] the half-life of $^{109}$Xe is 13$\pm$2~ms.'' This half-life is the currently accepted value.

\subsection*{$^{110}$Xe}\vspace{0.0cm}
Schardt et al.\ discovered $^{110}$Xe as reported in ``Alpha decay of neutron-deficient isotopes with 52$\leq$Z$\leq$55, including the new isotopes $^{106}$Te($T_{1/2}$=60 $\mu$s) and $^{110}$Xe'' \cite{1981Sch01}. A $^{58}$Ni target was bombarded with a $^{58}$Ni beam from the GSI UNILAC accelerator forming $^{110}$Xe in the (2p4n) fusion-evaporation reaction. $^{110}$Xe was separated with the GSI on-line mass separator and subsequent $\alpha$ decays were measured with surface-barrier detector telescopes. ``The high-energy $\alpha$-lines in the mass-110 spectrum at 3737$\pm$30~keV and 4160$\pm$30~keV are ascribed to the new isotopes $^{110}$Xe and $^{106}$Te, respectively, based on the mass assignment from the magnetic separation and on $\alpha$-energy systematics.'' The half-life was not measured and only estimated to be about 0.2~s. The currently adopted value is 93(3)~ms.

\subsection*{$^{111}$Xe}\vspace{0.0cm}
In the 1979 paper ``Alpha decay studies of tellurium, iodine, xenon and cesium isotopes'' Schardt et al.\ described the observation of $^{111}$Xe \cite{1979Sch02}. A 290~MeV $^{58}$Ni beam from the GSI UNILAC accelerator bombarded a $^{58}$Ni target to produce $^{111}$Xe in the (2p3n) fusion-evaporation reaction. $^{111}$Xe was separated with the GSI on-line mass separator and subsequent $\alpha$ decay was measured with a detector telescope. ``The 3.463~MeV line and the 3.560~MeV line are both correlated with the 3.833~MeV line. Therefore we assign the first two lines to the decay of $^{111}$Xe and the 3.833~MeV line, which is fed by the two $^{111}$Xe decays, to $^{107}$Te... From the decay data for the 3.560 MeV $\alpha$-line of $^{111}$Xe shown in [the figure], a half-life of 0.9$\pm$0.2~s is determined.'' This value agrees with the currently adopted half-life of 740(20)~ms.

\subsection*{$^{112}$Xe}\vspace{0.0cm}
The discovery of $^{112}$Xe was announced by Roeckl et al.\ in the 1978 paper ``A new island of $\alpha$-emission: $\alpha$-decay energies and widths of neutron deficient tellurium, iodine and xenon isotopes \cite{1978Roe01}. A 5~MeV/u $^{58}$Ni beam from the GSI UNILAC bombarded a $^{58}$Ni target. Evaporation residues were separated with the GSI on-line mass separator and implanted into a carbon foil in front of a $\Delta$E-E telescope. ``Using on-line mass separation, eight $\alpha$-emitters (given with their $\alpha$-decay energies in keV) in the trans-tin region were investigated: $^{108}$Te, 3300(30); $^{109}$Te, 3080(15); $^{110}$I, 3424(15); $^{111}$I, 3150(30); $^{112}$I, 2866(50); $^{112}$Xe, 2990(30); $^{114}$Cs (or $^{114}$Ba), 3226(30).'' The measured half-life of 2.8(2)~s agrees with the currently adopted value of 2.7(8)~s.

\subsection*{$^{113}$Xe}\vspace{0.0cm}
In the 1977 paper ``Beta-delayed proton emitter $^{113}$Xe'', Hagberg et al.\ reported the discovery of $^{113}$Xe \cite{1973Hag01}. A hydrous cerium oxide target was bombarded with 600~MeV protons from the CERN synchrocyclotron. $^{113}$Xe was separated with the ISOLDE separator and subsequent proton emission was measured in Ilford K2 emulsion plates. ``The proton energy spectrum measured for $^{113}$Xe is shown in [the figure]. It is very similar to that of $^{115}$Xe, but the half-life shown in the inset (T$_{\frac{1}{2}}$=2.8$\pm$0.2~sec) clearly demonstrates that any contribution from $^{115}$Xe must be negligibly small.'' This half-life is consistent with the currently adopted value of 2.74(8)~s.

\subsection*{$^{114}$Xe}\vspace{0.0cm}
$^{114}$Xe was discovered in 1977 by Kirchner et al.\ and the results were described in ``New neutron-deficient tellurium, iodine, and xenon isotopes produced by reactions of 290~MeV $^{58}$Ni ions on $^{58}$Ni and $^{63}$Cu targets'' \cite{1977Kir01}. A $^{58}$Ni target was bombarded with 290~MeV $^{58}$Ni beams forming $^{114}$Xe in (2p) fusion-evaporation reactions. Beta-, gamma-, and X-rays, as well as protons and $\alpha$ particles were measured following mass separation with the GSI on-line mass separator facility. ``We wish to report in this letter the identification of the new neutron-deficient isotopes $^{108-110}$Te, $^{110-114}$I, and $^{114}$Xe.'' The measured 10.0(4)~s half-life corresponds to the presently accepted value.

\subsection*{$^{115-117}$Xe}\vspace{0.0cm}
Hansen et al.\ reported the first observation of $^{115}$Xe, $^{116}$Xe, and $^{117}$Xe in the paper ``Decay characteristics of short-lived radio-nuclides studied by on-line isotope separator techniques'' in 1969 \cite{1969Han01}. Protons of 600~MeV from the CERN synchrocyclotron bombarded a molten tin target and cadmium was separated using the ISOLDE facility. The paper summarized the ISOLDE program and did not contain details about the individual nuclei but the results were presented in a table. The measured half-lives of 19(5)~s for $^{115}$Xe and 55(3)~s for $^{116}$Xe agree with the currently adopted values of 18(4)~s and 61(2)~s, respectively. The half-life of 1.08(10)~min for $^{116}$Xe is included in the calculation of the currently adopted half-life.

\subsection*{$^{118-120}$Xe}\vspace{0.0cm}
$^{118}$Xe, $^{119}$Xe, and $^{120}$Xe were discovered by Andersson et al.\ in 1965 in ``Decay data on some Xe, I, and Te isotopes'' \cite{1965And01}. Lanthanum was bombarded with 600~MeV protons at CERN to produce $^{118}$Xe, $^{119}$Xe, and $^{120}$Xe in spallation reactions. Beta- and gamma-ray spectra were measured with a beta counter and a NaI(Tl) scintillation spectrometer, respectively, following chemical separation. ``From the decay of suitable peaks in consecutive scintillation gamma spectra, Xe$^{118}$ (peak at about 50~keV) and Xe$^{119}$ (peak at about 100~keV) were assigned the same half-life, 6$\pm$1~min, in both cases defined over six periods. In the same way a half-life of 40$\pm$1~min was found for Xe$^{120}$ (x-rays and 70~keV peak).'' The measured half-lives of 6(1)~min for $^{118}$Xe and 6(1)~min for $^{119}$Xe are close to the presently adopted values of 3.8(9)~min and 5.8(3)~min, respectively. The half-life of 40(1)~min for $^{120}$Xe is the currently accepted value. Six months later Butement and Qaim \cite{1965But01} independently assigned a half-life of 43(3)~min to $^{120}$Xe.

\subsection*{$^{121-123}$Xe}\vspace{0.0cm}
Dropesky and Wiig reported the first observation of $^{121}$Xe, $^{122}$Xe, and $^{123}$Xe in the 1952 article ``Three new neutron deficient xenon isotopes'' \cite{1952Dro01}. Anhydrous lithium iodide was irradiated with 240~MeV protons in the Rochester cyclotron forming $^{121}$Xe, $^{122}$Xe, and $^{123}$Xe in spallation reactions. Activities were measured with a NaI(Tl) phosphor and a halogen filled GM tube following chemical separation. ``From the yields of I$^{121}$ and I$^{123}$ versus the time of parent-daughter separations the half-lives of Xe$^{121}$ and Xe$^{123}$ were determined to be 40($\pm$2)~minutes and 1.7($\pm$0.2)~hours, respectively. The yields of I$^{122}$ proved to be completely compatible with the half-life of 20($\pm$1)~hours found with a GM tube for the xenon in the last trap of the series.'' These values agree with the currently adopted half-lives of 40.1(20)~min, 20.1(1)~h, and 2.08(2)~h for $^{121}$Xe, $^{122}$Xe, and $^{123}$Xe, respectively.

\subsection*{$^{124}$Xe}\vspace{0.0cm}
Aston discovered stable $^{124}$Xe in 1922 as reported in ``The isotopes of selenium and some other elements'' \cite{1922Ast03}. Xenon gas was used in the Cavendish mass spectrograph. ``During some work requiring very prolonged exposures with a gas containing xenon, two new isotopes of that element were discovered at 124, 126, making nine in all. The extreme faintness of both lines indicates that the proportion of these light isotopes in the element is minute.''

\subsection*{$^{125}$Xe}\vspace{0.0cm}
The first observation of $^{125}$Xe was published in the 1950 paper ``Radioactive xenon 125, and xenon 127'' by Anderson and Pool \cite{1950And01}. Enriched $^{122}$Te and $^{124}$Te targets were bombarded with alpha particles and $^{125}$Xe was formed in the reaction $^{122}$Te($\alpha$,n). Absorption spectra and decay curves were measured with a Wulf electrometer following chemical separation. ``From absorption measurements repeated at intervals it was possible to show that the intensity of the x-radiation decayed with a half-life of the order of 20~hours. This would seem to establish that the 20-hour Xe$^{125}$ activity decays, in large measure, by K-capture.'' The quoted half-life of 20(1)~h is close to the currently accepted value of 16.9(2)~h.

\subsection*{$^{126}$Xe}\vspace{0.0cm}
Aston discovered stable $^{126}$Xe in 1922 as reported in ``The isotopes of selenium and some other elements'' \cite{1922Ast03}. Xenon gas was used in the Cavendish mass spectrograph. ``During some work requiring very prolonged exposures with a gas containing xenon, two new isotopes of that element were discovered at 124, 126, making nine in all. The extreme faintness of both lines indicates that the proportion of these light isotopes in the element is minute.''

\subsection*{$^{127}$Xe}\vspace{0.0cm}
The first observation of $^{125}$Xe was published in the 1950 paper ``Radioactive xenon 125, and xenon 127'' by Anderson and Pool \cite{1950And01}. Enriched $^{122}$Te and $^{124}$Te targets were bombarded with alpha particles and $^{125}$Xe was formed in the reaction $^{124}$Te($\alpha$,n). Absorption spectra and decay curves were measured with a Wulf electrometer following chemical separation. ``The activity obtained by the reaction Te$^{124}$($\alpha$,n)Xe$^{127}$ has a half-life of 32$\pm$2~days.'' This value is in agreement with the currently adopted half-life of 36.4(1)~d. Ten years earlier, Creutz et al.\ reported a 34(2)~d half-life in proton induced reactions on $^{127}$I \cite{1940Cre01}. However, they did not explicitly state that they observed the (p,n) reaction rather than the (p,$\gamma$). In the 1948 Table of Isotopes \cite{1948Sea01} the assignment to $^{127}$Xe was made based on the results by Creutz et al.

\subsection*{$^{128}$Xe}\vspace{0.0cm}
The identification of stable $^{128}$Xe was reported by Aston in 1922 in ``The isotopes of tin'' \cite{1922Ast02}. The existence of $^{128}$Xe was demonstrated with the Cavendish mass spectrograph. ``Incidentally I may add that the presence of the two faint components of xenon 128 and 130 previously suspected has now been satisfactorily confirmed.'' Two years earlier Aston had tentatively reported the existence of 128 \cite{1920Ast01} but subsequently reassigned it to 129 \cite{1920Ast05}. In the same paper as well as in two other paper he again tentatively reported a xenon isotope with mass 128, quoting it in brackets \cite{1920Ast05,1921Ast05,1921Ast04}.

\subsection*{$^{129}$Xe}\vspace{0.0cm}
Stable $^{129}$Xe was identified by Aston in November 1920 in ``The constitution of the elements'' \cite{1920Ast05}. Xenon gas was used in the Cavendish mass spectrograph. In February of 1920 Aston had assigned 5 lines of xenon isotopes: ``The partial pressure, of xenon (atomic weight 130.2) in the gas used was only sufficient to show its singly charged lines clearly. These appear to follow the whole number rule, and rough provisional values for the five made out may be taken as 128, 130, 131, 133, and 135.'' \cite{1920Ast05}. In the November 1920 paper Aston reassigned the 128 line to mass 129 and stated: ``Owing to the kindness of Prof.\ Collie and Dr.\ Masson in providing me with a sample of gas rich in xenon, I have been able to identify two more probable isotopes of that element and obtain trustworthy values for the atomic weights of the five already found. The provisional figures given for these turn out to be too low. The values quoted below were obtained from the position of the second-order line 64.5. They should be trustworthy to about one·fifth of a unit.''

\subsection*{$^{130}$Xe}\vspace{0.0cm}
The identification of stable $^{130}$Xe was reported by Aston in 1922 in ``The isotopes of tin'' \cite{1922Ast02}. The existence of $^{130}$Xe was demonstrated with the Cavendish mass spectrograph. ``Incidentally I may add that the presence of the two faint components of xenon 128 and 130 previously suspected has now been satisfactorily confirmed.'' Two years earlier Aston had tentatively reported the existence of 130 \cite{1920Ast01} but subsequently reassigned it to 131 \cite{1920Ast05}. In the same paper as well as in two other paper he again tentatively reported a xenon isotope with mass 130, quoting it in brackets and a question mark \cite{1920Ast05,1921Ast05,1921Ast04}.

\subsection*{$^{131,132}$Xe}\vspace{0.0cm}
Stable $^{131}$Xe and $^{132}$Xe were identified by Aston in November 1920 in ``The constitution of the elements'' \cite{1920Ast05}. Xenon gas was used in the Cavendish mass spectrograph. In February of 1920 Aston had assigned 5 lines of xenon isotopes: ``The partial pressure, of xenon (atomic weight 130.2) in the gas used was only sufficient to show its singly charged lines clearly. These appear to follow the whole number rule, and rough provisional values for the five made out may be taken as 128, 130, 131, 133, and 135.'' \cite{1920Ast05}. In the November 1920 paper Aston reassigned the 130 and 131 lines to mass 131 and 132, respectively, and stated: ``Owing to the kindness of Prof.\ Collie and Dr.\ Masson in providing me with a sample of gas rich in xenon, I have been able to identify two more probable isotopes of that element and obtain trustworthy values for the atomic weights of the five already found. The provisional figures given for these turn out to be too low. The values quoted below were obtained from the position of the second-order line 64.5. They should be trustworthy to about one·fifth of a unit.''

\subsection*{$^{133}$Xe}\vspace{0.0cm}
Wu reported the identification of $^{133}$Xe in the 1940 article ``Identification of two radioactive xenons from uranium fission'' \cite{1940Wu01}. Barium and cesium targets were irradiated with neutrons produced by bombarding beryllium with 16 MeV deuterons forming $^{133}$Xe in the reactions $^{136}$Ba(n,$\alpha$) and $^{133}$Cs(n,p), respectively. Resulting activities were measured with an ionization chamber following chemical separation. ``Cesium has only one stable isotope with a mass number of 133, and since its bombardment gives only one radioactive xenon, we assume that the 5-day xenon is produced by the following reaction $_{55}$Cs$^{133}$(n,p)$_{54}$Xe$^{133}$ and assign to it the mass number 133. Although barium has seven known isotopes (130, 132, 134, 135, 136, 137 and 138), only three of them would be able to produce radioactive xenon (except for isomers of stable nuclei) by a (n,$\alpha$) reaction. These are 130, 136, and 138. We have seen that Ba gives by a (n,$\alpha$) reaction the 5-day xenon and we interpret this as $_{56}$Ba$^{136}$(n,$\alpha$)$_{54}$Xe$^{133}$.'' This half-life agrees with the currently adopted value of 5.2475(5)~d. Segre and Wu had reported the 5~d half-life in xenon without a mass assignment earlier in the year \cite{1940Seg01}. Also a 5.5~d half-life had previously been observed in noble gases from neutron fission of thorium \cite{1939Lan01} and a 4.3~d half-life was assigned to a xenon isotope with mass larger than 131 \cite{1940Dod01}. In addition, in 1920 Aston had incorrectly reported $^{133}$Xe to be a stable isotope \cite{1920Ast05}.

\subsection*{$^{134}$Xe}\vspace{0.0cm}
Stable $^{134}$Xe was identified by Aston in November 1920 in ``The constitution of the elements'' \cite{1920Ast05}. Xenon gas was used in the Cavendish mass spectrograph. In February of 1920 Aston had assigned 5 lines of xenon isotopes: ``The partial pressure, of xenon (atomic weight 130.2) in the gas used was only sufficient to show its singly charged lines clearly. These appear to follow the whole number rule, and rough provisional values for the five made out may be taken as 128, 130, 131, 133, and 135.'' \cite{1920Ast05}. In the November 1920 paper Aston reassigned the 133 line to mass 134 and stated: ``Owing to the kindness of Prof.\ Collie and Dr.\ Masson in providing me with a sample of gas rich in xenon, I have been able to identify two more probable isotopes of that element and obtain trustworthy values for the atomic weights of the five already found. The provisional figures given for these turn out to be too low. The values quoted below were obtained from the position of the second-order line 64.5. They should be trustworthy to about one·fifth of a unit.''

\subsection*{$^{135}$Xe}\vspace{0.0cm}
Wu reported the identification of $^{135}$Xe in the 1940 article ``Identification of two radioactive xenons from uranium fission'' \cite{1940Wu01}. Barium targets were irradiated with neutrons produced by bombarding beryllium with 16 MeV deuterons forming $^{135}$Xe in the reactions $^{138}$Ba(n,$\alpha$). Resulting activities were measured with an ionization chamber following chemical separation. ``Although barium has seven known isotopes (130, 132, 134, 135, 136, 137 and 138), only three of them would be able to produce radioactive xenon (except for isomers of stable nuclei) by a(n,$\alpha$) reaction. These are 130, 136, and 138. We have seen that Ba gives by a (n,$\alpha$) reaction the 5-day xenon and we interpret this as $_{56}$Ba$^{136}$(n,$\alpha$)$_{54}$Xe$^{133}$. Since the abundance of Ba$^{130}$ is only 1/700 of that of Ba$^{138}$, it is very likely that the barium of mass number 138 is responsible for the formation of the 9.4-hour xenon according to the reaction $_{56}$Ba$^{138}$(n,$\alpha$)$_{54}$Xe$^{135}$.'' This half-life agrees with the currently adopted value of 9.14(2)~h. Segre and Wu had reported the 9.4~h half-life in xenon without a mass assignment earlier in the year \cite{1940Seg01}. In 1920 Aston had incorrectly reported $^{135}$Xe to be a stable isotope \cite{1920Ast05}.

\subsection*{$^{136}$Xe}\vspace{0.0cm}
Stable $^{136}$Xe was identified by Aston in November 1920 in ``The constitution of the elements'' \cite{1920Ast05}. Xenon gas was used in the Cavendish mass spectrograph. In February of 1920 Aston had assigned 5 lines of xenon isotopes: ``The partial pressure, of xenon (atomic weight 130.2) in the gas used was only sufficient to show its singly charged lines clearly. These appear to follow the whole number rule, and rough provisional values for the five made out may be taken as 128, 130, 131, 133, and 135.'' \cite{1920Ast05}. In the November 1920 paper Aston reassigned the 135 line to mass 136 and stated: ``Owing to the kindness of Prof.\ Collie and Dr.\ Masson in providing me with a sample of gas rich in xenon, I have been able to identify two more probable isotopes of that element and obtain trustworthy values for the atomic weights of the five already found. The provisional figures given for these turn out to be too low. The values quoted below were obtained from the position of the second-order line 64.5. They should be trustworthy to about one·fifth of a unit.''

\subsection*{$^{137}$Xe}\vspace{0.0cm}
The identification of $^{137}$Xe was reported by Riezler in the 1947 article ``Aktivierung von Xenon durch Neutronen'' \cite{1943Rie01}. Xenon gas was irradiated with thermal neutrons and neutrons produced by bombarding beryllium and lithium with 7~MeV deuterons. The resulting activities were measured with a double-walled counter. ``Die Strahlung des 3.4-Minuten-K\"orpers ist sehr hart, 2~mm Aluminium lassen noch 25\% durch. Es ist anzunehmen, da\ss\ diese Aktivit\"at mit dem von H.\ J.\ Born und W.\ Seelman-Eggebert bei der Uranspaltung gefundenen 3.8-Minuten-K\"orper identisch ist. Als Massenzahl kommt dann nur 137 in Frage.'' [The radiation of the 3.4~min emitter is very hard; 25\% are transmitted through 2~mm aluminum. It can be assumed that this activity is identical to the 3.8~min emitter that H.\ J.\ Born and W.\ Seelman-Eggebert found in the fission of uranium. Only mass number 137 is then reasonable.] This half-life agrees with the currently accepted half-life of 3.818(13)~min. The quote referred to a1943 paper by Seelmann-Eggebert and Born who measured a 3.8~min half-life without a mass assignment \cite{1943See01}.

\subsection*{$^{138}$Xe}\vspace{0.0cm}
The identification of $^{138}$Xe was determined in the 1943 article ``\"Uber einige aktive Xenon-Isotope'' by Seelmann-Eggebrecht \cite{1943See03}. In 1939 Hahn and Strassmann observed a $\sim$15~min half-life of xenon decaying to a 33~min cesium activity in neutron-induced fission of uranium \cite{1939Hah03}. A year later these results were confirmed by Glasoe and Steigmen reporting a 17(1)~min xenon half-life decaying to 32.0(5)~min cesium isotope \cite{1940Gla02}. Seelmann-Eggebrecht used fast neutron irradiation of barium to identify the origin of this cesium activity \cite{1943See03}. ``Au\ss er den bei der Bestrahlung von Barium mit schnellen Neutronen durch n,$\alpha$-Proze\ss\ entstehenden Xenon-Isotopen entsteht, ebenfalls mit schnellen Neutronen, durch einen n,p-Proze\ss\ aus dem Barium ein C\"asium-Isotop von 33 Minuten Halbwertszeit. Sowohl nach seiner Halbwertszeit von 33 Minuten als auch nach der Absorptionskurve seiner $\beta$-Strahlen ist dieses mit dem bei der Uranspaltung nachgewiesenen 33-Minuten-C\"asium identisch. Nun hat W. Riezler das 3,8-Minuten-Xenon der Masse 137 zuordnen k\"onnen. Da dieses Xenon-Isotop jedoch keinen Folgek\"orper von 33 Minuten Halbswertzeit besitzt, mu\ss\ dieses C\"asium-Isotop der Masse 138 zugeordnet werden.'' [In addition to the xenon isotopes produced in (n,$\alpha$) processes with fast neutrons, a cesium isotope with a half-life of 33~min was produced in (n,p) processes also induced with fast neutrons. This isotope is according to the 33~min half-life as well as the $\beta$-ray absorption curve identical with the 33~min cesium isotope observed in uranium fission. W. Riezler was recently able to assign a 3.8~min xenon isotope to mass 137. The present cesium isotope must therefore be assigned to mass 138, because there is an isotope with a 33~min half-life in the decay chain of this xenon isotopes.] Although Seelmann-Eggebrecht did not measure or mention $^{138}$Xe directly his mass identification of $^{138}$Cs directly implied the identification of $^{138}$Xe. In 1948 Sugarman assumes a half-life of 17~min for $^{148}$Xe to be known without a reference. The reported half-lives of $\sim$15~min \cite{1939Hah03} and 17(1)~min \cite{1940Gla02} agree with the currently accepted value of 14.08(8)~min.

\subsection*{$^{139-141}$Xe}\vspace{0.0cm}
Dillard et al.\ reported the discovery of $^{139}$Xe, $^{140}$Xe, and $^{141}$Xe in 1951 as part of the Manhattan Project Technical Series: ``Determination of gas half-life by the charged-wire technique (II)'' \cite{1951Dil01}. A uranium salt solution was irradiated with neutrons in the Argonne Graphite Pile to produce  $^{139}$Xe, $^{140}$Xe, and $^{141}$Xe. Noble gases were swept out of the uranium solution during the irradiation and the half-lives were measured with the charged-wire technique. ``Xe$^{139}$:... [The figure] shows the distribution of 85 m Ba$^{139}$ activity on the wire 8.5~hr after the end of the irradiation. The activity fell to half value every 6.14~in. At the flow rate of 100~cc/min an average gas atom required 6.66~sec to travel 1~in. up in the tube, hence the half-life of Xe$^{139}$ is 41~sec... Xe$^{140}$:... [The figure] shows the distribution of 12.8~d Ba$^{140}$ activity 29~hr after the irradiation. The activity fell to half value every 5.6~in., indicating that the half-life of Xe$^{140}$ is 16.3~sec... Xe$^{141}$:... Analyses of the decay curves gave the distribution of 28~d Ce$^{141}$. A semilogarithmic graph of the activity of Ce$^{141}$ per unit of wire length against time indicates that the activity fell to half value every 1.8~in. At the flow rate used this gives a half-life of 2.7~sec. Averaged with other determinations the half-life is considered to be 3$\pm$1~sec.'' These half-lives of 41~s for $^{139}$Xe, 16.0(5)~s for $^{140}$Xe, and 3(1)~s for $^{141}$Xe are close to the currently adopted values of 39.68~s, 13.60(10)~s, and 1.73(1)~s, respectively.

\subsection*{$^{142}$Xe}\vspace{0.0cm}
In the 1960 article ``Identification of $^{142}$Xe and measurement of its cumulative yield from thermal-neutron fission of $^{235}$U'', Wolfsberg et al.\ reported the identification of $^{142}$Xe \cite{1960Wol01}. Uranyl nitrate enriched in $^{235}$U was irradiated with neutrons produced by bombarding beryllium with deuterons from the Washington University cyclotron. Activities were measured with a side-window, methane-flow, $\beta$-proportional counter following chemical separation. ``Since the lower emanating power at atmospheric pressure is due to the successful competition of the rate of decay of xenon with the rate of diffusion of xenon in air, the half-life of $^{142}$Xe must be only a little less than the half-life of $^{141}$Xe, say $\sim$1.5~sec.'' This half-life agrees with the currently adopted half-life of 1.22(2)~s.

\subsection*{$^{143}$Xe}\vspace{0.0cm}
Dillard et al.\ reported the discovery of  $^{143}$Xe in 1951 as part of the Manhattan Project Technical Series: ``Determination of gas half-life by the charged-wire technique (II)'' \cite{1951Dil01}. A uranium salt solution was irradiated with neutrons in the Argonne Graphite Pile to produce $^{143}$Xe. Noble gases were swept out of the uranium solution during the irradiation and the half-lives were measured with the charged-wire technique. ``Xe$^{143}$:... An analysis of the decay curves gives the Xe$^{143}$ distribution. The 33~h activity fell to half value every 1.25~in. Therefore the half-life of Xe$^{143}$ is 1.0~sec. This half-life is based on data that are inconclusive and cannot be considered to be more accurate than 1.0$\pm$0.5~sec.'' This half-life is close to the currently adopted value of 0.30(3)~s.

\subsection*{$^{144,145}$Xe}\vspace{0.0cm}
Bergmann et al.\ reported the identification of $^{144}$Xe and $^{145}$Xe in the 2003 paper ``Beta-decay properties of the neutron-rich $^{94-99}$Kr and $^{142-147}$Xe'' \cite{2003Ber01}. Uranium fission was induced with a pulsed beam of 1 or 1.4~GeV protons from the CERN proton synchrotron booster facility. Reaction products were separated with the on-line isotope separator ISOLDE and the half-lives were measured with a cylindrical 4$\pi$ long neutron-counter. ``There is only partial agreement between the results of the present experiment and the existing data. In particular, the half-lives from the earlier indirect radiochemical measurements (quoted by nuclear data evaluators for $^{95}$Kr and $^{144,145}$Xe) deviate considerably from our results, indicating that these identifications probably were not correct.'' The measured half-lives are listed in a table as 388(7)~ms for $^{144}$Xe and 188(4)~ms for $^{145}$Xe and correspond to the currently accepted values. The previous measurements mentioned in the quotes referred to half-lives of 1~s \cite{1962Wah01} and 1.15(20)~s \cite{1976Ahr01} for $^{144}$Xe and a 900(300)~ms half-life for $^{145}$Xe \cite{1971Wol01}. Another measurement of a 8.8(22)~s half-life for $^{144}$Xe \cite{1966Laz01} was evidently also incorrect.

\subsection*{$^{146}$Xe}\vspace{0.0cm}
In the 1989 paper ``Xenon isotopes far from stability studied by collisional ionization laser spectroscopy'' Borchers et al.\ reported the discovery of $^{146}$Xe \cite{1989Bor02}. Neutron rich isotopes were produced by proton-induced spallation of lanthanum or by uranium fission. Hyperfine structure and isotopes shifts of xenon isotopes were measured in collinear laser spectroscopy. ``The new collisional ionization scheme in collinear laser spectroscopy has enabled the study of hyperfine structures and isotope shifts of xenon isotopes over the large mass range A=116$-$146. The sensitivity of this method is demonstrated by the discovery of $^{146}$Xe.''

\subsection*{$^{147}$Xe}\vspace{0.0cm}
In 1994, Bernas et al.\ published the discovery of $^{147}$Xe in ``Projectile fission at relativistic velocities: a novel and powerful source of neutron-rich isotopes well suited for in-flight isotopic separation'' \cite{1994Ber01}. The isotopes were produced using projectile fission of $^{238}$U at 750 MeV/nucleon on a lead target at GSI, Germany. ``Forward emitted fragments from $^{80}$Zn up to $^{155}$Ce were analyzed with the Fragment Separator (FRS) and unambiguously identified by their energy-loss and time-of-flight.'' This experiment yielded 4 counts of $^{147}$Xe.

\subsection*{$^{148}$Xe}\vspace{0.0cm}
The discovery of $^{148}$Xe was reported in the 2010 article ``Identification of 45 new neutron-rich isotopes produced by in-flight fission of a $^{238}$U beam at 345 MeV/nucleon,'' by Ohnishi et al.\ \cite{2010Ohn01}. The experiment was performed at the RI Beam Factory at RIKEN, where the new isotopes were created by in-flight fission of a 345 MeV/nucleon $^{238}$U beam on a lead target. $^{148}$Xe was separated and identified with the BigRIPS superconducting in-flight separator. The results for the new isotopes discovered in this study were summarized in a table. One count of $^{148}$Xe was recorded.

\section{Summary}
The discoveries of the known palladium, antimony, tellurium, iodine, and xenon isotopes have been compiled and the methods of their production discussed. A summary of the individual discoveries is in Table 1. A chronological listing of the discoveries displayed in Table 2 shows the emergence of the different production methods and the predominance of the various laboratories and countries as a function of time. The discoveries are sorted by the submission date of the publications. In the few cases where a submission date was not listed the publication date was used.

The identification of the palladium isotopes was relatively straightforward. The half-lives of $^{109}$Pd and $^{111}$Pd had initially been reported without a mass assignment, the half-life of $^{98}$Pd was at first assigned to either $^{96}$Pd or $^{98}$Pd, and the half-life of $^{107}$Pd was at first assigned to an isomeric state of $^{105}$Pd.

The half-lives of six antimony isotopes ($^{122,124,125,127,128,132}$Sb) had initially been measured without mass assignments, the assignment of $^{126}$Sb was at first incorrect and the half-life of $^{128}$Sb had initially been assigned to $^{130}$Sn or to $^{130}$Sb.

The original assignments of $^{107}$Te and $^{108}$Te were corrected to $^{108}$Te and $^{109}$Te, respectively. In addition, the half-lives of $^{132-134}$Te were at first published without mass assignment and the half-life of $^{129}$Te had been assigned to either $^{129}$Te or $^{131}$Te.

$^{137}$I was identified indirectly by identifying the daughter nucleus $^{137}$Xe where the genetic relationship between the half-lives of $^{137}$I and $^{137}$Xe had been determined earlier. The half-lives of $^{120}$I and $^{125}$I had initially been assigned to $^{119}$I or $^{120}$I and to $^{129}$Te, respectively, and the half-life reported for $^{117}$I was at first incorrect. In addition, the half-lives of six other iodine isotopes ($^{131-136}$I) had been reported without unique mass assignments.

The identification of the stable xenon isotopes is a rare example where Aston initially assigned incorrect mass numbers. The assignments of the five isotopes to 128, 130, 131, 133, and 135 was later shifted by one to 129, 131, 132, 134, and 136, respectively. $^{128}$Xe was identified indirectly by identifying the daughter nucleus $^{128}$Cs where the genetic relationship between the half-lives of $^{128}$Xe and$^{128}$Cs had been determined earlier. Two xenon isotopes ($^{144,145}$Xe) were at first misidentified and the half-lives of four others ($^{127,133,135,137}$Xe) were initially measured without mass assignment.

\ack

The main research on the individual elements was performed by JK (palladium, antimony, tellurium, and iodine) and CF (xenon). This work was supported by the National Science Foundation under grants No. PHY06-06007 (NSCL) and PHY10-62410 (REU).

\bibliography{../isotope-discovery-references}

\begin{thebibliography}{100}
\expandafter\ifx\csname url\endcsname\relax
  \def\url#1{\texttt{#1}}\fi
\expandafter\ifx\csname urlprefix\endcsname\relax\def\urlprefix{URL }\fi

\bibitem{2009Gin01}
G.~Q. Ginepro, J.~Snyder, M.~Thoennessen, At. Data Nucl. Data Tables 95 (2009)
  805.

\bibitem{2003Aud01}
G.~Audi, O.~Bersillon, J.~Blachot, A.~H. Wapstra, Nucl. Phys. A 729 (2003) 3.

\bibitem{2008ENS01}
{ http://www.nndc.bnl.gov/ensdf/ ENSDF, Evaluated Nuclear Structure Data File,
  maintained by the National Nuclear Data Center at Brookhaven National
  Laboratory, published in Nuclear Data Sheets (Academic Press, Elsevier
  Science) }.

\bibitem{1970Joh01}
W.~John, F.~W. Guy, J.~J. Wesolowski, Phys. Rev. C 2 (1970) 1451.

\bibitem{1946TPP01}
{ The Plutonium Project, J. Am. Chem. Soc. 68 (1946) 2411; Rev. Mod. Phys. 18
  (1946) 513 }.

\bibitem{1951Cor01}
{ C. D. Coryell and N. Sugarman (Editors), Radiochemical Studies: The Fission
  Products, National Nuclear Energy Series IV, 9, (McGraw-Hill, New York 1951)
  }.

\bibitem{2008NSR01}
{ http://www.nndc.bnl.gov/nsr/ NSR, Nuclear Science References, maintained by
  the National Nuclear Data Center at Brookhaven National Laboratory }.

\bibitem{1940Liv01}
J.~J. Livingood, G.~T. Seaborg, Rev. Mod. Phys. 12 (1940) 30.

\bibitem{1944Sea01}
G.~T. Seaborg, Rev. Mod. Phys. 16 (1944) 1.

\bibitem{1948Sea01}
G.~Seaborg, I.~Perlman, Rev. Mod. Phys. 20 (1948) 585.

\bibitem{1953Hol02}
J.~M. Hollander, I.~Perlman, G.~T. Seaborg, Rev. Mod. Phys. 25 (1953) 469.

\bibitem{1958Str01}
D.~Strominger, J.~M. Hollander, G.~T. Seaborg, Rev. Mod. Phys. 30 (1958) 585.

\bibitem{1967Led01}
{ C. M. Lederer, J. M. Hollander, I. Perlman, Table of Isotopes, 6$^{th}$
  Edition, John Wiley \& Sons 1967 }.

\bibitem{1942Ast01}
{ F. W. Aston, Mass Spectra and Isotopes, 2$^{nd}$ Edition, Longmans, Green \&
  Co., New York 1942 }.

\bibitem{2007Gor01}
S.~Goriely, M.~Samyn, J.~M. Pearson, Phys. Rev. C 75 (2007) 064312.

\bibitem{1995Ryk01}
K.~Rykaczewski, R.~Anne, G.~Auger, D.~Bazin, C.~Borcea, V.~Borrel, J.~M. Corre,
  T.~D\"orfler, A.~Fomichov, R.~Grzywacz, D.~Guillemaud-Mueller, R.~Hue,
  M.~Huyse, Z.~Janas, H.~Keller, M.~Lewitowicz, S.~Lukyanov, A.~C. Mueller,
  Y.~Penionzhkevich, M.~Pf\"utzner, F.~Pougheon, M.~G. Saint-Laurent,
  K.~Schmidt, W.~D. Schmidt-Ott, O.~Sorlin, J.~Szerypo, O.~B. Tarasov,
  J.~Wauters, J.~Zylicz, Phys. Rev. C 52 (1995) 2310.

\bibitem{1994Hen01}
M.~Hencheck, R.~N. Boyd, M.~Hellstr\"om, D.~J. Morrissey, M.~J. Balbes, F.~R.
  Choupek, M.~Fauerbach, C.~A. Mitchell, R.~Pfaff, C.~F. Powell, G.~Raimann,
  B.~M. Sherrill, M.~Steiner, J.~Vandegriff, S.~J. Yennello, Phys. Rev. C 50
  (1994) 2219.

\bibitem{1982Kur01}
W.~Kurcewicz, E.~F. Zganjar, R.~Kirchner, O.~Klepper, E.~Roeckl, P.~Komninos,
  E.~Nolte, D.~Schardt, P.~Tidemand-Petersson, Z. Phys. A 308 (1982) 21.

\bibitem{1980Nol01}
E.~Nolte, H.~Hick, Phys. Lett. B 97 (1980) 55.

\bibitem{1980Ara01}
N.~K. Aras, P.~W. Gallagher, W.~B. Walters, J. Phys. G 6 (1980) L195.

\bibitem{1969Ate01}
A.~H.~W. {Aten Jr.}, J.~C. Kapteijn, Radiochim. Acta 12 (1969) 218.

\bibitem{1955Ate01}
A.~H.~W. {Aten Jr.}, T.~{De Vries-Hamerling}, Physica 21 (1955) 597.

\bibitem{1953Ate01}
A.~H.~W. {Aten Jr.}, T.~{De Vries-Hamerling}, Physica 19 (1953) 1200.

\bibitem{1948Lin02}
M.~Lindner, I.~Perlman, Phys. Rev. 73 (1948) 1202.

\bibitem{1935Dem02}
A.~J. Dempster, Nature 136 (1935) 65.

\bibitem{1950Mei02}
J.~Y. Mei, C.~M. Huddleston, A.~C.~G. Mitchell, Phys. Rev. 79 (1950) 429.

\bibitem{1958Sch01}
U.~Schindewolf, Phys. Rev. 109 (1958) 1280.

\bibitem{1952Fla02}
A.~Flammersfeld, Z. Naturforsch. 7a (1952) 296.

\bibitem{1949Par02}
{ G. W. Parker, G. E.Creek, G. M. Hebert, andP.M.Lantz, reported in Oak Ridge
  National Laboratory Classified Report ORNL-336 (May 1949) }.

\bibitem{1937Kra01}
J.~D. Kraus, J.~M. Cork, Phys. Rev. 52 (1937) 763.

\bibitem{1935Ama01}
E.~Amaldi, O.~D'Agostino, F.~Rasetti, E.~Segre, Proc. Roy. Soc. A 149 (1935)
  522.

\bibitem{1951Sei01}
J.~A. Seiler, Radiochemical Studies: The Fission Products, Paper 119, p. 860,
  National Nuclear Energy Series IV, 9, McGraw-Hill, New York, 1951.

\bibitem{1940Nis02}
Y.~Nishina, T.~Yasaki, K.~Kimura, M.~Ikawa, Phys. Rev. 58 (1940) 660.

\bibitem{1941Seg01}
E.~Segr\`e, G.~T. Seaborg, Phys. Rev. 59 (1941) 212.

\bibitem{1954Hic01}
H.~G. Hicks, R.~S. Gilbert, Phys. Rev. 94 (1954) 371.

\bibitem{1958Ale01}
J.~M. Alexander, U.~Schindewolf, C.~D. Coryell, Phys. Rev. 111 (1958) 228.

\bibitem{1970Che01}
E.~Cheifetz, R.~C. Jared, S.~G. Thompson, J.~B. Wilhelmy, Phys. Rev. Lett. 25
  (1970) 38.

\bibitem{1970Aro01}
P.~O. Aronsson, E.~Ehn, J.~Rydberg, Phys. Rev. Lett. 25 (1970) 590.

\bibitem{1968Wei02}
H.~V. Weiss, J.~L. Elzie, J.~M. Fresco, Phys. Rev. 172 (1968) 1269.

\bibitem{1969Wei01}
H.~V. Weiss, N.~E. Ballou, J.~L. Elzie, J.~M. Fresco, Phys. Rev. 188 (1969)
  1893.

\bibitem{1991Pen02}
H.~Penttil\"a, J.~\"Ayst\"o, K.~Eskola, Z.~Janas, P.~P. Jauho, A.~Jokinen,
  M.~Leino, J.~M. Parmonen, P.~Taskinen, Z. Phys. A 338 (1991) 291.

\bibitem{1993Jan01}
Z.~Janas, J.~\"Ayst\"o, K.~Eskola, P.~P. Jauho, A.~Jokinen, J.~Kownacki,
  M.~Leino, J.~M. Parmonen, H.~Penttil\"a, J.~Szerypo, J.~Zylicz, Nucl. Phys. A
  552 (1993) 340.

\bibitem{1994Ber01}
M.~Bernas, S.~Czajkowski, P.~Armbruster, H.~Geissel, P.~Dessagne, C.~Donzaud,
  H.~R. Faust, E.~Hanelt, A.~Heinz, M.~Hesse, C.~Kozhuharov, C.~Mieh\'e,
  G.~M\"unzenberg, M.~Pf\"utzner, C.~R\"ohl, K.-H. Schmidt, W.~Schwab,
  C.~St\'ephan, K.~S\"ummerer, L.~Tassan-Got, B.~Voss, Phys. Lett. B 331 (1994)
  19.

\bibitem{1997Ber01}
M.~Bernas, C.~Engelmann, P.~Armbruster, S.~Czajkowski, F.~Ameil,
  C.~B\"ockstiegel, P.~Dessagne, C.~Donzaud, H.~Geissel, A.~Heinz, Z.~Janas,
  C.~Kozhuharov, C.~Mieh\'e, G.~M\"unzenberg, M.~Pf\"utzner, W.~Schwab,
  C.~St\'ephan, K.~S\"ummerer, L.~Tassan-Got, B.~Voss, Phys. Lett. B 415 (1997)
  111.

\bibitem{2008Ohn01}
T.~Ohnishi, T.~Kubo, K.~Kusaka, A.~Yoshida, K.~Yoshida, N.~Fukuda, M.~Ohtake,
  Y.~Yanagisawa, H.~Takeda, D.~Kameda, Y.~Yamaguchi, N.~Aoi, K.-I. Yoneda,
  H.~Otsu, S.~Takeuchi, T.~Sugimoto, Y.~Kondo, H.~Scheit, Y.~Gono, H.~Sakurai,
  T.~Motobayashi, H.~Suzuki, T.~Nakao, H.~Kimura, Y.~Mizoi, M.~Matsushita,
  K.~Ieki, T.~Kuboki, T.~Yamaguchi, T.~Suzuki, A.~Ozawa, T.~Moriguchi,
  Y.~Yasuda, T.~Nakamura, T.~Nannichi, T.~Shimamura, Y.~Nakayama, H.~Geissel,
  H.~Weick, J.~A. Nolen, O.~B. Tarasov, A.~S. Nettleton, D.~Bazin, B.~M.
  Sherrill, D.~J. Morrissey, W.~Mittig, J. Phys. Soc. Japan 77 (2008) 083201.

\bibitem{2010Ohn01}
T.~Ohnishi, T.~Kubo, K.~Kusaka, A.~Yoshida, K.~Yoshida, M.~Ohtake, N.~Fukuda,
  H.~Takeda, D.~Kameda, K.~Tanaka, N.~Inabe, Y.~Yanagisawa, Y.~Gono,
  H.~Watanabe, H.~Otsu, H.~Baba, T.~Ichihara, Y.~Yamaguchi, M.~Takechi,
  S.~Nishimura, H.~Ueno, A.~Yoshimi, H.~Sakurai, T.~Motobayashi, T.~Nakao,
  Y.~Mizoi, M.~Matsushita, K.~Ieki, N.~Kobayashi, K.~Tanaka, Y.~Kawada,
  N.~Tanaka, S.~Deguchi, Y.~Satou, Y.~Kondo, T.~Nakamura, K.~Yoshinaga,
  C.~Ishii, H.~Yoshii, Y.~Miyashita, N.~Uematsu, Y.~Shiraki, T.~Sumikama,
  J.~Chiba, E.~Ideguchi, A.~Saito, T.~Yamaguchi, I.~Hachiuma, T.~Suzuki,
  T.~Moriguchi, A.~Ozawa, T.~Ohtsubo, M.~A. Famiano, H.~Geissel, A.~S.
  Nettleton, O.~B. Tarasov, D.~Bazin, B.~M. Sherrill, S.~L. Manikonda, J.~A.
  Nolen, J. Phys. Soc. Japan 79 (2010) 073201.

\bibitem{2004Tho01}
M.~Thoennessen, Rep. Prog. Phys. 67 (2004) 1187.

\bibitem{1994Tig01}
R.~J. Tighe, D.~M. Moltz, J.~C. Batchelder, T.~J. Ognibene, M.~W. Rowe,
  J.~Cerny, Phys. Rev. C 49 (1994) R2871.

\bibitem{1987Gil01}
A.~Gillitzer, T.~Faestermann, K.~Hartel, P.~Kienle, E.~Nolte, Z. Phys. A 326
  (1987) 107.

\bibitem{1981Plo01}
A.~Plochocki, J.~Zylicz, R.~Kirchner, O.~Klepper, E.~Roeckl,
  P.~Tidemand-Petersson, I.~S. Grant, P.~Misaelides, W.-D. Schmidt-Ott, Phys.
  Lett. B 106 (1981) 285.

\bibitem{1994Sew01}
D.~Seweryniak, J.~Cederkall, B.~Cederwall, J.~Blomqvist, C.~Fahlander,
  A.~Johnson, L.-O. Norlin, J.~Nyberg, A.~Atac, A.~Kerek, J.~Kownacki, R.~Wyss,
  E.~Adamides, H.~Grawe, E.~Ideguchi, R.~Julin, S.~Juutinen, W.~Karczmarczyk,
  S.~Mitarai, M.~Piiparinen, R.~Schubart, G.~Sletten, S.~Tormanen, A.~Virtanen,
  Phys. Lett. B 321 (1994) 323.

\bibitem{1979Sch02}
D.~Schardt, R.~Kirchner, O.~Klepper, W.~Reisdorf, E.~Roeckl,
  P.~Tidemand-Petersson, G.~T. Ewan, E.~Hagberg, B.~Jonson, S.~Mattsson,
  G.~Nyman, Nucl. Phys. A 326 (1979) 65.

\bibitem{1997Shi01}
M.~Shibata, Z.~Hu, J.~Agramunt, D.~Cano-Ott, R.~Collatz, M.~Gorska, H.~Grawe,
  M.~Hellstr\"om, Z.~Janas, M.~Karny, R.~Kirchner, O.~Klepper, A.~Plochocki,
  E.~Roeckl, K.~Rykaczewski, K.~Schmidt, A.~Weber, J.~Zylicz, Phys. Rev. C 55
  (1997) 1715.

\bibitem{1976Oxo01}
K.~Oxorn, A.~J. Houdayer, S.~K. Mark, Z. Phys. A 279 (1976) 289.

\bibitem{1972Miy01}
K.~Miyano, H.~Nakahara, C.~Gil, J. Phys. Soc. Japan 33 (1972) 1505.

\bibitem{1970Sun01}
{ J. W. Sunier, M. Singh, R. M. DeVries and G. E. Thompson: Abstr. of Conf. on
  Prop. of Nuclei Far from Region of Beta-stability. (Leysim, Switzerland,
  1970) p. 1015. }.

\bibitem{1972Sin01}
M.~Singh, J.~W. Sunier, R.~M. Devries, G.~E. Thompson, Nucl. Phys. A 193 (1972)
  449.

\bibitem{1959Sel01}
I.~P. Selinov, Y.~A. Grits, Y.~P. Kushakevich, Y.~A. Bliodze, S.~S. Vasilev,
  T.~N. Mikhaleva, Sov. At. Energy 7 (1959) 1011.

\bibitem{1958Sel01}
I.~P. Selinov, Y.~A. Grits, D.~E. Khulelidze, E.~E. Baroni, Y.~A. Bliodze,
  A.~G. Demin, Y.~P. Kushakevich, Sov. At. Energy 5 (1958) 1605.

\bibitem{1957Rho01}
{ A. Rhodes, Master's Thesis, University of California Radiation Laboratory
  Report UCRL-3879, August, 1957 (unpublished) }.

\bibitem{1949Tem01}
G.~M. Temmer, Phys. Rev. 76 (1949) 424.

\bibitem{1947Col01}
K.~D. Coleman, M.~L. Pool, Phys. Rev. 72 (1947) 1070.

\bibitem{1937Bot01}
W.~Bothe, W.~Gentner, Naturwiss. 25 (1937) 126.

\bibitem{1922Ast04}
F.~W. Aston, Nature 110 (1922) 732.

\bibitem{1939Liv04}
J.~J. Livingood, G.~T. Seaborg, Phys. Rev. 55 (1939) 414.

\bibitem{1937Liv03}
J.~J. Livingood, G.~T. Seaborg, Phys. Rev. 52 (1937) 135.

\bibitem{1951Sta01}
C.~W. Stanley, L.~E. Glendenin, Radiochemical Studies: The Fission Products,
  Paper 134, p. 947, National Nuclear Energy Series IV, 9, McGraw-Hill, New
  York, 1951.

\bibitem{1951Cam01}
G.~W. Campbell, N.~R. Sleight, W.~H. Sullivan, Radiochemical Studies: The
  Fission Products, Paper 132, p. 931, National Nuclear Energy Series IV, 9,
  McGraw-Hill, New York, 1951.

\bibitem{1951Lea01}
G.~R. Leader, W.~H. Sullivan, Radiochemical Studies: The Fission Products,
  Paper 133, p. 934, National Nuclear Energy Series IV, 9, McGraw-Hill, New
  York, 1951.

\bibitem{1949Ker01}
B.~D. Kern, A.~C.~G. Mitchell, D.~J. Zaffarano, Phys. Rev. 76 (1949) 94.

\bibitem{1956Fra02}
I.~Fr\"anz, R.~Radicella, J.~Rodriguez, Z. Naturforsch. 11a (1956) 1038.

\bibitem{1951Bar02}
J.~W. Barnes, A.~J. Freedman, Phys. Rev. 84 (1951) 365.

\bibitem{1939Abe01}
P.~Abelson, Phys. Rev. 55 (1939) 876.

\bibitem{1956Fra01}
I.~Fr\"anz, J.~Rodriguez, R.~Radicella, Z. Naturforsch. 11a (1956) 1037.

\bibitem{1955Fra01}
I.~Fr\"anz, J.~Rodriguez, H.~Carminatti, Z. Naturforsch. 10a (1955) 82.

\bibitem{1939Abe02}
P.~Abelson, Phys. Rev. 55 (1939) 670.

\bibitem{1962Hag01}
E.~Hagebo, A.~Kjelberg, A.~C. Pappas, J. Inorg. Nucl. Chem. 24 (1962) 117.

\bibitem{1956Pap01}
A.~C. Pappas, D.~R. Wiles, J. Inorg. Nucl. Chem. 2 (1956) 69.

\bibitem{1951Coo01}
{ G. B. Cook, British Atomic Energy Report AERE C/R-729 June, 1951
  (unpublished) }.

\bibitem{1953Pap01}
{ A. C. Pappas (September 1953) Laboratory for Nuclear Science, Technical
  Report No. 63, Massachusetts Institute of Technology }.

\bibitem{1954Pap01}
A.~C. Pappas, Z. Elektrochem. 58 (1954) 620.

\bibitem{1939Abe03}
P.~Abelson, Phys. Rev. 56 (1939) 1.

\bibitem{1966Str01}
P.~O. Strom, D.~L. Love, A.~E. Greendale, A.~A. Delucchi, D.~Sam, N.~E. Ballou,
  Phys. Rev. 144 (1966) 984.

\bibitem{1967Tom01}
L.~Tomlinson, M.~H. Hurdus, Phys. Lett. B 25 (1967) 545.

\bibitem{1968Del01}
A.~A. Delucchi, A.~E. Greendale, P.~O. Strom, Phys. Rev. 173 (1968) 1159.

\bibitem{1964Bem01}
C.~E. Bemis, G.~E. Gordon, C.~D. Coryell, J. Inorg. Nucl. Chem. 26 (1964) 213.

\bibitem{1976Lun01}
E.~Lund, G.~Rudstam, Phys. Rev. C 13 (1976) 1544.

\bibitem{2006Sew01}
D.~Seweryniak, K.~Starosta, C.~N. Davids, S.~Gros, A.~A. Hecht, N.~Hoteling,
  T.~L. Khoo, K.~Lagergren, G.~Lotay, D.~Peterson, A.~Robinson, C.~Vaman, W.~B.
  Walters, P.~J. Woods, S.~Zhu, Phys. Rev. C 73 (2006) 061301.

\bibitem{2006Lid01}
S.~N. Liddick, R.~Grzywacz, C.~Mazzocchi, R.~D. Page, K.~Rykaczewski, J.~C.
  Batchelder, C.~R. Bingham, I.~G. Darby, G.~Drafta, C.~Goodin, C.~J. Gross,
  J.~H. Hamilton, A.~A. Hecht, J.~K. Hwang, S.~Ilyushkin, D.~T. Joss,
  A.~Korgul, W.~Krolas, K.~Lagergren, K.~Li, M.~N. Tantawy, J.~Thomson, J.~A.
  Winger, Phys. Rev. Lett. 97 (2006) 802501.

\bibitem{1981Sch01}
D.~Schardt, T.~Batsch, R.~Kirchner, O.~Klepper, W.~Kurcewicz, E.~Roeckl,
  P.~Tidemand-Petersson, Nucl. Phys. A 368 (1981) 153.

\bibitem{1965Mac02}
R.~D. Macfarlane, A.~Siivola, Phys. Rev. Lett. 14 (1965) 114.

\bibitem{1974Bog02}
D.~D. Bogdanov, V.~A. Karnaukhov, L.~A. Petrov, Sov. J. Nucl. Phys. 18 (1974)
  1.

\bibitem{1977Kir01}
R.~Kirchner, O.~Klepper, G.~Nyman, W.~Reisdorf, E.~Roeckl, D.~Schardt,
  N.~Kaffrell, P.~Peuser, K.~Schneeweiss, Phys. Lett. B 70 (1977) 150.

\bibitem{1965Sii02}
A.~Siivola, Phys. Rev. Lett. 14 (1965) 142.

\bibitem{1973Bog01}
D.~D. Bogdanov, V.~A. Karnaukhov, L.~A. Petrov, Sov. J. Nucl. Phys. 17 (1973)
  233.

\bibitem{1967Kar01}
V.~A. Karnaukhov, G.~M. Ter-Akopian, L.~S. Vertogradov, L.~A. Petrov, Sov. J.
  Nucl. Phys. 4 (1967) 327.

\bibitem{1967Kar02}
V.~A. Karnaukhov, G.~M. Ter-Akopian, L.~S. Vertogradov, L.~A. Petrov, Nucl.
  Phys. 90 (1967) 23.

\bibitem{1976Wig01}
M.~E.~J. Wigmans, R.~J. Heynis, P.~M.~A. {van der Kam}, H.~Verheul, Phys. Rev.
  C 14 (1976) 243.

\bibitem{1974Cha01}
A.~Charvet, R.~Chery, R.~Duffait, J. Phys. (Paris) 35 (1974) L--41.

\bibitem{1974Bur02}
V.~P. Burminskii, B.~G. Kiselev, O.~D. Kovrigin, Bull. Acad. Sci. USSR Phys.
  Ser. 8 (1974) 4.

\bibitem{1968Rah01}
O.~Rahmouni, J. Phys. (Paris) 29 (1968) 550.

\bibitem{1961Sel01}
I.~P. Selinov, N.~A. Vartanov, D.~E. Khulelidze, Y.~A. Bliodze, N.~G. Zaitseva,
  V.~A. Khalkin, Sov. Phys. JETP 11 (1961) 1191.

\bibitem{1958Kuz01}
M.~Y. Kuznetsova, V.~N. Mekhedov, V.~N. Rybakov, V.~A. Khalkin, Sov. At. Energy
  4 (1958) 766.

\bibitem{1961Var01}
N.~A. Vartanov, Y.~A. Ryukhin, I.~P. Selinov, V.~L. Chikhladze, D.~E.
  Khulelidze, Sov. Phys. JETP 14 (1962) 215.

\bibitem{1948Lin01}
M.~Lindner, I.~Perlman, Phys. Rev. 73 (1948) 1124.

\bibitem{1936Dem02}
A.~J. Dempster, Phys. Rev. 50 (1936) 186.

\bibitem{1939Sea02}
G.~T. Seaborg, J.~J. Livingood, J.~W. Kennedy, Phys. Rev. 55 (1939) 794.

\bibitem{1938Tap01}
G.~F. Tape, J.~M. Cook, Phys. Rev. 53 (1938) 676.

\bibitem{1932Bai01}
K.~T. Bainbridge, Phys. Rev. 39 (1932) 1021.

\bibitem{1931Ast05}
F.~W. Aston, Proc. Roy. Soc. A 132 (1931) 487.

\bibitem{1924Ast03}
F.~W. Aston, Nature 114 (1924) 717.

\bibitem{1938Liv06}
J.~J. Livingood, G.~T. Seaborg, Phys. Rev. 53 (1938) 1015.

\bibitem{1937Bot03}
W.~Bothe, W.~Gentner, Naturwiss. 25 (1937) 191.

\bibitem{1948Kat01}
S.~Katcoff, J.~A. Miskel, C.~W. Stanley, Phys. Rev. 74 (1948) 631.

\bibitem{1940Pol02}
A.~Polessitsky, N.Nemerovsky, Compt. Rend. (Doklady) Acad. Sci. URSS 28 (1940)
  217.

\bibitem{1944Jol02}
F.~Joliot, Compt. Rend. Acad. Sci. 218 (1944) 488.

\bibitem{1947Suz01}
F.~Suzor, Compt. Rend. Acad. Sci. 224 (1947) 1155.

\bibitem{1947Suz02}
F.~Suzor, J. Phys. Radium 8 (1947) 39.

\bibitem{1940Wu01}
C.~S. Wu, Phys. Rev. 58 (1940) 926.

\bibitem{1940Seg01}
E.Segre, C.~S. Wu, Phys. Rev. 57 (1940) 552.

\bibitem{1969Den01}
H.~O. Denschlag, J. Inorg. Nucl. Chem. 31 (1969) 1873.

\bibitem{1964Her01}
G.~Herrmann, Radiochim. Acta 3 (1964) 169.

\bibitem{1962Gre01}
{ A. E. Greendale, P. O. Strom, D. L. Love and D. Sam, Paper presented at Disc.
  Nucl. Chem. Oxford (1962) }.

\bibitem{1974Gra01}
B.~Grapengiesser, E.~Lund, G.~Rudstam, J. Inorg. Nucl. Chem. 36 (1974) 2409.

\bibitem{1969Sch01}
{ H. D. Sch\"ussler, H. Ahrens, H. Folger, H. Franz, W. Grimm, G. Herrmann, J.
  V. Kratz and K. L. Kratz, Proceedings of the Second Symposium on Physics and
  Chemistry of Fission, p. 591, I.A.E.A. Vienna (1969). }.

\bibitem{1949Sta01}
C.~W. Stanley, S.~Katcoff, J. Chem. Phys. 17 (1949) 653.

\bibitem{1967Wun01}
F.~Wunderlich, Radiochim. Acta 7 (1967) 105.

\bibitem{1975Asg01}
M.~Asghar, J.~P. Gautheron, G.~Bailleul, J.~P. Bocquet, J.~Greif, H.~Schrader,
  G.~Siegert, C.~Ristori, J.~Crancon, G.~I. Crawford, Nucl. Phys. A 247 (1975)
  359.

\bibitem{1991Pag01}
R.~D. Page, P.~J. Woods, S.~J. Bennett, M.~Freer, B.~R. Fulton, R.~A.
  Cunningham, J.~Groves, M.~A.~C. Hotchkis, A.~N. James, Z. Phys. A 338 (1991)
  295.

\bibitem{1984Fae01}
T.~Faestermann, A.~Gillitzer, K.~Hartel, P.~Kienle, E.~Nolte, Phys. Lett. B 137
  (1984) 23.

\bibitem{1969Han01}
P.~G. Hansen, P.~Hornshoj, H.~L. Nielsen, K.~Wilsky, H.~Kugler, G.~Astner,
  E.~Hagebo, J.~Hudis, A.~Kjelberg, F.~M\"unnich, P.~Patzelt, M.~Alpsten,
  G.~Andersson, A.~Appelqvist, B.~Bengtsson, R.~A. Naumann, O.~B. Nielsen,
  E.~Beck, R.~Foucher, P.~Husson, J.~Jastrzebski, A.~Johnson, J.~Alstad,
  T.~Jahnsen, A.~C. Pappas, T.~Tunaal, R.~Henck, P.~Siffert, G.~Rudstam, Phys.
  Lett. B 28 (1969) 415.

\bibitem{1976Gow01}
G.~M. Gowdy, A.~C. Xenoulis, J.~L. Wood, K.~R. Baker, R.~W. Fink, J.~L. Weil,
  B.~D. Kern, K.~J. Hofstetter, E.~H. Spejewski, R.~L. Mlekodaj, H.~K. Carter,
  W.~D. Schmidt-Ott, J.~Lin, C.~R. Bingham, L.~L. Riedinger, E.~F. Zganjar,
  K.~S.~R. Sastry, A.~V. Ramayya, J.~H. Hamilton, Phys. Rev. C 13 (1976) 1601.

\bibitem{1974Ham01}
J.~H. Hamilton, E.~H. Spejewski, R.~L. Mlekodaj, W.-D. Schmidt-Ott, R.~W. Fink,
  A.~Xenoulis, K.~R. Baker, J.~L. Wood, G.~Gowdy, H.~K. Carter, B.~D. Kern,
  K.~J. Hofstetter, J.~L. Weil, E.~F. Zganjar, K.~S.~R. Sastry, F.~T. Avignone,
  C.~R. Bingham, L.~L. Riedinger, L.~Harwood, F.~Turner, D.~J.~P. I.~A.~Sellin,
  J.~Lin, A.~V. Ramayya, S.~Lee, G.~Garcia-Bermudez, E.~Bosworth, K.~S. Toth,
  N.~R. Johnson, Bull. Acad. Sci. USSR Phys. Ser. 38, No. 10 (1974) 22.

\bibitem{1969Spe01}
E.~H. Spejewski, P.~K. Hopke, F.~W. {Loeser, Jr}, Phys. Rev. 186 (1969) 1270.

\bibitem{1960Zai01}
N.~G. Zaitseva, Radiokhimiya 2 (1960) 451.

\bibitem{1965But01}
F.~D.~S. Butement, S.~M. Qaim, J. Inorg. Nucl. Chem. 27 (1965) 1729.

\bibitem{1965And01}
G.~Andersson, G.~Rudstam, G.~Sorensen, Ark. Fysik 28 (1965) 37.

\bibitem{1969Ser01}
H.~Sergolle, G.~Albouy, J.-M. Lagrange, M.~Pautrat, J.~{Van Horenbeeck}, Compt.
  Rend. Acad. Sci. 268 (1969) 701.

\bibitem{1969Lad01}
I.~M. Ladenbauer-Bellis, H.~Bakhru, A.~Luzzati, Phys. Rev. 187 (1969) 1739.

\bibitem{1957Aag01}
P.~Aagaard, G.~Andersson, J.~O. Burgman, A.~C. Pappas, J. Inorg. Nucl. Chem. 5
  (1957) 105.

\bibitem{1954Ros01}
G.~B. Rossi, W.~B. Jones, J.~M. Hollander, J.~G. Hamilton, Phys. Rev. 93 (1954)
  256.

\bibitem{1950Mar01}
L.~Marquez, I.~Perlman, Phys. Rev. 78 (1950) 189.

\bibitem{1949Mit01}
A.~C.~G. Mitchell, J.~Y. Mei, F.~C. Maienschein, C.~L. Peacock, Phys. Rev. 76
  (1949) 1450.

\bibitem{1938Liv05}
J.~J. Livingood, G.~T. Seaborg, Phys. Rev. 54 (1938) 775.

\bibitem{1947Gle01}
L.~E. Glendenin, R.~R. Edwards, Phys. Rev. 71 (1947) 742.

\bibitem{1946Rei01}
A.~F. Reid, A.~S. Keston, Phys. Rev. 70 (1946) 987.

\bibitem{1920Ast05}
F.~W. Aston, Nature 106 (1920) 468.

\bibitem{1934Fer01}
E.~Fermi, E.~Amaldi, O.~D'Agostino, F.~Rasetti, E.~Segre, Proc. Roy. Soc. A 146
  (1934) 483.

\bibitem{1951Kat06}
S.~Katcoff, O.~A. Schaeffer, J.~M.Hastings, Phys. Rev. 82 (1951) 688.

\bibitem{1951Lea02}
G.~R. Leader, W.~H. Sullivan, Radiochemical Studies: The Fission Products,
  Paper 144, p. 1023, National Nuclear Energy Series IV, 9, McGraw-Hill, New
  York, 1951.

\bibitem{1947Kat02}
S.~Katcoff, Phys. Rev. 71 (1947) 826.

\bibitem{1949Par03}
{ G. W. Parker, G. E.Creek, G. M. Hebert, P.M.Lantz, and W. J. Martin, Oak
  Ridge National Laboratory Classified Report No. 286, p. 53 1949 (unpublished)
  }.

\bibitem{1939Hah04}
O.~Hahn, F.~Strassmann, S.~Fl\"ugge, Naturwiss. 27 (1939) 544.

\bibitem{1940Str01}
F.~Strassmann, O.~Hahn, Naturwiss. 28 (1940) 817.

\bibitem{1943Rie01}
W.~Riezler, Naturwiss. (1943) 326.

\bibitem{1943See01}
W.~Seelmann-Eggebert, H.~J. Born, Naturwiss. 31 (1943) 59.

\bibitem{1947Sne01}
A.~H. Snell, J.~S. Levinger, E.~P. {Meiners Jr.}, M.~B. Sampson, R.~G.
  Wilkinson, Phys. Rev. 72 (1947) 545.

\bibitem{1949Sug01}
N.~Sugarman, J. Chem. Phys. 17 (1949) 11.

\bibitem{1972Sch01}
H.-D. Schussler, G.~Herrmann, Radiochim. Acta 18 (1972) 123.

\bibitem{1974Kra01}
K.-L. Kratz, G.~Herrmann, Nucl. Phys. A 229 (1974) 179.

\bibitem{1974Rud01}
G.~Rudstam, S.~Shalev, O.~C. Jonsson, Nucl. Instrum. Meth. 120 (1974) 333.

\bibitem{1973Tom01}
L.~Tomlinson, At. Data Nucl. Data Tables 12 (1973) 179.

\bibitem{1973Rud01}
{ G. Rudstam, B. Grapengiesser and E. Lund, Contribution to the IAEA Panel on
  Fission Product Nuclear Data, Bologna 26-30 November 1973 }.

\bibitem{1970Her01}
{ G. Herrmann, N. Kaffrell, N. Trautmann, R. Denig, W. Herzog, D. H\"ubscher,
  and K.L. Kratz, Report No. CERN 70-30 p. 985. }.

\bibitem{1973Sch03}
{ H. D. Sch\"ussler and G.Herrmann, Radiochim. Acta, to bepublished }.

\bibitem{1975Kra01}
K.-L. Kratz, W.~Lauppe, G.~Herrmann, Inorg. Nucl. Chem. Lett. 11 (1975) 331.

\bibitem{1978Roe01}
E.~Roeckl, R.~Kirchner, O.~Klepper, G.~Nyman, W.~Reisdorf, D.~Schardt, K.~Wien,
  R.~Fass, S.~Mattsson, Phys. Lett. B 78 (1978) 393.

\bibitem{1973Hag01}
E.~Hagberg, P.~G. Hansen, B.~Jonson, B.~G.~G. Jorgensen, E.~Kugler,
  T.~Mowinckel, Nucl. Phys. A 208 (1973) 309.

\bibitem{1952Dro01}
B.~Dropesky, E.~O. Wiig, Phys. Rev. 88 (1952) 683.

\bibitem{1922Ast03}
F.~W. Aston, Nature 110 (1922) 664.

\bibitem{1950And01}
D.~L. Anderson, M.~L. Pool, Phys. Rev. 77 (1950) 142.

\bibitem{1940Cre01}
E.~C. Creutz, L.~A. Delsasso, R.~B. Sutton, M.~G. White, W.~Barkas, Phys. Rev.
  58 (1940) 481.

\bibitem{1922Ast02}
F.~W. Aston, Nature 109 (1922) 813.

\bibitem{1920Ast01}
F.~W. Aston, Nature 105 (1920) 8.

\bibitem{1921Ast05}
F.~W. Aston, Nature 107 (1921) 334.

\bibitem{1921Ast04}
F.~W. Aston, Phil. Mag. 42 (1921) 140.

\bibitem{1939Lan01}
A.~{Langsdorf Jr.}, Phys. Rev. 56 (1939) 205.

\bibitem{1940Dod01}
R.~W. Dodson, R.~D. Fowler, Phys. Rev. 57 (1940) 966.

\bibitem{1943See03}
W.~Seelmann-Eggebert, Naturwiss. 31 (1943) 491.

\bibitem{1939Hah03}
O.~Hahn, F.~Strassmann, Naturwiss. 27 (1939) 529.

\bibitem{1940Gla02}
G.~N. Glasoe, J.~Steigman, Phys. Rev. 58 (1940) 1.

\bibitem{1951Dil01}
C.~R. Dillard, R.~M. Adams, H.~Finston, A.~Turkevich, Radiochemical Studies:
  The Fission Products, Paper 68, p. 624, National Nuclear Energy Series IV, 9,
  McGraw-Hill, New York, 1951.

\bibitem{1960Wol01}
D.~R. Wolfsberg, D.~R. Nethaway, H.~P. Malan, A.~C. Wahl, J. Inorg. Nucl. Chem.
  12 (1960) 201.

\bibitem{2003Ber01}
U.~C. Bergmann, C.~A. Diget, K.~Riisager, L.~Weissman, G.~Aub\"ock,
  J.~Cederkall, L.~M. Fraile, H.~O.~U. Fynbo, H.~Gausemel, H.~Jeppesen,
  U.~K\"oster, K.-L. Kratz, P.~M\"oller, T.~Nilsson, B.~Pfeiffer, H.~Simon,
  K.~{Van de Vel}, J.~\"Ayst\"o, {the ISOLDE Collaboration}, Nucl. Phys. A 714
  (2003) 21.

\bibitem{1962Wah01}
A.~C. Wahl, R.~L. Ferguson, D.~R. Nethaway, D.~E. Troutner, K.~Wolfsberg, Phys.
  Rev. 126 (1962) 1112.

\bibitem{1976Ahr01}
H.~Ahrens, P.~Patzelt, G.~Herrmann, J. Inorg. Nucl. Chem. 38 (1976) 191.

\bibitem{1971Wol01}
K.~Wolfsberg, J. Inorg. Nucl. Chem. 33 (1971) 587.

\bibitem{1966Laz01}
E.~Lazzarini, M.~Terrani, J. Appl. Radiat. Isot. 17 (1966) 554.

\bibitem{1989Bor02}
W.~Borchers, E.~Arnold, W.~Neu, R.~Neugart, K.~Wendt, G.~Ulm, {the ISOLDE
  Collaboration}, Phys. Lett. B 216 (1989) 7.

\end{thebibliography}

\newpage

\newpage

\TableExplanation

\bigskip
\renewcommand{\arraystretch}{1.0}

\section{Table 1.\label{tbl1te} Discovery of palladium, antimony, tellurium, iodine, and xenon isotopes }
\begin{tabular*}{0.95\textwidth}{@{}@{\extracolsep{\fill}}lp{5.5in}@{}}
\multicolumn{2}{p{0.95\textwidth}}{ }\\

Isotope & palladium, antimony, tellurium, iodine, and xenon isotope \\
Author & First author of refereed publication \\
Journal & Journal of publication \\
Ref. & Reference \\
Method & Production method used in the discovery: \\

  & FE: fusion evaporation \\
  & LP: light-particle reactions (including neutrons) \\
  & MS: mass spectroscopy \\
  & NC: neutron capture reactions \\
  & NF: neutron induced fission \\
  & CPF: charged-particle induced fission \\
  & SP: spallation \\
  & PF: projectile fragmentation of fission \\

Laboratory & Laboratory where the experiment was performed\\
Country & Country of laboratory\\
Year & Year of discovery \\
\end{tabular*}
\label{tableI}

\section{Table 2.\label{tbl2te} Chronological listing of the discovery of palladium, antimony, tellurium, iodine, and xenon isotopes }
\begin{tabular*}{0.95\textwidth}{@{}@{\extracolsep{\fill}}lp{5.5in}@{}}
\multicolumn{2}{p{0.95\textwidth}}{ }\\

Submitted & Submission date of publication \\
    &   \\
& The other columns are the same as in Table 1 \\
\end{tabular*}
\label{tableII}

\datatables 



\setlength{\LTleft}{0pt}
\setlength{\LTright}{0pt}


\setlength{\tabcolsep}{0.5\tabcolsep}

\renewcommand{\arraystretch}{1.0}

\footnotesize 

\begin{longtable}{@{\extracolsep\fill}llllllll@{}}
\caption{Discovery of palladium, antimony, tellurium, iodine, and xenon isotopes. See page\ \pageref{tbl1te} for Explanation of Tables}
Isotope & Author & Journal & Ref. & Method & Laboratory & Country & Year\\
\hline\\
\endfirsthead\\
\caption[]{(continued)}
Isotope & Author & Journal & Ref. & Method & Laboratory & Country & Year\\
\hline\\
\endhead
$^{91}$Pd & K. Rykaczewski & Phys. Rev. C &\cite{1995Ryk01}& PF & GANIL & France &1995 \\
$^{92}$Pd & M. Hencheck & Phys. Rev. C &\cite{1994Hen01}& PF & Michigan State & USA &1994 \\
$^{93}$Pd & M. Hencheck & Phys. Rev. C &\cite{1994Hen01}& PF & Michigan State & USA &1994 \\
$^{94}$Pd & W. Kurcewicz & Z. Phys. A &\cite{1982Kur01}& FE & Darmstadt & Germany &1982 \\
$^{95}$Pd & E. Nolte & Phys. Lett. B &\cite{1980Nol01}& FE & Munich & Germany &1980 \\
$^{96}$Pd & N.K. Aras & J. Phys. G &\cite{1980Ara01}& LP & Maryland & USA &1980 \\
$^{97}$Pd & A.H.W. Aten & Radiochim. Acta &\cite{1969Ate01}& LP & Amsterdam & Netherlands &1969 \\
$^{98}$Pd & A.H.W. Aten & Physica &\cite{1955Ate01}& LP & Amsterdam & Netherlands &1955 \\
$^{99}$Pd & A.H.W. Aten & Physica &\cite{1955Ate01}& LP & Amsterdam & Netherlands &1955 \\
$^{100}$Pd & M. Lindner & Phys. Rev. &\cite{1948Lin02}& SP & Berkeley & USA &1948 \\
$^{101}$Pd & M. Lindner & Phys. Rev. &\cite{1948Lin02}& SP & Berkeley & USA &1948 \\
$^{102}$Pd & A.J. Dempster & Nature &\cite{1935Dem02}& MS & Chicago & USA &1935 \\
$^{103}$Pd & J. Y. Mei & Phys. Rev. &\cite{1950Mei02}& LP & Indiana & USA &1950 \\
$^{104}$Pd & A.J. Dempster & Nature &\cite{1935Dem02}& MS & Chicago & USA &1935 \\
$^{105}$Pd & A.J. Dempster & Nature &\cite{1935Dem02}& MS & Chicago & USA &1935 \\
$^{106}$Pd & A.J. Dempster & Nature &\cite{1935Dem02}& MS & Chicago & USA &1935 \\
$^{107}$Pd & U. Schindewolf & Phys. Rev. &\cite{1958Sch01}& NC & MIT & USA &1958 \\
$^{108}$Pd & A.J. Dempster & Nature &\cite{1935Dem02}& MS & Chicago & USA &1935 \\
$^{109}$Pd & J.D. Kraus & Phys. Rev. &\cite{1937Kra01}& LP & Michigan & USA &1937 \\
$^{110}$Pd & A.J. Dempster & Nature &\cite{1935Dem02}& MS & Chicago & USA &1935 \\
$^{111}$Pd & J.D. Kraus & Phys. Rev. &\cite{1937Kra01}& LP & Michigan & USA &1937 \\
$^{112}$Pd & J.A. Seiler & Nat. Nucl. Ener. Ser. &\cite{1951Sei01}& NF & Argonne & USA &1951 \\
$^{113}$Pd & H. G. Hicks & Phys. Rev. &\cite{1954Hic01}& CPF & Berkeley & USA &1954 \\
$^{114}$Pd & J. M. Alexander & Phys. Rev. &\cite{1958Ale01}& CPF & MIT & USA &1958 \\
$^{115}$Pd & J. M. Alexander & Phys. Rev. &\cite{1958Ale01}& CPF & MIT & USA &1958 \\
$^{116}$Pd & E. Cheifetz & Phys. Rev. Lett. &\cite{1970Che01}& SF & Berkeley & USA &1970 \\
$^{117}$Pd & H.V. Weiss & Phys. Rev. &\cite{1968Wei02}& NF & Naval Radiological Defense Laboratory & USA &1968 \\
$^{118}$Pd & H.V. Weiss & Phys. Rev. &\cite{1969Wei01}& NF & Naval Radiological Defense Laboratory & USA &1969 \\
$^{119}$Pd & H. Penttil\"a & Z. Phys. A &\cite{1991Pen02}& CPF & Jyv\"askyl\"a & Finland &1991 \\
$^{120}$Pd & Z. Janas & Nucl. Phys. A &\cite{1993Jan01}& CPF & Jyv\"askyl\"a & Finland &1993 \\
$^{121}$Pd & M. Bernas & Phys. Lett. B &\cite{1994Ber01}& PF & Darmstadt & Germany &1994 \\
$^{122}$Pd & M. Bernas & Phys. Lett. B &\cite{1994Ber01}& PF & Darmstadt & Germany &1994 \\
$^{123}$Pd & M. Bernas & Phys. Lett. B &\cite{1994Ber01}& PF & Darmstadt & Germany &1994 \\
$^{124}$Pd & M. Bernas & Phys. Lett. B &\cite{1997Ber01}& PF & Darmstadt & Germany &1997 \\
$^{125}$Pd & T. Ohnishi & J. Phys. Soc. Japan &\cite{2008Ohn01}& PF & RIKEN & Japan &2008 \\
$^{126}$Pd & T. Ohnishi & J. Phys. Soc. Japan &\cite{2008Ohn01}& PF & RIKEN & Japan &2008 \\
$^{127}$Pd & T. Ohnishi & J. Phys. Soc. Japan &\cite{2010Ohn01}& PF & RIKEN & Japan &2010 \\
$^{128}$Pd & T. Ohnishi & J. Phys. Soc. Japan &\cite{2010Ohn01}& PF & RIKEN & Japan &2010 \\
 & & & & & & & \\
 & & & & & & & \\
$^{103}$Sb& K. Rykaczewski & Phys. Rev. C &\cite{1995Ryk01}& PF & GANIL & France &1995 \\
$^{104}$Sb & K. Rykaczewski & Phys. Rev. C &\cite{1995Ryk01}& PF & GANIL & France &1995 \\
$^{105}$Sb & R.J. Tighe & Phys. Rev. C &\cite{1994Tig01}& FE & Berkeley & USA &1994 \\
$^{106}$Sb & A. Plochocki & Phys. Lett. B &\cite{1981Plo01}& FE & Darmstadt & Germany &1981 \\
$^{107}$Sb & D. Seweryniak & Phys. Lett. B &\cite{1994Sew01}& FE & Roskilde & Denmark &1994 \\
$^{108}$Sb & K. Oxorn & Z. Phys. A &\cite{1976Oxo01}& LP & McGill & Canada &1976 \\
$^{109}$Sb & K. Oxorn & Z. Phys. A &\cite{1976Oxo01}& LP & McGill & Canada &1976 \\
$^{110}$Sb & K. Miyano & J. Phys. Soc. Japan &\cite{1972Miy01}& LP & Tokyo & Japan &1972 \\
$^{111}$Sb & M. Singh & Nucl. Phys. A &\cite{1972Sin01}& LP & UCLA & USA &1972 \\
$^{112}$Sb & I.B. Selinov & Sov. At. Energy &\cite{1959Sel01}& LP & Moscow & Russia &1959 \\
$^{113}$Sb & I.B. Selinov & Sov. At. Energy &\cite{1958Sel01}& LP & Moscow & Russia &1958 \\
$^{114}$Sb & I.B. Selinov & Sov. At. Energy &\cite{1959Sel01}& LP & Moscow & Russia &1959 \\
$^{115}$Sb & I.B. Selinov & Sov. At. Energy &\cite{1958Sel01}& LP & Moscow & Russia &1958 \\
$^{116}$Sb & G.M. Temmer & Phys. Rev. &\cite{1949Tem01}& LP & Berkeley & USA &1949 \\
$^{117}$Sb & K.D. Coleman& Phys. Rev. &\cite{1947Col01}& LP & Ohio State & USA &1947 \\
$^{118}$Sb & K.D. Coleman& Phys. Rev. &\cite{1947Col01}& LP & Ohio State & USA &1947 \\
$^{119}$Sb & K.D. Coleman& Phys. Rev. &\cite{1947Col01}& LP & Ohio State & USA &1947 \\
$^{120}$Sb & W. Bothe & Naturwiss. &\cite{1937Bot01}& NC & Heidelberg & Germany &1937 \\
$^{121}$Sb & F.W. Aston & Nature &\cite{1922Ast04}& MS & Cambridge & UK &1922 \\
$^{122}$Sb & J.J. Livingood & Phys. Rev. &\cite{1939Liv04}& NC & Berkeley & USA &1939 \\
$^{123}$Sb & F.W. Aston & Nature &\cite{1922Ast04}& MS & Cambridge & UK &1922 \\
$^{124}$Sb & J.J. Livingood & Phys. Rev. &\cite{1939Liv04}& LP & Berkeley & USA &1939 \\
$^{125}$Sb & C.W. Stanley & Nat. Nucl. Ener. Ser. &\cite{1951Sta01}& NF & Los Alamos & USA &1951 \\
$^{126}$Sb & I. Franz & Z. Naturforsch. &\cite{1956Fra02}& NF & Buenos Aires & Argentina &1956 \\
$^{127}$Sb & P. Abelson & Phys. Rev. &\cite{1939Abe01}& NF & Berkeley & USA &1939 \\
$^{128}$Sb & I. Franz & Z. Naturforsch. &\cite{1956Fra01}& NF & Buenos Aires & Argentina &1956 \\
$^{129}$Sb & P. Abelson & Phys. Rev. &\cite{1939Abe01}& NF & Berkeley & USA &1939 \\
$^{130}$Sb & E. Hagebo& J. Inorg. Nucl. Chem. &\cite{1962Hag01}& CPF & Amsterdam & Netherlands &1962 \\
$^{131}$Sb & A.C. Pappas& J. Inorg. Nucl. Chem. &\cite{1956Pap01}& NF & Oslo & Norway &1956 \\
$^{132}$Sb & A.C. Pappas& J. Inorg. Nucl. Chem. &\cite{1956Pap01}& NF & Oslo & Norway &1956 \\
$^{133}$Sb & P.O. Strom & Phys. Rev. &\cite{1966Str01}& NF & Naval Radiological Defense Laboratory & USA &1966 \\
$^{134}$Sb & L. Tomlinson & Phys. Lett. B &\cite{1967Tom01}& NF & Harwell & UK &1967 \\
$^{135}$Sb & C.E. Bemis & J. Inorg. Nucl. Chem. &\cite{1964Bem01}& NF & MIT & USA &1964 \\
$^{136}$Sb & E. Lund & Phys. Rev. C &\cite{1976Lun01}& NF & Studsvik & Sweden &1976 \\
$^{137}$Sb & M. Bernas & Phys. Lett. B &\cite{1994Ber01}& PF & Darmstadt & Germany &1994 \\
$^{138}$Sb & M. Bernas & Phys. Lett. B &\cite{1994Ber01}& PF & Darmstadt & Germany &1994 \\
$^{139}$Sb & M. Bernas & Phys. Lett. B &\cite{1994Ber01}& PF & Darmstadt & Germany &1994 \\
$^{140}$Sb& T. Ohnishi & J. Phys. Soc. Japan &\cite{2010Ohn01}& PF & RIKEN & Japan &2010 \\
 & & & & & & & \\
 & & & & & & & \\
$^{105}$Te& D. Seweryniak & Phys. Rev. C &\cite{2006Sew01}& FE & Argonne & USA &2006 \\
$^{106}$Te& D. Schardt & Nucl. Phys. A &\cite{1981Sch01}& FE & Darmstadt & Germany &1981 \\
$^{107}$Te & D. Schardt & Nucl. Phys. A &\cite{1979Sch02}& FE & Darmstadt & Germany &1979 \\
$^{108}$Te & D.D. Bogdanov & Sov. J. Nucl. Phys. &\cite{1974Bog02}& FE & Dubna & Russia &1974 \\
$^{109}$Te & V.A.Karnaukhov & Sov. J. Nucl. Phys. &\cite{1967Kar01}& FE & Dubna & Russia &1967 \\
$^{110}$Te & R. Kirchner & Phys. Lett. B &\cite{1977Kir01}& FE & Darmstadt & Germany &1977 \\
$^{111}$Te & V.A.Karnaukhov & Sov. J. Nucl. Phys. &\cite{1967Kar01}& FE & Dubna & Russia &1967 \\
$^{112}$Te & M.E.J. Wigmans & Phys. Rev. C &\cite{1976Wig01}& LP & Amsterdam & Netherlands &1976 \\
$^{113}$Te & A. Charvet & J. Phys. (Paris) &\cite{1974Cha01}& LP & Lyon & France &1974 \\
$^{114}$Te & O. Rahmouni & J. Phys. (Paris) &\cite{1968Rah01}& LP & Orsay & France &1968 \\
$^{115}$Te & I.P. Selinov& Sov. Phys. JETP &\cite{1961Sel01}& LP & Moscow & Russia &1961 \\
$^{116}$Te & M.Y. Kuznetsova & Sov. At. Energy &\cite{1958Kuz01}& SP & Dubna & Russia &1958 \\
$^{117}$Te & M.Y. Kuznetsova & Sov. At. Energy &\cite{1958Kuz01}& SP & Dubna & Russia &1958 \\
$^{118}$Te & M. Lindner & Phys. Rev. &\cite{1948Lin01}& LP & Berkeley & USA &1948 \\
$^{119}$Te & M. Lindner & Phys. Rev. &\cite{1948Lin01}& LP & Berkeley & USA &1948 \\
$^{120}$Te & A.J. Dempster & Phys. Rev. &\cite{1936Dem02}& MS & Chicago & USA &1936 \\
$^{121}$Te & G.T. Seaborg & Phys. Rev. &\cite{1939Sea02}& LP & Berkeley & USA &1939 \\
$^{122}$Te & K.T. Bainbridge& Phys. Rev. &\cite{1932Bai01}& MS & Bartol Research Foundation & USA &1932 \\
$^{123}$Te & K.T. Bainbridge& Phys. Rev. &\cite{1932Bai01}& MS & Bartol Research Foundation & USA &1932 \\
$^{124}$Te & K.T. Bainbridge& Phys. Rev. &\cite{1932Bai01}& MS & Bartol Research Foundation & USA &1932 \\
$^{125}$Te & F.W. Aston & Nature &\cite{1931Ast05}& MS & Cambridge & UK &1931 \\
$^{126}$Te & F.W. Aston & Nature &\cite{1924Ast03}& MS & Cambridge & UK &1924 \\
$^{127}$Te & J.J. Livingood & Phys. Rev. &\cite{1938Liv06}& LP & Berkeley & USA &1938 \\
$^{128}$Te & F.W. Aston & Nature &\cite{1924Ast03}& MS & Cambridge & UK &1924 \\
$^{129}$Te & G.T. Seaborg & Phys. Rev. &\cite{1939Sea02}& LP & Berkeley & USA &1939 \\
$^{130}$Te & F.W. Aston & Nature &\cite{1924Ast03}& MS & Cambridge & UK &1924 \\
$^{131}$Te & G.T. Seaborg & Phys. Rev. &\cite{1939Sea02}& LP & Berkeley & USA &1939 \\
$^{132}$Te & S. Katcoff & Phys. Rev. &\cite{1948Kat01}& NF & Los Alamos & USA &1948 \\
$^{133}$Te & C.S. Wu & Phys. Rev. &\cite{1940Wu01}& NF & Berkeley & USA &1940 \\
$^{134}$Te & S. Katcoff & Phys. Rev. &\cite{1948Kat01}& NF & Los Alamos & USA &1948 \\
$^{135}$Te & H.O. Denschlag & J. Inorg. Nucl. Chem. &\cite{1969Den01}& NF & Mainz & Germany &1969 \\
$^{136}$Te & B. Grapengiesser & J. Inorg. Nucl. Chem. &\cite{1974Gra01}& NF & Studsvik & Sweden &1974 \\
$^{137}$Te & M. Asghar & Nucl. Phys. A &\cite{1975Asg01}& NF & Grenoble & France &1975 \\
$^{138}$Te & M. Asghar & Nucl. Phys. A &\cite{1975Asg01}& NF & Grenoble & France &1975 \\
$^{139}$Te & M. Bernas & Phys. Lett. B &\cite{1994Ber01}& PF & Darmstadt & Germany &1994 \\
$^{140}$Te & M. Bernas & Phys. Lett. B &\cite{1994Ber01}& PF & Darmstadt & Germany &1994 \\
$^{141}$Te & M. Bernas & Phys. Lett. B &\cite{1994Ber01}& PF & Darmstadt & Germany &1994 \\
$^{142}$Te & M. Bernas & Phys. Lett. B &\cite{1994Ber01}& PF & Darmstadt & Germany &1994 \\
$^{143}$Te& T. Ohnishi & J. Phys. Soc. Japan &\cite{2010Ohn01}& PF & RIKEN & Japan &2010 \\
 & & & & & & & \\
 & & & & & & & \\
$^{108}$I& R.D. Page & Z. Phys. A &\cite{1991Pag01}& FE & Daresbury & UK &1991 \\
$^{109}$I & T. Faestermann & Phys. Lett. B &\cite{1984Fae01}& FE & Munich & Germany &1984 \\
$^{110}$I & R. Kirchner & Phys. Lett. B &\cite{1977Kir01}& FE & Darmstadt & Germany &1977 \\
$^{111}$I & R. Kirchner & Phys. Lett. B &\cite{1977Kir01}& FE & Darmstadt & Germany &1977 \\
$^{112}$I & R. Kirchner & Phys. Lett. B &\cite{1977Kir01}& FE & Darmstadt & Germany &1977 \\
$^{113}$I & R. Kirchner & Phys. Lett. B &\cite{1977Kir01}& FE & Darmstadt & Germany &1977 \\
$^{114}$I & R. Kirchner & Phys. Lett. B &\cite{1977Kir01}& FE & Darmstadt & Germany &1977 \\
$^{115}$I & P.G. Hansen & Phys. Lett. B &\cite{1969Han01}& SP & CERN & Switzerland &1969 \\
$^{116}$I & G.M. Gowdy & Phys. Rev. C &\cite{1976Gow01}& FE & Oak Ridge & USA &1976 \\
$^{117}$I & P.G. Hansen & Phys. Lett. B &\cite{1969Han01}& SP & CERN & Switzerland &1969 \\
$^{118}$I & P. Aagaard & J. Inorg. Nucl. Chem. &\cite{1957Aag01}& SP & Uppsala & Sweden &1957 \\
$^{119}$I & G.B. Rossi& Phys. Rev. &\cite{1954Ros01}& FE & Berkeley & USA &1954 \\
$^{120}$I & P. Aagaard & J. Inorg. Nucl. Chem. &\cite{1957Aag01}& SP & Uppsala & Sweden &1957 \\
$^{121}$I & L. Marquez & Phys. Rev. &\cite{1950Mar01}& LP & Berkeley & USA &1950 \\
$^{122}$I & L. Marquez & Phys. Rev. &\cite{1950Mar01}& LP & Berkeley & USA &1950 \\
$^{123}$I & A.C.G. Mitchell & Phys. Rev. &\cite{1949Mit01}& LP & Indiana & USA &1949 \\
$^{124}$I & J.J. Livingood & Phys. Rev. &\cite{1938Liv05}& LP & Berkeley & USA &1938 \\
$^{125}$I & L.E. Glendenin & Phys. Rev. &\cite{1947Gle01}& LP & MIT & USA &1947 \\
$^{126}$I & J.J. Livingood & Phys. Rev. &\cite{1938Liv06}& LP & Berkeley & USA &1938 \\
$^{127}$I & F.W. Aston & Nature &\cite{1920Ast05}& MS & Cambridge & UK &1920 \\
$^{128}$I & E. Fermi & Proc. Roy. Soc. A &\cite{1934Fer01}& LP & Rome & Italy &1934 \\
$^{129}$I & S. Katcoff & Phys. Rev. &\cite{1951Kat06}& NF & Oak Ridge & USA &1951 \\
$^{130}$I & J.J. Livingood & Phys. Rev. &\cite{1938Liv05}& LP & Berkeley & USA &1938 \\
$^{131}$I & G.T. Seaborg & Phys. Rev. &\cite{1939Sea02}& LP & Berkeley & USA &1939 \\
$^{132}$I & S. Katcoff & Phys. Rev. &\cite{1948Kat01}& NF & Los Alamos & USA &1948 \\
$^{133}$I & C.S. Wu & Phys. Rev. &\cite{1940Wu01}& NF & Berkeley & USA &1940 \\
$^{134}$I & S. Katcoff & Phys. Rev. &\cite{1948Kat01}& NF & Los Alamos & USA &1948 \\
$^{135}$I & C.S. Wu & Phys. Rev. &\cite{1940Wu01}& NF & Berkeley & USA &1940 \\
$^{136}$I & C.W. Stanley & Phys. Rev. &\cite{1949Sta01}& NF & Los Alamos & USA &1949 \\
$^{137}$I & W. Riezler & Naturwiss. &\cite{1943Rie01}& NF & Grundlsee & Austria &1943 \\
$^{138}$I & N. Sugarman & J. Chem. Phys. &\cite{1949Sug01}& NF & Argonne & USA &1949 \\
$^{139}$I & N. Sugarman & J. Chem. Phys. &\cite{1949Sug01}& NF & Argonne & USA &1949 \\
$^{140}$I & H.-D. Schussler & Radiochim. Acta &\cite{1972Sch01}& NF & Mainz & Germany &1972 \\
$^{141}$I & K.L. Kratz & Nucl. Phys. A &\cite{1974Kra01}& NF & Mainz & Germany &1974 \\
$^{142}$I & K.L. Kratz & Inorg. Nucl. Chem. Lett. &\cite{1975Kra01}& NF & Mainz & Germany &1975 \\
$^{143}$I & M. Bernas & Phys. Lett. B &\cite{1994Ber01}& PF & Darmstadt & Germany &1994 \\
$^{144}$I & M. Bernas & Phys. Lett. B &\cite{1994Ber01}& PF & Darmstadt & Germany &1994 \\
$^{145}$I& T. Ohnishi & J. Phys. Soc. Japan &\cite{2010Ohn01}& PF & RIKEN & Japan &2010 \\
 & & & & & & & \\
 & & & & & & & \\
$^{109}$Xe& S.N. Liddick & Phys. Rev. Lett. &\cite{2006Lid01}& FE & Oak Ridge & USA &2006 \\
$^{110}$Xe& D. Schardt & Nucl. Phys. A &\cite{1981Sch01}& FE & Darmstadt & Germany &1981 \\
$^{111}$Xe & D. Schardt & Nucl. Phys. A &\cite{1979Sch02}& FE & Darmstadt & Germany &1979 \\
$^{112}$Xe & E. Roeckl & Phys. Lett. B &\cite{1978Roe01}& FE & Darmstadt & Germany &1978 \\
$^{113}$Xe & E. Hagberg & Nucl. Phys. A &\cite{1973Hag01}& SP & CERN & Switzerland &1973 \\
$^{114}$Xe & R. Kirchner & Phys. Lett. B &\cite{1977Kir01}& FE & Darmstadt & Germany &1977 \\
$^{115}$Xe & P.G. Hansen & Phys. Lett. B &\cite{1969Han01}& SP & CERN & Switzerland &1969 \\
$^{116}$Xe & P.G. Hansen & Phys. Lett. B &\cite{1969Han01}& SP & CERN & Switzerland &1969 \\
$^{117}$Xe & P.G. Hansen & Phys. Lett. B &\cite{1969Han01}& SP & CERN & Switzerland &1969 \\
$^{118}$Xe & G. Andersson & Ark. Fysik &\cite{1965And01}& SP & CERN & Switzerland &1965 \\
$^{119}$Xe & G. Andersson & Ark. Fysik &\cite{1965And01}& SP & CERN & Switzerland &1965 \\
$^{120}$Xe & G. Andersson & Ark. Fysik &\cite{1965And01}& SP & CERN & Switzerland &1965 \\
$^{121}$Xe & B. Dropesky & Phys. Rev. &\cite{1952Dro01}& SP & Rochester & USA &1952 \\
$^{122}$Xe & B. Dropesky & Phys. Rev. &\cite{1952Dro01}& SP & Rochester & USA &1952 \\
$^{123}$Xe & B. Dropesky & Phys. Rev. &\cite{1952Dro01}& SP & Rochester & USA &1952 \\
$^{124}$Xe & F.W. Aston & Nature &\cite{1922Ast03}& MS & Cambridge & UK &1922 \\
$^{125}$Xe & D.L. Anderson & Phys. Rev. &\cite{1950And01}& LP & Harvard & USA &1950 \\
$^{126}$Xe & F.W. Aston & Nature &\cite{1922Ast03}& MS & Cambridge & UK &1922 \\
$^{127}$Xe & D.L. Anderson & Phys. Rev. &\cite{1950And01}& LP & Harvard & USA &1950 \\
$^{128}$Xe & F.W. Aston & Nature &\cite{1922Ast02}& MS & Cambridge & UK &1922 \\
$^{129}$Xe & F.W. Aston & Nature &\cite{1920Ast05}& MS & Cambridge & UK &1920 \\
$^{130}$Xe & F.W. Aston & Nature &\cite{1922Ast02}& MS & Cambridge & UK &1922 \\
$^{131}$Xe & F.W. Aston & Nature &\cite{1920Ast05}& MS & Cambridge & UK &1920 \\
$^{132}$Xe & F.W. Aston & Nature &\cite{1920Ast05}& MS & Cambridge & UK &1920 \\
$^{133}$Xe & C.S. Wu & Phys. Rev. &\cite{1940Wu01}& LP & Berkeley & USA &1940 \\
$^{134}$Xe & F.W. Aston & Nature &\cite{1920Ast05}& MS & Cambridge & UK &1920 \\
$^{135}$Xe & C.S. Wu & Phys. Rev. &\cite{1940Wu01}& LP & Berkeley & USA &1940 \\
$^{136}$Xe & F.W. Aston & Nature &\cite{1920Ast05}& MS & Cambridge & UK &1920 \\
$^{137}$Xe & W. Riezler & Naturwiss. &\cite{1943Rie01}& NF & Grundlsee & Austria &1943 \\
$^{138}$Xe & W. Seelmann-Eggebert & Naturwiss. &\cite{1943See03}& NF & Berlin & Germany &1943 \\
$^{139}$Xe & C.R. Dillard & Nat. Nucl. Ener. Ser. &\cite{1951Dil01}& NF & Argonne & USA &1951 \\
$^{140}$Xe & C.R. Dillard & Nat. Nucl. Ener. Ser. &\cite{1951Dil01}& NF & Argonne & USA &1951 \\
$^{141}$Xe & C.R. Dillard & Nat. Nucl. Ener. Ser. &\cite{1951Dil01}& NF & Argonne & USA &1951 \\
$^{142}$Xe & D.R. Wolfsberg & J. Inorg. Nucl. Chem. &\cite{1960Wol01}& NF & St. Louis & USA &1960 \\
$^{143}$Xe & C.R. Dillard & Nat. Nucl. Ener. Ser. &\cite{1951Dil01}& NF & Argonne & USA &1951 \\
$^{144}$Xe & U.C. Bergmann & Nucl. Phys. A &\cite{2003Ber01}& SP & CERN & Switzerland &2003 \\
$^{145}$Xe & U.C. Bergmann & Nucl. Phys. A &\cite{2003Ber01}& SP & CERN & Switzerland &2003 \\
$^{146}$Xe & W. Borchers & Phys. Lett. B &\cite{1989Bor02}& SP & CERN & Switzerland &1989 \\
$^{147}$Xe & M. Bernas & Phys. Lett. B &\cite{1994Ber01}& PF & Darmstadt & Germany &1994 \\
$^{148}$Xe& T. Ohnishi & J. Phys. Soc. Japan &\cite{2010Ohn01}& PF & RIKEN & Japan &2010 \\
\end{longtable}

\begin{longtable}{@{\extracolsep\fill}rllllllll@{}}
\caption{Chronological listing of the discovery of palladium, antimony, tellurium, iodine, and xenon isotopes. See page\ \pageref{tbl2te} for Explanation of Tables}
Submitted & Isotope & Author & Journal & Ref. & Method & Laboratory & Country & Year\\
\hline\\
\endfirsthead\\
\caption[]{(continued)}
Submitted & Isotope & Author & Journal & Ref. & Method & Laboratory & Country & Year\\
\hline\\
\endhead
11/30/1920 & $^{127}$I & F.W. Aston & Nature & \cite{1920Ast05} & MS & Cambridge & UK & 1920 \\
11/30/1920 & $^{129}$Xe & F.W. Aston & Nature & \cite{1920Ast05} & MS & Cambridge & UK & 1920 \\
11/30/1920 & $^{131}$Xe & F.W. Aston & Nature & \cite{1920Ast05} & MS & Cambridge & UK & 1920 \\
11/30/1920 & $^{132}$Xe & F.W. Aston & Nature & \cite{1920Ast05} & MS & Cambridge & UK & 1920 \\
11/30/1920 & $^{134}$Xe & F.W. Aston & Nature & \cite{1920Ast05} & MS & Cambridge & UK & 1920 \\
11/30/1920 & $^{136}$Xe & F.W. Aston & Nature & \cite{1920Ast05} & MS & Cambridge & UK & 1920 \\
6/7/1922 & $^{128}$Xe & F.W. Aston & Nature & \cite{1922Ast02} & MS & Cambridge & UK & 1922 \\
6/7/1922 & $^{130}$Xe & F.W. Aston & Nature & \cite{1922Ast02} & MS & Cambridge & UK & 1922 \\
11/6/1922 & $^{124}$Xe & F.W. Aston & Nature & \cite{1922Ast03} & MS & Cambridge & UK & 1922 \\
11/6/1922 & $^{126}$Xe & F.W. Aston & Nature & \cite{1922Ast03} & MS & Cambridge & UK & 1922 \\
11/16/1922 & $^{121}$Sb & F.W. Aston & Nature & \cite{1922Ast04} & MS & Cambridge & UK & 1922 \\
11/16/1922 & $^{123}$Sb & F.W. Aston & Nature & \cite{1922Ast04} & MS & Cambridge & UK & 1922 \\
11/4/1924 & $^{126}$Te & F.W. Aston & Nature & \cite{1924Ast03} & MS & Cambridge & UK & 1924 \\
11/4/1924 & $^{128}$Te & F.W. Aston & Nature & \cite{1924Ast03} & MS & Cambridge & UK & 1924 \\
11/4/1924 & $^{130}$Te & F.W. Aston & Nature & \cite{1924Ast03} & MS & Cambridge & UK & 1924 \\
5/29/1931 & $^{125}$Te & F.W. Aston & Nature & \cite{1931Ast05} & MS & Cambridge & UK & 1931 \\
3/4/1932 & $^{122}$Te & K.T. Bainbridge & Phys. Rev. & \cite{1932Bai01} & MS & Bartol Research Found. & USA & 1932 \\
3/4/1932 & $^{123}$Te & K.T. Bainbridge & Phys. Rev. & \cite{1932Bai01} & MS & Bartol Research Found. & USA & 1932 \\
3/4/1932 & $^{124}$Te & K.T. Bainbridge & Phys. Rev. & \cite{1932Bai01} & MS & Bartol Research Found. & USA & 1932 \\
7/25/1934 & $^{128}$I & E. Fermi & Proc. Roy. Soc. A & \cite{1934Fer01} & LP & Rome & Italy & 1934 \\
6/12/1935 & $^{102}$Pd & A.J. Dempster & Nature & \cite{1935Dem02} & MS & Chicago & USA & 1935 \\
6/12/1935 & $^{104}$Pd & A.J. Dempster & Nature & \cite{1935Dem02} & MS & Chicago & USA & 1935 \\
6/12/1935 & $^{105}$Pd & A.J. Dempster & Nature & \cite{1935Dem02} & MS & Chicago & USA & 1935 \\
6/12/1935 & $^{106}$Pd & A.J. Dempster & Nature & \cite{1935Dem02} & MS & Chicago & USA & 1935 \\
6/12/1935 & $^{108}$Pd & A.J. Dempster & Nature & \cite{1935Dem02} & MS & Chicago & USA & 1935 \\
6/12/1935 & $^{110}$Pd & A.J. Dempster & Nature & \cite{1935Dem02} & MS & Chicago & USA & 1935 \\
6/23/1936 & $^{120}$Te & A.J. Dempster & Phys. Rev. & \cite{1936Dem02} & MS & Chicago & USA & 1936 \\
2/9/1937 & $^{120}$Sb & W. Bothe & Naturwiss. & \cite{1937Bot01} & NC & Heidelberg & Germany & 1937 \\
8/23/1937 & $^{109}$Pd & J.D. Kraus & Phys. Rev. & \cite{1937Kra01} & LP & Michigan & USA & 1937 \\
8/23/1937 & $^{111}$Pd & J.D. Kraus & Phys. Rev. & \cite{1937Kra01} & LP & Michigan & USA & 1937 \\
6/1/1938 & $^{127}$Te & J.J. Livingood & Phys. Rev. & \cite{1938Liv06} & LP & Berkeley & USA & 1938 \\
6/1/1938 & $^{126}$I & J.J. Livingood & Phys. Rev. & \cite{1938Liv06} & LP & Berkeley & USA & 1938 \\
9/7/1938 & $^{124}$I & J.J. Livingood & Phys. Rev. & \cite{1938Liv05} & LP & Berkeley & USA & 1938 \\
9/7/1938 & $^{130}$I & J.J. Livingood & Phys. Rev. & \cite{1938Liv05} & LP & Berkeley & USA & 1938 \\
1/30/1939 & $^{122}$Sb & J.J. Livingood & Phys. Rev. & \cite{1939Liv04} & NC & Berkeley & USA & 1939 \\
1/30/1939 & $^{124}$Sb & J.J. Livingood & Phys. Rev. & \cite{1939Liv04} & LP & Berkeley & USA & 1939 \\
3/31/1939 & $^{121}$Te & G.T. Seaborg & Phys. Rev. & \cite{1939Sea02} & LP & Berkeley & USA & 1939 \\
3/31/1939 & $^{129}$Te & G.T. Seaborg & Phys. Rev. & \cite{1939Sea02} & LP & Berkeley & USA & 1939 \\
3/31/1939 & $^{131}$Te & G.T. Seaborg & Phys. Rev. & \cite{1939Sea02} & LP & Berkeley & USA & 1939 \\
3/31/1939 & $^{131}$I & G.T. Seaborg & Phys. Rev. & \cite{1939Sea02} & LP & Berkeley & USA & 1939 \\
4/11/1939 & $^{127}$Sb & P. Abelson & Phys. Rev. & \cite{1939Abe01} & NF & Berkeley & USA & 1939 \\
4/11/1939 & $^{129}$Sb & P. Abelson & Phys. Rev. & \cite{1939Abe01} & NF & Berkeley & USA & 1939 \\
10/17/1940 & $^{133}$Te & C.S. Wu & Phys. Rev. & \cite{1940Wu01} & NF & Berkeley & USA & 1940 \\
10/17/1940 & $^{133}$I & C.S. Wu & Phys. Rev. & \cite{1940Wu01} & NF & Berkeley & USA & 1940 \\
10/17/1940 & $^{135}$I & C.S. Wu & Phys. Rev. & \cite{1940Wu01} & NF & Berkeley & USA & 1940 \\
10/17/1940 & $^{133}$Xe & C.S. Wu & Phys. Rev. & \cite{1940Wu01} & LP & Berkeley & USA & 1940 \\
10/17/1940 & $^{135}$Xe & C.S. Wu & Phys. Rev. & \cite{1940Wu01} & LP & Berkeley & USA & 1940 \\
4/25/1943 & $^{137}$I & W. Riezler & Naturwiss. & \cite{1943Rie01} & NF & Grundlsee & Austria & 1943 \\
4/25/1943 & $^{137}$Xe & W. Riezler & Naturwiss. & \cite{1943Rie01} & NF & Grundlsee & Austria & 1943 \\
8/25/1943 & $^{138}$Xe & W. Seelmann-Eggebert & Naturwiss. & \cite{1943See03} & NF & Berlin & Germany & 1943 \\
1/1/1945 & $^{112}$Pd & J.A. Seiler & Nat. Nucl. Ener. Ser. & \cite{1951Sei01} & NF & Argonne & USA & 1951 \\
1/11/1945 & $^{139}$Xe & C.R. Dillard & Nat. Nucl. Ener. Ser. & \cite{1951Dil01} & NF & Argonne & USA & 1951 \\
1/11/1945 & $^{140}$Xe & C.R. Dillard & Nat. Nucl. Ener. Ser. & \cite{1951Dil01} & NF & Argonne & USA & 1951 \\
1/11/1945 & $^{141}$Xe & C.R. Dillard & Nat. Nucl. Ener. Ser. & \cite{1951Dil01} & NF & Argonne & USA & 1951 \\
1/11/1945 & $^{143}$Xe & C.R. Dillard & Nat. Nucl. Ener. Ser. & \cite{1951Dil01} & NF & Argonne & USA & 1951 \\
3/1/1946 & $^{125}$Sb & C.W. Stanley & Nat. Nucl. Ener. Ser. & \cite{1951Sta01} & NF & Los Alamos & USA & 1951 \\
4/9/1947 & $^{125}$I & L.E. Glendenin & Phys. Rev. & \cite{1947Gle01} & LP & MIT & USA & 1947 \\
8/8/1947 & $^{117}$Sb & K.D. Coleman & Phys. Rev. & \cite{1947Col01} & LP & Ohio State & USA & 1947 \\
8/8/1947 & $^{118}$Sb & K.D. Coleman & Phys. Rev. & \cite{1947Col01} & LP & Ohio State & USA & 1947 \\
8/8/1947 & $^{119}$Sb & K.D. Coleman & Phys. Rev. & \cite{1947Col01} & LP & Ohio State & USA & 1947 \\
3/10/1948 & $^{118}$Te & M. Lindner & Phys. Rev. & \cite{1948Lin01} & LP & Berkeley & USA & 1948 \\
3/10/1948 & $^{119}$Te & M. Lindner & Phys. Rev. & \cite{1948Lin01} & LP & Berkeley & USA & 1948 \\
4/5/1948 & $^{100}$Pd & M. Lindner & Phys. Rev. & \cite{1948Lin02} & SP & Berkeley & USA & 1948 \\
4/5/1948 & $^{101}$Pd & M. Lindner & Phys. Rev. & \cite{1948Lin02} & SP & Berkeley & USA & 1948 \\
4/23/1948 & $^{138}$I & N. Sugarman & J. Chem. Phys. & \cite{1949Sug01} & NF & Argonne & USA & 1949 \\
4/23/1948 & $^{139}$I & N. Sugarman & J. Chem. Phys. & \cite{1949Sug01} & NF & Argonne & USA & 1949 \\
5/28/1948 & $^{132}$Te & S. Katcoff & Phys. Rev. & \cite{1948Kat01} & NF & Los Alamos & USA & 1948 \\
5/28/1948 & $^{134}$Te & S. Katcoff & Phys. Rev. & \cite{1948Kat01} & NF & Los Alamos & USA & 1948 \\
5/28/1948 & $^{132}$I & S. Katcoff & Phys. Rev. & \cite{1948Kat01} & NF & Los Alamos & USA & 1948 \\
5/28/1948 & $^{134}$I & S. Katcoff & Phys. Rev. & \cite{1948Kat01} & NF & Los Alamos & USA & 1948 \\
12/2/1948 & $^{136}$I & C.W. Stanley & Phys. Rev. & \cite{1949Sta01} & NF & Los Alamos & USA & 1949 \\
8/1/1949 & $^{116}$Sb & G.M. Temmer & Phys. Rev. & \cite{1949Tem01} & LP & Berkeley & USA & 1949 \\
8/15/1949 & $^{123}$I & A.C.G. Mitchell & Phys. Rev. & \cite{1949Mit01} & LP & Indiana & USA & 1949 \\
11/17/1949 & $^{125}$Xe & D.L. Anderson & Phys. Rev. & \cite{1950And01} & LP & Harvard & USA & 1950 \\
11/17/1949 & $^{127}$Xe & D.L. Anderson & Phys. Rev. & \cite{1950And01} & LP & Harvard & USA & 1950 \\
1/13/1950 & $^{121}$I & L. Marquez & Phys. Rev. & \cite{1950Mar01} & LP & Berkeley & USA & 1950 \\
1/13/1950 & $^{122}$I & L. Marquez & Phys. Rev. & \cite{1950Mar01} & LP & Berkeley & USA & 1950 \\
4/26/1950 & $^{103}$Pd & J. Y. Mei & Phys. Rev. & \cite{1950Mei02} & LP & Indiana & USA & 1950 \\
2/15/1951 & $^{129}$I & S. Katcoff & Phys. Rev. & \cite{1951Kat06} & NF & Oak Ridge & USA & 1951 \\
9/17/1952 & $^{121}$Xe & B. Dropesky & Phys. Rev. & \cite{1952Dro01} & SP & Rochester & USA & 1952 \\
9/17/1952 & $^{122}$Xe & B. Dropesky & Phys. Rev. & \cite{1952Dro01} & SP & Rochester & USA & 1952 \\
9/17/1952 & $^{123}$Xe & B. Dropesky & Phys. Rev. & \cite{1952Dro01} & SP & Rochester & USA & 1952 \\
10/9/1953 & $^{113}$Pd & H. G. Hicks & Phys. Rev. & \cite{1954Hic01} & CPF & Berkeley & USA & 1954 \\
11/25/1953 & $^{119}$I & G.B. Rossi & Phys. Rev. & \cite{1954Ros01} & FE & Berkeley & USA & 1954 \\
5/5/1955 & $^{98}$Pd & A.H.W. Aten & Physica & \cite{1955Ate01} & LP & Amsterdam & Netherlands & 1955 \\
5/5/1955 & $^{99}$Pd & A.H.W. Aten & Physica & \cite{1955Ate01} & LP & Amsterdam & Netherlands & 1955 \\
6/9/1955 & $^{131}$Sb & A.C. Pappas & J. Inorg. Nucl. Chem. & \cite{1956Pap01} & NF & Oslo & Norway & 1956 \\
6/9/1955 & $^{132}$Sb & A.C. Pappas & J. Inorg. Nucl. Chem. & \cite{1956Pap01} & NF & Oslo & Norway & 1956 \\
11/17/1956 & $^{126}$Sb & I. Franz & Z. Naturforsch. & \cite{1956Fra02} & NF & Buenos Aires & Argentina & 1956 \\
11/17/1956 & $^{128}$Sb & I. Franz & Z. Naturforsch. & \cite{1956Fra01} & NF & Buenos Aires & Argentina & 1956 \\
4/16/1957 & $^{118}$I & P. Aagaard & J. Inorg. Nucl. Chem. & \cite{1957Aag01} & SP & Uppsala & Sweden & 1957 \\
4/16/1957 & $^{120}$I & P. Aagaard & J. Inorg. Nucl. Chem. & \cite{1957Aag01} & SP & Uppsala & Sweden & 1957 \\
9/23/1957 & $^{107}$Pd & U. Schindewolf & Phys. Rev. & \cite{1958Sch01} & NC & MIT & USA & 1958 \\
12/11/1957 & $^{116}$Te & M.Y. Kuznetsova & Sov. At. Energy & \cite{1958Kuz01} & SP & Dubna & Russia & 1958 \\
12/11/1957 & $^{117}$Te & M.Y. Kuznetsova & Sov. At. Energy & \cite{1958Kuz01} & SP & Dubna & Russia & 1958 \\
1/31/1958 & $^{114}$Pd & J. M. Alexander & Phys. Rev. & \cite{1958Ale01} & CPF & MIT & USA & 1958 \\
1/31/1958 & $^{115}$Pd & J. M. Alexander & Phys. Rev. & \cite{1958Ale01} & CPF & MIT & USA & 1958 \\
9/4/1958 & $^{113}$Sb & I.B. Selinov & Sov. At. Energy & \cite{1958Sel01} & LP & Moscow & Russia & 1958 \\
9/4/1958 & $^{115}$Sb & I.B. Selinov & Sov. At. Energy & \cite{1958Sel01} & LP & Moscow & Russia & 1958 \\
2/7/1959 & $^{112}$Sb & I.B. Selinov & Sov. At. Energy & \cite{1959Sel01} & LP & Moscow & Russia & 1959 \\
2/7/1959 & $^{114}$Sb & I.B. Selinov & Sov. At. Energy & \cite{1959Sel01} & LP & Moscow & Russia & 1959 \\
2/17/1959 & $^{142}$Xe & D.R. Wolfsberg & J. Inorg. Nucl. Chem. & \cite{1960Wol01} & NF & St. Louis & USA & 1960 \\
3/28/1960 & $^{115}$Te & I.P. Selinov & Sov. Phys. JETP & \cite{1961Sel01} & LP & Moscow & Russia & 1961 \\
7/24/1961 & $^{130}$Sb & E. Hagebo & J. Inorg. Nucl. Chem. & \cite{1962Hag01} & CPF & Amsterdam & Netherlands & 1962 \\
5/16/1963 & $^{135}$Sb & C.E. Bemis & J. Inorg. Nucl. Chem. & \cite{1964Bem01} & NF & MIT & USA & 1964 \\
6/3/1964 & $^{118}$Xe & G. Andersson & Ark. Fysik & \cite{1965And01} & SP & CERN & Switzerland & 1965 \\
6/3/1964 & $^{119}$Xe & G. Andersson & Ark. Fysik & \cite{1965And01} & SP & CERN & Switzerland & 1965 \\
6/3/1964 & $^{120}$Xe & G. Andersson & Ark. Fysik & \cite{1965And01} & SP & CERN & Switzerland & 1965 \\
12/6/1965 & $^{133}$Sb & P.O. Strom & Phys. Rev. & \cite{1966Str01} & NF & Naval Rad. Defense Lab. & USA & 1966 \\
1/3/1966 & $^{109}$Te & V.A.Karnaukhov & Sov. J. Nucl. Phys. & \cite{1967Kar01} & FE & Dubna & Russia & 1967 \\
1/3/1966 & $^{111}$Te & V.A.Karnaukhov & Sov. J. Nucl. Phys. & \cite{1967Kar01} & FE & Dubna & Russia & 1967 \\
10/12/1967 & $^{134}$Sb & L. Tomlinson & Phys. Lett. B & \cite{1967Tom01} & NF & Harwell & UK & 1967 \\
1/11/1968 & $^{114}$Te & O. Rahmouni & J. Phys. (Paris) & \cite{1968Rah01} & LP & Orsay & France & 1968 \\
3/25/1968 & $^{117}$Pd & H.V. Weiss & Phys. Rev. & \cite{1968Wei02} & NF & Naval Rad. Defense Lab.  & USA & 1968 \\
11/22/1968 & $^{115}$I & P.G. Hansen & Phys. Lett. B & \cite{1969Han01} & SP & CERN & Switzerland & 1969 \\
11/22/1968 & $^{117}$I & P.G. Hansen & Phys. Lett. B & \cite{1969Han01} & SP & CERN & Switzerland & 1969 \\
11/22/1968 & $^{115}$Xe & P.G. Hansen & Phys. Lett. B & \cite{1969Han01} & SP & CERN & Switzerland & 1969 \\
11/22/1968 & $^{116}$Xe & P.G. Hansen & Phys. Lett. B & \cite{1969Han01} & SP & CERN & Switzerland & 1969 \\
11/22/1968 & $^{117}$Xe & P.G. Hansen & Phys. Lett. B & \cite{1969Han01} & SP & CERN & Switzerland & 1969 \\
12/12/1968 & $^{135}$Te & H.O. Denschlag & J. Inorg. Nucl. Chem. & \cite{1969Den01} & NF & Mainz & Germany & 1969 \\
6/12/1969 & $^{97}$Pd & A.H.W. Aten & Radiochim. Acta & \cite{1969Ate01} & LP & Amsterdam & Netherlands & 1969 \\
7/1/1969 & $^{118}$Pd & H.V. Weiss & Phys. Rev. & \cite{1969Wei01} & NF & Naval Rad. Defense Lab.  & USA & 1969 \\
5/12/1970 & $^{116}$Pd & E. Cheifetz & Phys. Rev. Lett. & \cite{1970Che01} & SF & Berkeley & USA & 1970 \\
1/26/1972 & $^{140}$I & H.-D. Schussler & Radiochim. Acta & \cite{1972Sch01} & NF & Mainz & Germany & 1972 \\
4/14/1972 & $^{110}$Sb & K. Miyano & J. Phys. Soc. Japan & \cite{1972Miy01} & LP & Tokyo & Japan & 1972 \\
6/12/1972 & $^{111}$Sb & M. Singh & Nucl. Phys. A & \cite{1972Sin01} & LP & UCLA & USA & 1972 \\
1/9/1973 & $^{108}$Te & D.D. Bogdanov & Sov. J. Nucl. Phys. & \cite{1974Bog02} & FE & Dubna & Russia & 1974 \\
3/22/1973 & $^{113}$Xe & E. Hagberg & Nucl. Phys. A & \cite{1973Hag01} & SP & CERN & Switzerland & 1973 \\
1/10/1974 & $^{136}$Te & B. Grapengiesser & J. Inorg. Nucl. Chem. & \cite{1974Gra01} & NF & Studsvik & Sweden & 1974 \\
1/18/1974 & $^{113}$Te & A. Charvet & J. Phys. (Paris) & \cite{1974Cha01} & LP & Lyon & France & 1974 \\
3/6/1974 & $^{141}$I & K.L. Kratz & Nucl. Phys. A & \cite{1974Kra01} & NF & Mainz & Germany & 1974 \\
12/19/1974 & $^{142}$I & K.L. Kratz & Inorg. Nucl. Chem. Lett. & \cite{1975Kra01} & NF & Mainz & Germany & 1975 \\
2/17/1975 & $^{137}$Te & M. Asghar & Nucl. Phys. A & \cite{1975Asg01} & NF & Grenoble & France & 1975 \\
2/17/1975 & $^{138}$Te & M. Asghar & Nucl. Phys. A & \cite{1975Asg01} & NF & Grenoble & France & 1975 \\
9/8/1975 & $^{116}$I & G.M. Gowdy & Phys. Rev. C & \cite{1976Gow01} & FE & Oak Ridge & USA & 1976 \\
10/21/1975 & $^{136}$Sb & E. Lund & Phys. Rev. C & \cite{1976Lun01} & NF & Studsvik & Sweden & 1976 \\
10/28/1975 & $^{112}$Te & M.E.J. Wigmans & Phys. Rev. C & \cite{1976Wig01} & LP & Amsterdam & Netherlands & 1976 \\
8/27/1976 & $^{108}$Sb & K. Oxorn & Z. Phys. A & \cite{1976Oxo01} & LP & McGill & Canada & 1976 \\
8/27/1976 & $^{109}$Sb & K. Oxorn & Z. Phys. A & \cite{1976Oxo01} & LP & McGill & Canada & 1976 \\
7/8/1977 & $^{110}$Te & R. Kirchner & Phys. Lett. B & \cite{1977Kir01} & FE & Darmstadt & Germany & 1977 \\
7/8/1977 & $^{110}$I & R. Kirchner & Phys. Lett. B & \cite{1977Kir01} & FE & Darmstadt & Germany & 1977 \\
7/8/1977 & $^{111}$I & R. Kirchner & Phys. Lett. B & \cite{1977Kir01} & FE & Darmstadt & Germany & 1977 \\
7/8/1977 & $^{112}$I & R. Kirchner & Phys. Lett. B & \cite{1977Kir01} & FE & Darmstadt & Germany & 1977 \\
7/8/1977 & $^{113}$I & R. Kirchner & Phys. Lett. B & \cite{1977Kir01} & FE & Darmstadt & Germany & 1977 \\
7/8/1977 & $^{114}$I & R. Kirchner & Phys. Lett. B & \cite{1977Kir01} & FE & Darmstadt & Germany & 1977 \\
7/8/1977 & $^{114}$Xe & R. Kirchner & Phys. Lett. B & \cite{1977Kir01} & FE & Darmstadt & Germany & 1977 \\
7/11/1978 & $^{112}$Xe & E. Roeckl & Phys. Lett. B & \cite{1978Roe01} & FE & Darmstadt & Germany & 1978 \\
4/17/1979 & $^{107}$Te & D. Schardt & Nucl. Phys. A & \cite{1979Sch02} & FE & Darmstadt & Germany & 1979 \\
4/17/1979 & $^{111}$Xe & D. Schardt & Nucl. Phys. A & \cite{1979Sch02} & FE & Darmstadt & Germany & 1979 \\
6/30/1980 & $^{96}$Pd & N.K. Aras & J. Phys. G & \cite{1980Ara01} & LP & Maryland & USA & 1980 \\
8/11/1980 & $^{95}$Pd & E. Nolte & Phys. Lett. B & \cite{1980Nol01} & FE & Munich & Germany & 1980 \\
3/30/1981 & $^{106}$Te & D. Schardt & Nucl. Phys. A & \cite{1981Sch01} & FE & Darmstadt & Germany & 1981 \\
3/30/1981 & $^{110}$Xe & D. Schardt & Nucl. Phys. A & \cite{1981Sch01} & FE & Darmstadt & Germany & 1981 \\
8/26/1981 & $^{106}$Sb & A. Plochocki & Phys. Lett. B & \cite{1981Plo01} & FE & Darmstadt & Germany & 1981 \\
6/7/1982 & $^{94}$Pd & W. Kurcewicz & Z. Phys. A & \cite{1982Kur01} & FE & Darmstadt & Germany & 1982 \\
12/16/1983 & $^{109}$I & T. Faestermann & Phys. Lett. B & \cite{1984Fae01} & FE & Munich & Germany & 1984 \\
10/11/1988 & $^{146}$Xe & W. Borchers & Phys. Lett. B & \cite{1989Bor02} & SP & CERN & Switzerland & 1989 \\
5/14/1990 & $^{108}$I & R.D. Page & Z. Phys. A & \cite{1991Pag01} & FE & Daresbury & UK & 1991 \\
8/20/1990 & $^{119}$Pd & H. Penttil\"a & Z. Phys. A & \cite{1991Pen02} & CPF & Jyv\"askyl\"a & Finland & 1991 \\
8/6/1992 & $^{120}$Pd & Z. Janas & Nucl. Phys. A & \cite{1993Jan01} & CPF & Jyv\"askyl\"a & Finland & 1993 \\
9/14/1993 & $^{107}$Sb & D. Seweryniak & Phys. Lett. B & \cite{1994Sew01} & FE & Roskilde & Denmark & 1994 \\
11/1/1993 & $^{105}$Sb & R.J. Tighe & Phys. Rev. C & \cite{1994Tig01} & FE & Berkeley & USA & 1994 \\
3/3/1994 & $^{121}$Pd & M. Bernas & Phys. Lett. B & \cite{1994Ber01} & PF & Darmstadt & Germany & 1994 \\
3/3/1994 & $^{122}$Pd & M. Bernas & Phys. Lett. B & \cite{1994Ber01} & PF & Darmstadt & Germany & 1994 \\
3/3/1994 & $^{123}$Pd & M. Bernas & Phys. Lett. B & \cite{1994Ber01} & PF & Darmstadt & Germany & 1994 \\
3/3/1994 & $^{137}$Sb & M. Bernas & Phys. Lett. B & \cite{1994Ber01} & PF & Darmstadt & Germany & 1994 \\
3/3/1994 & $^{138}$Sb & M. Bernas & Phys. Lett. B & \cite{1994Ber01} & PF & Darmstadt & Germany & 1994 \\
3/3/1994 & $^{139}$Sb & M. Bernas & Phys. Lett. B & \cite{1994Ber01} & PF & Darmstadt & Germany & 1994 \\
3/3/1994 & $^{139}$Te & M. Bernas & Phys. Lett. B & \cite{1994Ber01} & PF & Darmstadt & Germany & 1994 \\
3/3/1994 & $^{140}$Te & M. Bernas & Phys. Lett. B & \cite{1994Ber01} & PF & Darmstadt & Germany & 1994 \\
3/3/1994 & $^{141}$Te & M. Bernas & Phys. Lett. B & \cite{1994Ber01} & PF & Darmstadt & Germany & 1994 \\
3/3/1994 & $^{142}$Te & M. Bernas & Phys. Lett. B & \cite{1994Ber01} & PF & Darmstadt & Germany & 1994 \\
3/3/1994 & $^{143}$I & M. Bernas & Phys. Lett. B & \cite{1994Ber01} & PF & Darmstadt & Germany & 1994 \\
3/3/1994 & $^{144}$I & M. Bernas & Phys. Lett. B & \cite{1994Ber01} & PF & Darmstadt & Germany & 1994 \\
3/3/1994 & $^{147}$Xe & M. Bernas & Phys. Lett. B & \cite{1994Ber01} & PF & Darmstadt & Germany & 1994 \\
4/15/1994 & $^{92}$Pd & M. Hencheck & Phys. Rev. C & \cite{1994Hen01} & PF & Michigan State & USA & 1994 \\
4/15/1994 & $^{93}$Pd & M. Hencheck & Phys. Rev. C & \cite{1994Hen01} & PF & Michigan State & USA & 1994 \\
10/25/1994 & $^{91}$Pd & K. Rykaczewski & Phys. Rev. C & \cite{1995Ryk01} & PF & GANIL & France & 1995 \\
10/25/1994 & $^{103}$Sb & K. Rykaczewski & Phys. Rev. C & \cite{1995Ryk01} & PF & GANIL & France & 1995 \\
10/25/1994 & $^{104}$Sb & K. Rykaczewski & Phys. Rev. C & \cite{1995Ryk01} & PF & GANIL & France & 1995 \\
7/25/1997 & $^{124}$Pd & M. Bernas & Phys. Lett. B & \cite{1997Ber01} & PF & Darmstadt & Germany & 1997 \\
5/14/2002 & $^{144}$Xe & U.C. Bergmann & Nucl. Phys. A & \cite{2003Ber01} & SP & CERN & Switzerland & 2003 \\
5/14/2002 & $^{145}$Xe & U.C. Bergmann & Nucl. Phys. A & \cite{2003Ber01} & SP & CERN & Switzerland & 2003 \\
3/9/2006 & $^{105}$Te & D. Seweryniak & Phys. Rev. C & \cite{2006Sew01} & FE & Argonne & USA & 2006 \\
5/16/2006 & $^{109}$Xe & S.N. Liddick & Phys. Rev. Lett. & \cite{2006Lid01} & FE & Oak Ridge & USA & 2006 \\
6/4/2008 & $^{125}$Pd & T. Ohnishi & J. Phys. Soc. Japan & \cite{2008Ohn01} & PF & RIKEN & Japan & 2008 \\
6/4/2008 & $^{126}$Pd & T. Ohnishi & J. Phys. Soc. Japan & \cite{2008Ohn01} & PF & RIKEN & Japan & 2008 \\
3/19/2010 & $^{127}$Pd & T. Ohnishi & J. Phys. Soc. Japan & \cite{2010Ohn01} & PF & RIKEN & Japan & 2010 \\
3/19/2010 & $^{128}$Pd & T. Ohnishi & J. Phys. Soc. Japan & \cite{2010Ohn01} & PF & RIKEN & Japan & 2010 \\
3/19/2010 & $^{140}$Sb & T. Ohnishi & J. Phys. Soc. Japan & \cite{2010Ohn01} & PF & RIKEN & Japan & 2010 \\
3/19/2010 & $^{143}$Te & T. Ohnishi & J. Phys. Soc. Japan & \cite{2010Ohn01} & PF & RIKEN & Japan & 2010 \\
3/19/2010 & $^{145}$I & T. Ohnishi & J. Phys. Soc. Japan & \cite{2010Ohn01} & PF & RIKEN & Japan & 2010 \\
3/19/2010 & $^{148}$Xe & T. Ohnishi & J. Phys. Soc. Japan & \cite{2010Ohn01} & PF & RIKEN & Japan & 2010 \\
\end{longtable}

\end{document}